\documentclass[showpacs,prl,onecolumn,aps,superscriptaddress,preprintnumbers,nofootinbib]{revtex4}
\usepackage[T1]{fontenc}
\usepackage[latin9]{inputenc}
\usepackage{graphicx,hyperref}

\usepackage{amsmath,amssymb}
\usepackage{epsfig}
\usepackage{graphicx}
\usepackage{amsmath}
\usepackage{amsfonts}
\usepackage{mathrsfs}
\usepackage{pifont}
\usepackage{relsize}
\usepackage{mathtools}
\def\slashchar#1{\setbox0=\hbox{$#1$}     		% set a box for #1
   \dimen0=\wd0                                 	% and get its size
   \setbox1=\hbox{/} \dimen1=\wd1               	% get size of /
   \ifdim\dimen0>\dimen1                        	% #1 is bigger
      \rlap{\hbox to \dimen0{\hfil/\hfil}}      	% so center / in box
      #1                                        	% and print #1
   \else                                        	% / is bigger
      \rlap{\hbox to \dimen1{\hfil$#1$\hfil}}   	% so center #1
      /                                         	% and print /
   \fi}

\renewcommand{\vec}{\boldsymbol}
\newcommand{\beq}{\begin{equation}}
\newcommand{\eeq}{\end{equation}}
\newcommand{\bea}{\begin{eqnarray}}
\newcommand{\eea}{\end{eqnarray}}
\newcommand{\baa}{\begin{array}}
\newcommand{\eaa}{\end{array}}

\def\eq#1{{Eq.~(\ref{#1})}}

\def\fig#1{{Fig.~\ref{#1}}}
\newcommand{\intl}{\int\limits}

\newcommand{\bas}{\bar{\alpha}_S}

\newcommand{\nn}{\nonumber}

\newcommand{\h}{\frac{1}{2}}

\newcommand{\x}{\vec{x}}
\newcommand{\vb}{\vec{b}}

\newcommand{\Lb}{\left(}
\newcommand{\Rb}{\right)}
\def\pom{{I\!\!P}}

\newcommand{\pp}{\partial}

\renewcommand{\vec}[1]{\boldsymbol{#1}}

\newcommand{\dY}{\delta \tilde{Y}}

\numberwithin{equation}{section}

\begin{document}
%%%%%%%%%%%%%%%%%%%%%%%%%%%%%%%%%%%%%%%%%%%%%%%%%%%%%%%%%%%%%%%%%%%%%%%
\title{  Multiplicity distribution of produced gluons in deep inelastic scattering: main equations and their homotopy solutions  for heavy nuclei}
\author{Carlos Contreras}
\email{carlos.contreras@usm.cl}
\affiliation{Departamento de F\'isica, Universidad T\'ecnica Federico Santa Mar\'ia,  Avenida Espa\~na 1680, Casilla 110-V, Valpara\'iso, Chile}
\author{Jos\'e Garrido}
\email{jose.garridom@sansano.usm.cl}
\affiliation{Departamento de F\'isica, Universidad T\'ecnica Federico Santa Mar\'ia,  Avenida Espa\~na 1680, Casilla 110-V, Valpara\'iso, Chile}
\affiliation{Instituto de F\'isica, Pontificia Universidad Cat\'olica de Valpara\'iso, Avenida Universidad 330, Curauma, Valpara\'iso, Chile}
\author{ Eugene Levin}
\email{leving@tauex.tau.ac.il}
\affiliation{Department of Particle Physics, School of Physics and Astronomy,
Raymond and Beverly Sackler
 Faculty of Exact Science, Tel Aviv University, Tel Aviv, 69978, Israel}
\date{\today}

%%%%%%%%%%%%%%%%%%%%%%%%%%%%%%%%%%%%%%%%%%%%%%%%%%%%%%%%%%%%%%%%

\begin{abstract}
In this paper we discuss the multiplicity distribution in the deep inelastic processes in the frame work of high energy QCD. We obtained three results. First, we get the new derivation of the equations for the cross sections of productions of $n$-cut Pomerons in the final states ($\sigma_n$). These equations coincide with the equations that have been derived using the Abramovsky, Gribov and Kancheli (AGK) cutting rules but based on the dipole approach to QCD. Second, we developed the homotopy approach for finding the solutions to these equations. It consists with the analytic solution for the first iteration and the converge procedure of calculating the next iterations using computing. Third, we found the analytical solution for  $\sigma_n$ at large  $n\,\gtrsim\,N(z) = 2 N_0 \,z\,\exp( z^2/(2 \kappa))$ with $z = \ln( r^2\,Q^2_s )$. Using this solution we calculate the entropy of the produced gluons at large $z$: $S_E = \ln \left( N(z)\right)$,  where the saturation momentum $Q_s$ and  all constants are discussed in the text. 

 \end{abstract}
\maketitle

\vspace{-0.5cm}
\tableofcontents

%\flushbottom

%\pagestyle{empty}

%\mbox{}

%\pagestyle{plain}

%\setcounter{page}{1}
%%%%%%%%%%%%%%%%%%%%%%%%%%%%%%%%%%%%%%%%%%%%%%%%%%%%%%%%%%
\section{Introduction}

%%%%%%%%%%%%%%%%%%%%%%%%%%%%%%%%%%%%%%%%%%%%%%%%%%%%%%%%%%
We continue\cite{CLMNEW,CGLM} to develop the  approach for the  solution of the non-linear evolution equations that govern the deep inelastic processes (DIS) in QCD. In this paper we are going to discuss the multiplicity distribution of the produced gluons in QCD. This distribution has become a hot subject during the past several years\cite{KUT,PES,KOLU1,PESE,KHLE,BAKH,BFV,HHXY,KOV1,GOLE1,GOLE2,KOV2,NEWA,LIZA,FPV,TKU,KOV3,KOV4,DVA1} especially because of the new view on the entropy in DIS. In Ref.\cite{KHLE} it was proposed that the entropy in DIS is closely related to the entropy of entanglement between the spatial region probed by deep inelastic scattering and the rest of the proton.

\begin{figure}
 	\begin{center}
 	\leavevmode
 		\includegraphics[width=14cm]{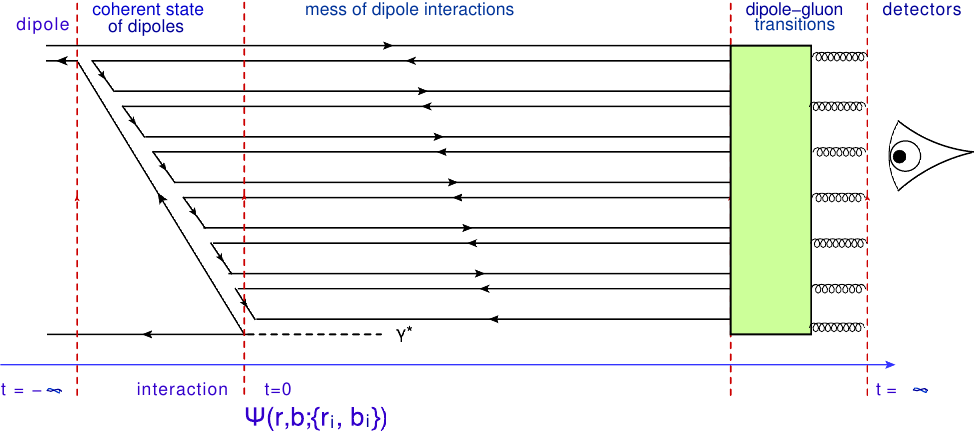}
 	\end{center}
 	\caption{The interaction of fast hadron (dipole) with the virtual photon ($\gamma^*$). The coherence of the partonic wave function of the fast hadron is destroyed at $t=0$, while the gluons can be measured at $t = \infty$.
}	
 	\label{parcas}
 \end{figure}

Our approach is based on two main ideas: the homotopy approach to  the solution of non-linear equations\cite{HE1,HE2}; and AGK cutting rules\cite{AGK}. The AGK cutting rules allow us\cite{KLP} to relate the gluons  that come to detectors ($t=\infty$ in  \fig{parcas}) with the dipoles that have been in the wave function of the fast hadron  $\Psi\Lb r,b,\{ r_i,b_i\}\Rb$.  The coherence of this wave function  is destroyed at time $t=0$ in \fig{parcas}. We will discuss these ideas in the next section, while in the introduction we summarize our theoretical understanding of DIS processes (see Ref.\cite{KOLEB} for the review).

{\bf 1.}
  The scattering amplitude of the colourless dipole with the size $x_{01}$ which determines the DIS cross section, satisfies the Balitsky-Kovchegov (BK) non-linear equation\cite{BK}:
\beq \label{BK}
 \frac{\partial}{\partial Y}\,N_{01}\Lb Y\Rb\,\,=\,\,\frac{\bas}{2\,\pi}\int d^2x_{2}\,\frac{ x^2_{01}}{x^2_{02}\,x^2_{12}}\,\bigg[\,N_{02}\Lb Y\Rb \,+\, N_{12}\Lb Y\Rb \,-\, N_{01}\Lb Y\Rb\,-\, N_{02}\Lb Y\Rb N_{12}\Lb Y\Rb\,\bigg]
\eeq
 where $N_{ik}\Lb Y\Rb=N\Lb Y, \vec{x}_{ik},\vec{b}\Rb$ is the scattering amplitude of the dipoles with size $x_{ik}\,=\,|\vec{x}_{ik}|\,=\,|\vec{x}_{i}\,-\,\vec{x}_k|$ and with rapidity $Y$ at the impact parameter $\vec{b}\,=\,\Lb\vec{x}_{0}+\vec{x}_1\Rb/2$.

 {\bf 2.} In Refs.\cite{GLR,MUT,MUPE} it has been shown that \eq{BK} leads to a new dimensional scale: saturation momentum\cite{GLR}  which has the following $Y$ dependence:
 \beq \label{QS}
 Q^2_s\Lb Y, \vec{b}\Rb\,\,=\,\,Q^2_s\Lb Y=Y_0, \vec{b}\Rb \,e^{\bas\,\kappa \,Y\,-\,\,\frac{3}{2\,(1-\gamma_{cr})} \,\ln \Lb \bas\,Y\Rb }
 \eeq 
 where $Y_0$ is the initial value of rapidity and $\kappa$ and $\gamma_{cr}$   are determined by the following equations\footnote{$\chi\Lb \gamma\Rb$ is the BFKL kernel\cite{BFKL} in anomalous dimension ($\gamma$) representation. $\psi$ is the Euler psi -function (see Ref.\cite{RY} formula {\bf 8.36}). }:
 \beq \label{GACR}
\kappa \,\,\equiv\,\, \frac{\chi\Lb \gamma_{cr}\Rb}{1 - \gamma_{cr}}\,\,=\,\, - \frac{d \chi\Lb \gamma_{cr}\Rb}{d \gamma_{cr}}~~~\,\,\,\mbox{and}\,\,\,~~~\chi\Lb \gamma\Rb\,=\,\,2\,\psi\Lb 1 \Rb\,-\,\psi\Lb \gamma\Rb\,-\,\psi\Lb 1 - \gamma\Rb
\eeq
  
%%%%%%%%%%%%%%%%%%%%%%%%%%%%%%%%%%%%%%%%%%%%%%%
 
 %%%%%%%%%%%%%%%%%%%%%%%%%%%%%%%%%%%%%%%%%%%%%%%%%%%%%%%%  
  {\bf 3.} It is shown\cite{MUT} (see also Ref.\cite{MUPE}) that in the vicinity  of the saturation scale the scattering amplitude takes the following form: 
  \beq \label{VQS}
 N\Lb  z\Rb\,\,\,=\,\,\,\mbox{Const} \Lb x^2_{01}\,Q^2_s\Lb Y,\vec{b}\Rb\Rb^{\bar \gamma}\,\,=\,\,\mbox{Const}\,\, e^{\bar\gamma\,z} \eeq
 with $\bar \gamma = 1 - \gamma_{cr}$. In \eq{VQS} we introduce a new variable $z$, which is equal to:
  \beq \label{z}
 z\,\,=\,\,\ln\Lb x^2_{01}\,Q^2_s\Lb Y, \vec{b}\Rb\Rb\,\, =\,\,\bas\,\kappa \,\Lb Y\,-\,Y_A\Rb\,\,+\,\,\xi
 \eeq  
 where $\xi = \ln \Lb x^2_{01}\,Q^2_s\Lb Y=Y_0, \vec{b}\Rb\Rb$. 
 
 {\bf 4.}  Inside the saturation region: $x^2_{01}\,Q^2_s\Lb Y\Rb \,>\,1$ the scattering amplitude $ N_{01}\Lb Y,x_{01},b\Rb$ is a function of one variable $x^2_{01}\,Q^2_s\Lb Y\Rb$ (geometric scaling behaviour), viz:
 \beq \label{GS}
   N\Lb Y,\vec{x}_{01},\vec{b}\Rb = N\Lb x^2_{01}\,Q^2_s\Lb Y,\vec{b}\Rb \Rb
   \eeq
   
    It is important to note, that this behaviour has been proven on general theoretical grounds\cite{BALE}  and has been seen in the experimental data on DIS\cite{GS}.

 {\bf 5.}
 Finally, in Ref.\cite{LETU}  it is shown that the solution to \eq{BK}  deep into saturation region for $z \,\gg\,1$  has the following form:
  \beq \label{BKS1}
N\Lb z\Rb\,\,=\,\,1\,\,-\,\,C\Lb z\Rb\,\exp\Big(  - \frac{z^2}{2\,\kappa}\Big)
 \eeq 
 where $C\Lb z\Rb$ is a smooth function of $z$.

\begin{figure}
 	\begin{center}
 	\leavevmode
 		\includegraphics[width=10cm]{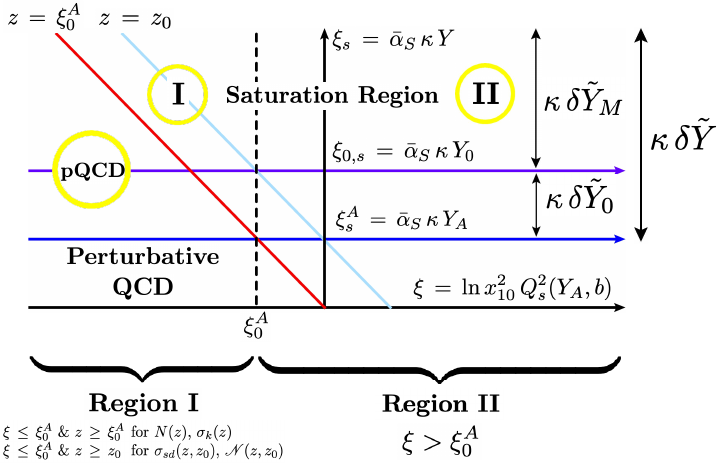}
 	\end{center}
 	\caption{  Saturation region of QCD for elastic amplitude. The critical line (z=0) is shown in red. The initial condition for scattering with the dilute system of partons (with proton) is given at $\xi_s = 0$. For heavy nuclei the initial conditions are placed at $Y_A = (1/3)\ln \,A \gg\,1$, where $A$ is the number of nucleon in a nucleus. Two blue lines show the   kinematic regions for the initial conditions: the upper one for nuclei and the low one for proton.}
 	\label{sat}
 \end{figure}

 {\bf 6.} The saturation region has been defined as 
  $ x^2_{01}\,Q^2_s\Lb Y,b\Rb\,>\,1$ (see \fig{sat}). However, in Refs.\cite{GOST,BEST} it has been noted that actually for very large impact parameters the non-linear corrections
 become small
  and we have to solve linear BFKL equation. This feature can be seen  directly
 from the eigenfunction of this equation.  Indeed, the eigenfunction  has the
 following form for any kernel which satisfies the conformal symmetry\cite{LIP,LIPREV}
\beq \label{EIGENF}
\phi_\gamma\Lb \vec{r} , \vec{R}, \vec{b}\Rb\,\,\,=\,\,\,\Lb \frac{
 r^2\,R^2}{\Lb \vec{b}  + \h(\vec{r} - \vec{R})\Rb^2\,\Lb \vec{b} 
 -  \h(\vec{r} - \vec{R})\Rb^2}\Rb^\gamma\,\,\xrightarrow{b\,\gg\,r,rR}\,\,\Lb \frac{ r^2\,R^2}{b^4}\Rb^\gamma\,\,\equiv\,\,e^{\gamma\,\xi}~~\mbox{with}\,\,0 \,<\,\mathrm{Re}\,\gamma\,<\,1\eeq
where $\xi\,=\,\ln \Lb \frac{r^2\,R^2}{b^4}\Rb$. In \eq{EIGENF} $R$ is the size of the initial dipole at $Y=0$  while $r\equiv x_{01}$ is the size of the dipole with rapidity $Y$.  However, we consider  the DIS with nuclei and for such target in Ref.\cite{CLMNEW} it has been shown that we can absorb all dependence on the impact parameter in the $b$ dependence of the saturation scale. 
	
   %%%%%%%%%%%%%%%%%%%%%%%%%%%%%%%%%%%%%%%%%%%%%%%%%%%%%%
 
 %%%%%%%%%%%%%%%%%%%%%%%%%%%%%%%%%%%%%%%%%%%%%%%%%%%%%%%%

{\bf 7.} \fig{sat} shows the typical kinematic regions in DIS. As has been mentioned we are dealing with the saturation region in this paper. However, this region for the elastic amplitude has to be divided in two parts: region I and region II  (shown in \fig{sat}). In the region I we expect the geometric scaling behaviour of the scattering amplitude. This amplitude  in the vicinity of the saturation scale (red curve in \fig{sat}) has the behaviour of \eq{VQS}. We   use this behaviour to put the boundary condition to the saturation region on the red line. They have the form:
  \begin{subequations}   \bea 
  N\Lb z= \xi_0^A\Rb &=& N_0;\label{BC1}\\
  \dfrac{d\, \ln  N\Lb z\Rb  }{d z}\Bigg{|}_{z= \xi_0^A}\,\,&=&\,\,\bar{\gamma} = 1 - \gamma_{cr};\label{BC2}
  \eea
   \end{subequations}   
   where $Y_A = \ln A^{1/3}$ and $A$ is the number of the nucleons in a nucleus. 
   
  For $\xi > \xi^A_0$ we use the McLerran-Venugopalan formula 
  for  the initial condition at $Y=Y_A$  for DIS with nuclei\footnote{ We use the simplified form of McLerran-Venugopalan formula given in Ref.\cite{GBW}. The exact form of this formula can be found in Ref.\cite{MV}} (see \fig{sat}):
\beq \label{MVF}
N\Lb Y = Y_A, \vec{x}_{01}, \vec{b} \Rb\,\,\,=\,\,1\,\,\,-\,\,\,\exp\Big( - r^2\,Q^2_s\Lb Y=Y_A, \vec{b}\Rb/4\Big)\,\,=\,\,1\,\,\,-\,\,\exp\Lb -\frac{1}{4}\, e^\xi\Rb
\eeq
where $r = x_{01}$ and we will use this notation for the size of the dipole below.  

In the next section we will discuss our approach.  It is based on two main ideas. The first is   
the derivation of the amplitude given by the BK equation using only the wave function of the dipoles $\Psi\Lb r,b,\{ r_i,b_i\}\Rb$. The second is the AGK \footnote{AGK stands for Abramovsky,Gribov and Kancheli \cite{AGK}.} approach to the gluon production.  In these approaches  we are dealing with  the cross sections of production of $n$-cut Pomerons ($\sigma^{\mbox{\tiny AGK}}_n$) which we introduce in this section.
We derive the evolution equation for $\sigma^{\mbox{\tiny AGK}}_n$. Our derivation does not use the AGK cutting rules in the explicit form, as it was done in Ref.\cite{KLP}, based on the main features of the BFKL cascade. We generalized the approach  for the process of diffraction production of Refs.\cite{KOLE,KOLEB} to the general case and show that this approach leads to the same set of equations as the AGK cutting rules. In particular, in this section we discuss how to calculate the multiplicity distributions of the produced gluons using $\sigma^{\mbox{\tiny AGK}}_n$ and the Poisson distributions for the produced gluons for the BFKL Pomeron.

 In section III we develop the homotopy approach for solutions of these equations. This method we have discussed in our previous papers (see Refs.\cite{CLMNEW,CGLM}). It is based on the idea \cite{HE1,HE2} to divide the general nonlinear equation in two parts:
 \beq \label{HOM1}
\mathscr{L}[u] +  \mathscr{N_{L}}[u]=0
\eeq 
 where $\mathscr{L}[u] $  include the linear evolution and part of the nonlinear corrections, which can be treated analytically (or almost analytically).
 The  non-linear part $
 \mathscr{N_{L}}[u] $ has an arbitrary form. As a solution, we introduce  the following  equation for the homotopy function $ {\mathscr H}\Lb p,u\Rb$:
 \beq \label{HOM2}
 {\mathscr H}\Lb p,u\Rb\,\,=\,\,\mathscr{L}[u_p] \,+ \,  p\, \mathscr{N_{L}}[u_p] \,\,=\,\,0
 \eeq
 
 Solving \eq{HOM2} we reconstruct the function
 \beq \label{HOM3}
 u_p\Lb Y,  \x_{10},  \vb\Rb\,\,=\,\, u_0\Lb Y,  \x_{10},  \vb\Rb\,\,+\,\,p\, u_1\Lb Y,  \x_{10},  \vb\Rb \,+\,p^2\, u_2\Lb Y,  \x_{10},  \vb\Rb \,\,+\,\,\dots
 \eeq
 with $\mathscr{L}[u_0] = 0$. \eq{HOM3}  gives  the solution to the non-linear equation at $ p = 1$.  The hope is that several  terms in series of \eq{HOM3} will give a good  approximation in the solution of the non-linear  equation. In section III we discuss the particular  form of this approach and show that we can choose  $\mathscr{L}[u] $ as the Balitsky-Kovchegov (BK) equation with the simplified  leading twist BFKL kernel.
 In section IV we solve the equations for $\sigma_n$ at large $n$:  $n\,\gtrsim\,N\Lb z\Rb = 2 N_0 \,z\,\exp\Lb z^2/(2 \kappa)\Rb$.  We show that these cross sections has the KNO\cite{KNO,KNO1,KNO2}\footnote{KNO stands for Koba,  Nielsen and  Olesen scaling behaviour \cite{KNO,KNO1,KNO2}} scaling behaviour. We calculate the entropy ($S_E$)  of the produced gluons and show that  $S_E = \ln N_{DIS}$ where $N_{DIS} $ is the multiplicity in  the deep inelastic scattering process, confirming the main ideas of Ref.\cite{KHLE}. In the conclusion we discuss our main results.
 
  %%%%%%%%%%%%%%%%%%%%%%%%%%%%%%%%%%%%%%%%%%%%%%%%%%%%%%%%%%
\section{Main Equations}
\subsection{Setting the problem}
%%%%%%%%%%%%%%%%%%%%%%%%%%%%%%%%%%%%%%%%%%%%%%%%%%%%%%%%%%

As has been mentioned in the introduction we discuss the multiplicity distributions of produced gluons at $t =+\infty$ in \fig{parcas}. The distribution of dipoles at $t=0$ is given by the dipole wave function of the fast dipole with size $r$ and rapidity $Y$.
From $t=0$ and $t=\infty$ the  dipoles can interact. We use the AGK cutting rules which give us an economical way to treat a mess of dipole interactions during  propagation of the dipole  cascade (see \fig{parcas}) from the moment of interaction of wee dipole  with the target ($t = 0$ in \fig{parcas})  till it reaches the detectors ($t = \infty$ in \fig{parcas}).

  In our treatment of the multiplicity distributions  we are going to explore that  the 
  scattering amplitudes  at high energy in QCD,  which can be found as the solution to BK equation (see \eq{BK}),  on the other hand is the sum of the `fan'   
  diagrams of  the  BFKL Pomeron calculus\cite{BFKL}.  Generally the scattering amplitude can be represented as the sum of multi-pomeron exchanges:
  \beq \label{I0} 
  N\Lb  Y, \vec{r},  \vec{b}\Rb\,\, =\,\, \sum^\infty_{k=1} \Lb -1\Rb^{k+1}\underbrace{ C_k\Lb \vec{r}, \vec{b} \Rb \Lb {\rm Im}\,G_{\pom} \Lb Y, \vec{r},  \vec{b}\Rb\Rb^k}_{F_k\Lb Y, \vec{r}, \vec{b}\Rb}
  \eeq 
where $F_k$ is the contribution of the exchange of $k$-Pomerons to  the cross section. The total cross section $\sigma_{tot}$ is therefore
\beq
\sigma_{tot}\Lb Y, \vec{r},\vec{b}\Rb\,\,=\,\,2 \sum_{k=1}^\infty (-1)^{k+1} \,F_k(Y, \vec{r},\vec{b}).\label{XS}
\eeq
  Our approach is based on  two main ideas. First, we recall that it is proven in Refs.\cite{BFKL,LEHP} that the $s$-channel unitarity for the BFKL Pomeron has the form:
  \beq \label{I1}
2\, { \rm Im} \,G_{\pom}\Lb Y, \vec{r},  \vec{b}\Rb\,=\,\,\sigma^{\mbox{\tiny BFKL}}_{in}(Y, \vec{r}, \vec{b})
\eeq
 where $G_{\pom}$ is the Green's function for the BFKL Pomeron and  $\sigma^{\mbox{\tiny BFKL}}_{in}$ is the inelastic cross sections of  produced gluons with mean multiplicity $\bar{n} = \Delta_{\pom}\,Y$, where $\Delta_{\pom}=4\,\bas\,\ln 2$ is the intercept of the BFKL Pomeron. We also know that produced gluons have the Poisson distribution  with this mean multiplicity (see Ref.\cite{LEHP} and appendix A in Ref.\cite{LEUTM}). \eq{I1} gives us connection between the Pomeron Green function and the cross section of produced gluons and determines so called cut Pomeron.  
 
 The next problem is to rewrite the scattering amplitude as the sum of contributions with $n$-cut Pomeron ($\sigma_n^{\mbox{\tiny AGK}}$). 
 The AGK cutting rules \cite{AGK}  solve this problem and 
allows us to calculate the contributions of $n$-cut Pomerons if we know $F_k$: the contribution of the exchange of $k$-Pomerons to the cross section. They take the form:
     \bea \label{AGKK}
\sigma_{n}^{\mbox{\tiny AGK}}\Lb Y,\vec{r},\vec{b}\Rb\,\,=\,\,\sum_{k=n}^{\infty}\sigma_n^k\Lb Y,\vec{r},\vec{b}\Rb;~~\mathrm{with} ~~
    \sigma_n^k \Lb Y, \vec{r},\vec{b}\Rb&=&
    \begin{cases}
        \displaystyle{\Lb -1\Rb^k \Big(2^k\,\,-\,\,2\Big) F_k(Y, \vec{r},\vec{b})}   &
        \text{ for\,  $n=0$}
        \\*[0.2cm]
        \displaystyle{(-1)^{k-n}\frac{k!}{(k - n)!\,n!}\,2^{k}\, F_k(Y, \vec{r},\vec{b} )}     &
        \text{ for\,  $n\geq 1$}
    \end{cases}
\eea
where $\sigma_0$ denotes the cross section with the multiplicity of produced dipoles  which is much less than $\Delta_{\pom}\,Y$. In other words, it is the cross section of the diffraction production.   

Finally, we can discuss  the multiplicity distributions of produced gluons, calculating 
the cross section of productions of $n$ gluons in the final state  ($\tilde{\sigma}_n$).
   Our master formula for   ($\tilde{\sigma}_n$)
takes the form of convolution for the cross section of produced $k$ cut Pomerons with the Poisson distributions of gluons in these $k$-Pomerons:
\beq \label{I10}
\underbrace{\sum_{n_i}}_{\sum_i^k  n_i=n}\prod^k_{i=1}\mathcal{P}_{n_i}^{\pom}\Lb \Delta_{\pom}\,Y\Rb\,\,=\,\,
 \underbrace{ \frac{\Lb k\,\Delta_{\pom}\,Y\Rb^n}{n!} \,e^{ - k\,\Delta_{\pom}\,Y}}_{\mbox{Poisson distribution}}\,\,=\,\,\mathcal{P}_n^{\pom}\Lb k\, \Delta_{\pom}\, Y\Rb
 \eeq
Finally,
 \beq \label{I2}
\tilde{ \sigma}_n\Lb Y,\vec{r},\vec{b} \Rb\,\,=\,\,\sum_{k=1}^{\infty}\underbrace{ \sigma_k^{\mbox{\tiny AGK}}\Lb Y, \vec{r}, \vec{b}  \Rb}_{ \propto\,\Lb\sigma^{\mbox{\tiny BFKL}}_{in}(Y, \vec{r}, \vec{b})
 \Rb^k}\,\mathcal{P}_n^{\pom}\Lb k\, \Delta_{\pom}\, Y\Rb\,\,\xrightarrow{Y \,\gg\,1} \,\, \sigma^{\mbox{\tiny AGK}}_{k = n/(\Delta_{\pom}\,Y)}\Lb  Y,\vec{r},\vec{b} \Rb
\eeq 
In the last equation we used that the Poisson distribution with average number of gluons: $k\,\Delta_{\pom}\,Y$ is much   steeper  function than $\sigma^{\mbox{\tiny AGK}}_k $ with typical $k $ of the order of $G_{\pom}$.

Let us illustrate the procedure  by obtaining the initial condition for $\sigma^{\mbox{\tiny AGK}}_n$. The initial conditions for the scattering amplitude of \eq{MVF} can be rewritten as:
\beq \label{I20}
\sigma_{tot}\Lb Y=Y_A, \vec{r},\vec{b}\Rb \,\,=\,\, 2 \sum^\infty_{k=1} \Lb -1\Rb^{k+1}\underbrace{ \frac{1}{k!}\Lb\frac{e^{\xi}}{4}\Rb^k}_{
F_k\Lb Y_A,\vec{r},\vec{b}\Rb}
\eeq
where $F_k\Lb Y_A,\vec{r},\vec{b}\Rb =\frac{1}{k!}\Lb{ \rm Im}\,G_{\pom}\Lb Y=Y_A,\vec{r},\vec{b}\Rb\Rb^k 
$ and $G_{\pom} \Lb Y=Y_A,
\vec{r},\vec{b}\Rb$ is the initial conditions for the BFKL Pomeron which is the scattering amplitude of the dipoles with exchange of two gluons. Using \eq{AGKK} we calculate the initial conditions for the cross section of $n$-cut Pomerons:
\beq 
\label{MVFXS}\sigma^{\mbox{\tiny AGK}}_n\Lb Y=Y_A, \vec{r},\vec{b}\Rb  \,\,=\,\,\sum_{k=n}^{\infty} (-1)^{k-n}\frac{k!}{(k - n)!\,n!}\,2^{k}\, F_k(Y_A,  \vec{r},\vec{b} ) =\,\,\dfrac{\Lb\h e^\xi\Rb^n}{n!}\, \exp\Lb -\frac{1}{2}\, e^\xi\Rb
 \eeq

Note, that in \eq{MVFXS} $\h e^\xi$ is $ 2 \sigma^{\rm BA}$ where $\sigma^{\rm BA}$ is the cross section of two dipoles scattering in the Born approximation of perturbative QCD. In the equation  $Y_A = \ln A^{1/3}$ and A is the number of the nucleons in a nucleus. For nuclei the $b$ dependence  of $Q_s(Y_A,\vec{b})$ is  determined by 
the optical width of the nucleus $T_A(\vec{b})$, which gives the number of nucleons at given value of impact parameter $b$: $Q_s(Y_A,\vec{b}) =  Q_0 \,T_A\Lb \vec{b} \Rb$.

The amplitude in the form of \eq{I0} as well as $\sigma^{\mbox{\tiny AGK}}_n$ have been found for the BK cascade in Ref.\cite{KLP}.

\begin{figure}
 	\begin{center}
 	\leavevmode
 		\includegraphics[width= \textwidth]{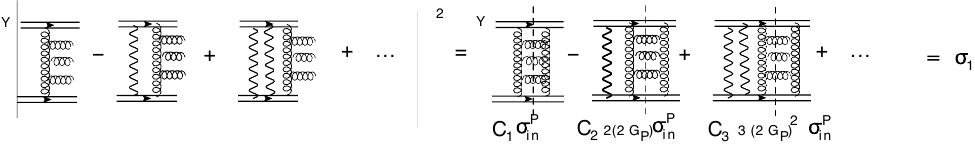}
 	\end{center}
 	\caption{ $\sigma^{\mbox{\tiny AGK}}_1 $ for the scattering amplitude of \eq{I0}. The vertical dashed lines show cut Pomerons.
 }
 	\label{xs1}
 \end{figure}

Therefore in this paper we calculate\footnote{We apologized  for any inconveniency that these notations could cost.} $\sigma_n \equiv \sigma^{\mbox{\tiny AGK}}_n$,  or, in other words, the cross section of $n$-cut Pomerons in the framework of the AGK approach.
For the sake of illustration we show in \fig{xs1} $\sigma^{\mbox{\tiny AGK}}_1 $ for the scattering amplitude of \eq{I0}.
	
   %%%%%%%%%%%%%%%%%%%%%%%%%%%%%%%%%%%%%%%%%%%%%%%%%%%%%%
 
 %%%%%%%%%%%%%%%%%%%%%%%%%%%%%%%%%%%%%%%%%%%%%%%%%%%%%%%%

  %%%%%%%%%%%%%%%%%%%%%%%%%%%%%%%%%%%%%%%%%%%%%%%%%%%%%%%%%%
\subsection{AGK equations without AGK cutting rules}
%%%%%%%%%%%%%%%%%%%%%%%%%%%%%%%%%%%%%%%%%%%%%%%%

The main equation for $\sigma_n$ has been written in Ref.\cite{KLP} where they were derived from the AGK cutting rules. In this section we are going to rewrite them in dipole approach to QCD \cite{MUDI} without addressing the AGK cutting rules in the explicit form.  We have two indications that we can do this. First,
it was shown in Refs.\cite{KOLE,KOLEB} that this can be done for the processes of diffraction production. The equations in Ref.\cite{KLP} include these processes  and this indicates that we can rewrite the result of Ref.\cite{KLP} in a such way. Second, in the simple but instructive one dimensional models for $\sigma_n$, the evolution equation  
 can be written for $\sigma_n$ which do not address the AGK cutting rules in an explicit way\cite{LEUTM}.

As has been discussed in Ref.\cite{KOLE,KOLEB} the derivation  of the  evolution equation for the diffractive production 
 is based on two principal ingredients. 
The first one is that     only dipoles in the parton wave function of the fast hadron at $t \to - \infty$ in \fig{parcas} can be produced at    $t \to + \infty $ and  measured by our detectors. This point was proven in Refs.\cite{KOLE,KOLEB} and the proof is used the results of Ref.\cite{CHMU}. In these papers were shown that neither production of the new partons nor annihilation of the partons in the hadronic wave function contribute to 
process of diffractive production. Actually the proof is general for any final state and does not depend on specific interaction between dipoles.
 It has been demonstrated\cite{KOLE}  that the AGK cutting rules provides these features in the framework of the  dipole approach for the diffractive production.

The equation for the S-matrix in the BFKL cascade has the following form \cite{MUDI} in the dipole approach to QCD:
\beq \label{ME1}
\frac{\pp }{\pp Y}\,S\Lb Y, \vec{r}, \vec{b} \Rb\,\,=\,\,\frac{\bas}{2\,\pi}\int d^2 r'  \underbrace{\frac{r^2 }{r'^2\,(\vec{r} - \vec{r}')^2 }}_{K\Lb \vec{r}, \vec{r}'\Rb}\bigg[\, S\Lb Y , \vec{r}', \vec{b} - \h(\vec{r} - \vec{r}')\Rb  \,S\Lb Y , \vec{r} - \vec{r}', \vec{b} - \h \vec{r}'\Rb\,\,-\,\,
S\Lb Y, \vec{r}, \vec{b} \Rb\,\bigg]
\eeq
The equation for the cross section of the diffractive production can be written in the same form as \eq{ME1}  but for \cite{KOLE}
\beq \label{ME2}
S^D\Lb Y, \vec{r}; \vec{b} \Rb \,\,=\,\,  1 \,-\, 2 \,N\Lb Y, \vec{r}; \vec{b} \Rb\,\,+\,\, \sigma_{sd}\Lb Y, \vec{r}; \vec{b} \Rb
\eeq
Indeed, plugging in \eq{ME1} $S^D$ and rewriting the equation for 
$ \sigma_{sd}\Lb Y, \vec{r}; \vec{b}\Rb \,=\, 1 \,-\, S^D\Lb Y,\vec{r};\vec{b}\Rb$ we obtain:
\bea \label{ME3}
\frac{\partial}{\partial Y}\,\sigma_{sd}\Lb Y, \vec{r}_{01}, \vec{b} \Rb\,\,&=&\,\,\frac{\bas}{2\,\pi}\,\int\! d^2 r_2 \,K\Lb \vec{r}_{01}\vert \vec{r}_{12}, \vec{r}_{02}\Rb \nn\\ &\times&\,\bigg[\,\sigma_{sd}\Lb Y, \vec{r}_{12}, \vec{b} \Rb\,\,+\,\,
\sigma_{sd}\Lb Y, \vec{r}_{02}, \vec{b} \Rb\,\,-\,
\,\sigma_{sd}\Lb Y, \vec{r}_{01}, \vec{b} \Rb \nn  \\
&+&\,\,\sigma_{sd}(Y,  \vec{r}_{12},\vec{b})  \,\sigma_{sd}(Y,  \vec{r}_{02}, \vec{b})\,\,-\,\,2\,\sigma_{sd}(Y,  \vec{r}_{12}, \vec{b} )\,N(Y,\vec{r}_{02}, \vec{b}) \nn\\&-&\,\,2\,
N(Y,\vec{r}_{12}, \vec{b})
\,\sigma_{sd}(Y, \vec{r}_{02}, \vec{b})\,\,+\,\, 2\,
N(Y, \vec{r}_{12}, \vec{b})
\,N(Y,\vec{r}_{02}, \vec{b})\,\bigg]
 \eea 
\eq{ME3} is the equation of Ref.\cite{KOLE} for the cross section of the diffractive production with  all possible  rapidity gaps for the same scattering process. $N(Y,\vec{r}_{01}, \vec{b}) = 1 - S(Y,\vec{r}_{01}, \vec{b}) $ is the imaginary part of the elastic scattering amplitude of the dipole with the size $\vec{r}_{01}\equiv \vec{r}$ and rapidity $Y$   at the impact parameter $b$.   We use notations: $\vec{r}_{02} \equiv \vec{r}'$ and $ \vec{r}_{12} \equiv \vec{r} - \vec{r}'$.
 
 We propose  \eq{ME1} for   $S_{in}\Lb Y, \vec{r}; \vec{b}\Rb$ which is equal to
  \beq \label{ME4}  
  S_{in} \Lb Y, \vec{r}, \vec{b}\Rb\,\,=\,\,1 \,\, -\,\, 2 \,N\Lb Y, \vec{r}, \vec{b}\Rb  \,\,+\,\,\sigma_{sd}\Lb Y, \vec{r}, \vec{b}\Rb\,\,+\,\,\sum_{n=1} \sigma_n\Lb Y, \vec{r}, \vec{b}\Rb
 \eeq 
 with additional selection 
  for finding $\sigma_n$: the fixed multiplicity, which is equal to   $ n \,\Delta_{\pom}\,Y$, for each term.

 The equation has the form:
 \beq \label{ME5}
\underbrace{ \frac{\pp }{\pp \,Y}\,S_{in} \Lb Y, \vec{r}, \vec{b}\Rb }_{\mbox{select fixed multiplicity terms}} \,=\, \frac{\bas}{2\,\pi} \intl\! d^2 r'  K\Lb \vec{r}, \vec{r}'\Rb\underbrace{
 \bigg[\, S_{in}\Lb Y, \vec{r}', \vec{b} - \h(\vec{r} - \vec{r}')\Rb  S_{in}\Lb Y , \vec{r} - \vec{r}', \vec{b} - \h \vec{r}'\Rb\,\,-\,\,
S_{in}\Lb Y, \vec{r}, \vec{b} \Rb\,\bigg]}_{\mbox{select fixed multiplicity terms}}
\eeq 
 %%%%%%%%%%%%%%%%%%%%%%%%%%%%%%%%%%%%%%%%%
     \begin{figure}[ht]
   \centering
  \leavevmode
      \includegraphics[width=18cm]{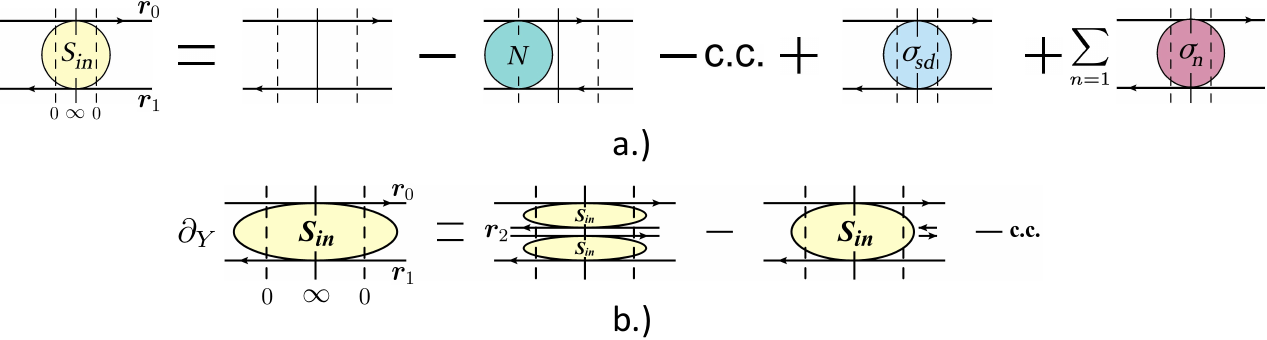}  
      \caption{a.) The graphic form of \eq{ME4} for the interaction of one dipole which  is denoted by two parallel horizontal lines  with opposite directions of arrows.   All notations in the caption of \fig{eveq}. b.) The evolution equation for $S_{in}$.}
\label{eveqgen}
   \end{figure}
%%%%%%%%%%%%%%%%%%%%%%%%%%%%%%%%%%%%%%%%%%    
	In \fig{eveqgen} a.) and b.) one can see the graphic form of \eq{ME4} and \eq{ME5} respectively. 
	
\begin{figure}[ht]
   \centering
  \leavevmode
      \includegraphics[width=18cm]{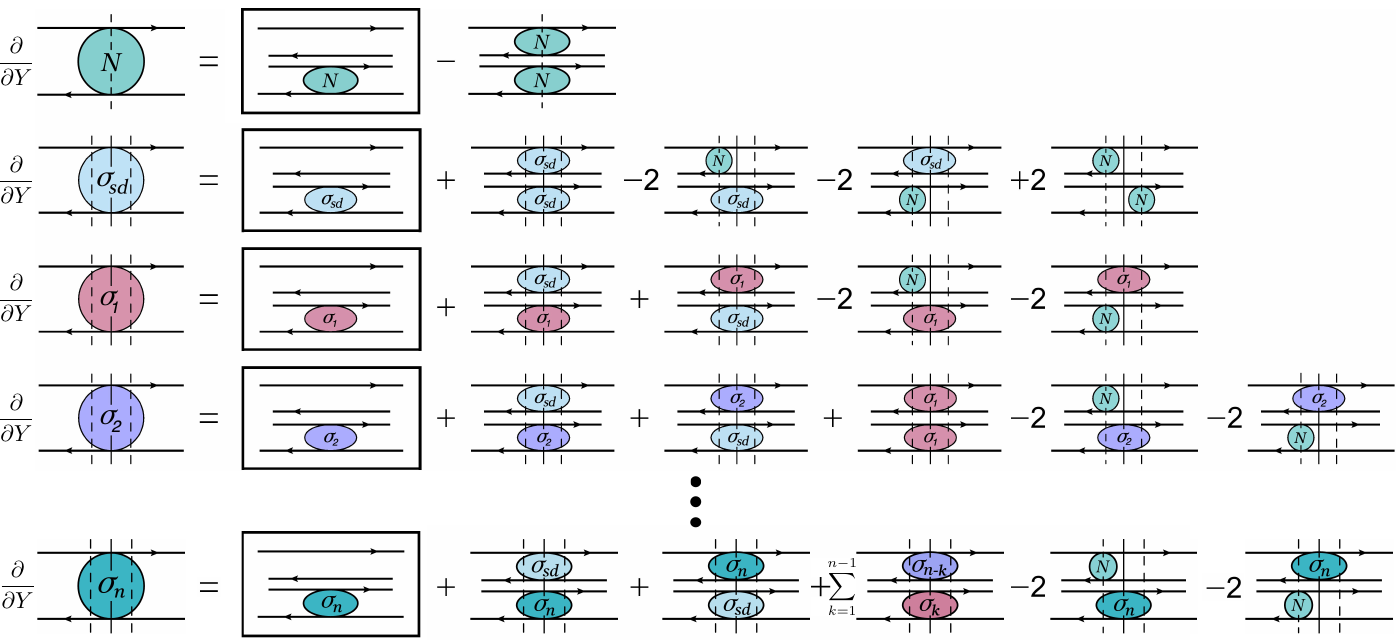}  
      \caption{The graphic form of \eq{ME7} for the interaction of one dipole which  is denoted by two parallel horizontal lines  with opposite directions of arrows.    The first stage of the interaction is the decay of one dipole to two , which is described by the kernel $K\Lb \vec{r},\vec{r}'\Rb$.      
       The vertical dashed lines denote  time t=0 in \fig{parcas} where the interaction with the target occurs in the amplitude and complex conjugated amplitude.  The vertical  line denotes  $t= +\infty$ where our detectors are placed.  The Pomeron intercept is absorbed in the definition of $Y$. The coefficients reproduce the AGK cutting rules. Note, that the linear term has the form of \eq{ME70}.}
\label{eveq}
   \end{figure}

The equation for $\sigma_n$  takes the following form from \eq{ME5}	
\bea \label{ME7}
   \frac{\partial}{\partial Y}\,\sigma_n\Lb Y,\vec{r}_{01},\vec{b}\Rb \,\,&=&\,\,
 \frac{\bas}{2 \,\pi}\int
\!d^2 r_2 \,K\Lb \vec{r}_{01}| \vec{r}_{12}, \vec{r}_{02}\Rb \nn\\ &\times&\,\,\bigg[\,\sigma_n\Lb Y,\vec{r}_{12},\vec{b}\Rb \,\,+\,\,\sigma_n\Lb Y,\vec{r}_{02},\vec{b}\Rb\,\,-\,\,\sigma_n\Lb Y,\vec{r}_{01},\vec{b}\Rb\,\,\nn\\
&+&\sigma_n\Lb Y,\vec{r}_{12},\vec{b}\Rb\,\sigma_{sd}\Lb Y,\vec{r}_{02},\vec{b}\Rb\,\,+\,\,\sigma_n\Lb Y,\vec{r}_{02},\vec{b}\Rb\,\sigma_{sd} \Lb Y,\vec{r}_{12},\vec{b}\Rb\,\,\nn\\
&+&\,\,\sum_{k=1}^{n-1}  \sigma_{n - k}\Lb Y,\vec{r}_{02},\vec{b}\Rb\,\sigma_{k}\Lb Y,\vec{r}_{12},\vec{b}\Rb  \nn\\
&-&\,\,2\,
\sigma_n\Lb Y,\vec{r}_{12},\vec{b}\Rb\,N\Lb Y,\vec{r}_{02},\vec{b}\Rb\,\,-\,\,2\,\sigma_n\Lb Y,\vec{r}_{02},\vec{b}\Rb\,N \Lb Y,\vec{r}_{12},\vec{b}\Rb\,\bigg]
\eea
$\sigma_{sd}$ is the cross section for the diffraction dissociation. 

\eq{ME7} coincides with the equations that have been derived in the BFKL Pomeron calculus in QCD from the AGK cutting rules\cite{KLP} (see also \cite{LEPRI}).  It is instructive to note that all these equations have a natural coefficients for each term. To illustrate this point   we present all these equation in \fig{eveq}. Note that all these coefficients reproduce the AGK cutting rules.

      %%%%%%%%%%%%%%%%%%%%%%%%%%%%%%%%%%%%%%%%%
%%%%%%%%%%%%%%%%%%%%%%%%%%%%%%%%%%%%%%%%%%    
	
The main point is that these coefficients are natural if you account for the dipoles in the initial wave function and only they  go to the detectors.
 Let us consider the equation of $\sigma_2$ (see \fig{eveq} and \eq{ME7} with $n=2$). The first stage is the decay of the dipole with the size $r_{01} \equiv  r$ into two dipoles $r_{02}\equiv r'$ and $r_{12}\equiv \vec{r} - \vec{r}'$ which describes by $K\Lb \vec{r},\vec{r}'\Rb$. The first terms  describe the linear evolution of $\sigma_2$. It should be noted that  we put coefficient 1 in front of this term\footnote{This coefficient is correct in the one dimensional model which could be viewed as QCD but with fixed dipole size\cite{LEUTM}.}. Actually this term has the form:
 \beq \label{ME70}
 \frac{\bas}{2\,\pi}\int
\!d^2 r_2 \,K\Lb \vec{r}_{01}| \vec{r}_{12}, \vec{r}_{02}\Rb \,\bigg[\,\sigma_n\Lb Y,\vec{r}_{02},\vec{b}\Rb \,\,+\,\,\sigma_n\Lb Y,\vec{r}_{12},\vec{b}\Rb\,\,-\,\,\sigma_n\Lb Y,\vec{r}_{01},\vec{b}\Rb \,\bigg]
\eeq
for any value of $n$ including elastic amplitude and the cross section of diffraction production.
 In the second and third term one dipole interacts with the target with $\sigma_{sd}$  while the second dipole interaction is still $\sigma_2$. Note, that the multiplicity of the final stage is $2\,\Delta_{\pom} \,Y$ which is the same as for $\sigma_2$.  $\Delta_{\mathrm{BFKL}} $ is the intercept of the BFKL Pomeron. We have two such diagrams  since any of two dipoles can interact with $\sigma_{sd}$.  The fourth term is the interaction of both dipoles with $\sigma_1$ which  again gives us the multiplicity $2 \,\Delta_{\pom} \,Y$. The fifth term is the shadowing in the initial state induced by the elastic amplitude. Factor 2 stems from the fact that any of two dipoles can interact in the initial state. The last term is the same shadowing but in the final state. Hence, you see that all these coefficients have a  simple meaning.

 %%%%%%%%%%%%%%%%%%%%%%%%%%%%%%%%%%%%%%%%%%%%%%%%%%%%%%%%%%
 
      %%%%%%%%%%%%%%%%%%%%%%%%%%%%%%%%%%%%%%%%%
     \begin{figure}[ht]
   \centering
  \leavevmode
      \includegraphics[width=14cm]{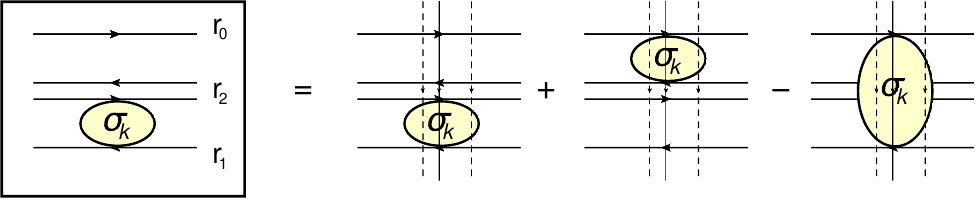}  
      \caption{The graphic form of the first term in each equations of \fig{eveq} (see \eq{ME5}-\eq{ME7}) for the interaction of one dipole which  is denoted by two parallel horizontal lines  with opposite directions of arrows    The first stage of the interaction is the decay of one dipole to two , which describes by kernel $K\Lb r_{01}| r_{12}, r_{02}\Rb$.      
       The vertical dashed lines denote  time t=0 in \fig{parcas} where the interaction with the target occurs.  The vertical line denotes  $t= +\infty$ where our detectors are placed.}
\label{eveq1}
   \end{figure}
%%%%%%%%%%%%%%%%%%%%%%%%%%%%%%%%%%%%%%%%%%
It is instructive to note that the equation for  the inelastic cross section: $\sigma_{in}
\Lb Y, \vec{r}_{01},\vec{b}\Rb  =\sum_{n=1}^\infty \sigma_n\Lb Y, \vec{r}_{01},\vec{b}\Rb$, which we obtain by summing $\sigma_n\Lb Y, \vec{r}_{01},\vec{b}\Rb$ of \eq{ME7}, has the following form:
\bea \label{ME71}
   &&\frac{\partial}{\partial Y}\,\sigma_{in}\Lb Y,\vec{r}_{01},\vec{b}\Rb \,\,=\,\,
 \frac{\bas}{2 \,\pi}\int
\!d^2 r_2 \,K\Lb \vec{r}_{01}| \vec{r}_{12}, \vec{r}_{02}\Rb \nn\\ &&\times\,\bigg[\,\sigma_{in}\Lb Y,\vec{r}_{12},\vec{b}\Rb \,\,+\,\,\sigma_{in}\Lb Y,\vec{r}_{02},\vec{b}\Rb\,\,-\,\,\sigma_{in}\Lb Y,\vec{r}_{01},\vec{b}\Rb\,\,+\,\,\sigma_{in}\Lb Y,\vec{r}_{12},\vec{b}\Rb\,\sigma_{in}\Lb Y,\vec{r}_{02},\vec{b}\Rb\nn\\
&&-\,\,\sigma_{in}\Lb Y,\vec{r}_{12},\vec{b}\Rb\,\bigg(2\,N\Lb Y,\vec{r}_{02},\vec{b}\Rb\,-\,
\sigma_{sd}\Lb Y,\vec{r}_{02},\vec{b}\Rb\bigg)\nn\\&&-\,\sigma_{in}\Lb Y,\vec{r}_{02},\vec{b}\Rb\,\bigg(2\,N\Lb Y,\vec{r}_{12},\vec{b}\Rb \,-\,
\sigma_{sd}\Lb Y,\vec{r}_{12},\vec{b}\Rb\bigg)\,\bigg]\eea
 From the $s$-channel unitarity constraints : $ 2 \,N\Lb Y,\vec{r}_{ik},\vec{b}\Rb\,=\,
 \sigma_{sd}\Lb Y,\vec{r}_{ik},\vec{b}\Rb\,+\,\sigma_{in}\Lb Y,\vec{r}_{ik},\vec{b}\Rb\,$, one can see that \eq{ME71} takes the form:
 \bea \label{ME72}
  &&\frac{\partial}{\partial Y}\,\sigma_{in}\Lb Y,\vec{r}_{01},\vec{b}\Rb \,\,=\,\, \frac{\bas}{2 \,\pi}\int
\!d^2 r_2 \,K\Lb \vec{r}_{01}| \vec{r}_{12}, \vec{r}_{02}\Rb \nn\\ &&\times\,\bigg[\,\sigma_{in}\Lb Y,\vec{r}_{12},\vec{b}\Rb \,\,+\,\,\sigma_{in}\Lb Y,\vec{r}_{02},\vec{b}\Rb\,\,-\,\,\sigma_{in}\Lb Y,\vec{r}_{01},\vec{b}\Rb\,\,-\,\, \sigma_{in}\Lb Y,\vec{r}_{12},\vec{b}\Rb \,\sigma_{in}\Lb Y,\vec{r}_{02},\vec{b}\Rb\,\bigg]\nn
\eea
Therefore, we obtain the BK equation for $\sigma_{in}$ as it is expected from the unitarity constraint.

Concluding this section we state that we found the evolution equations for $\sigma^{\mbox{\tiny AGK}}_n$ which reproduce the AGK cutting rules. However, it is well known that AGK cutting rules are violated in QCD \cite{KO01,KOTU01,JMKO,BRAGK,MAAGK,KOLUAGK,LEPRAGK}. We wish to recall that all  these violations are related to the emission of the measured gluon from the triple Pomeron vertex. Calculating  $\sigma_n \equiv \,\sigma^{\mbox{\tiny AGK}}_n$ we integrated over all momenta of emitted gluons and for such processes AGK cutting rules are correct.

  %%%%%%%%%%%%%%%%%%%%%% Concluding this section%%%%%%%%%%%%%%%%%%%

\section{Leading twist BFKL kernel: homotopy approach for the solutions}
%%%%%%%%%%%%%%%%%%%%%%%%%%%%%%%%%%%%%%%%%%
All equations for $\sigma_n$ depend on the amplitude $ \mathscr N\Lb Y, \vec{r}_{ik}, \vec{b}\Rb =   2\, N\Lb Y, \vec{r}_{ik}, \vec{b}\Rb\, -\, \sigma_{sd}(Y,  \vec{r}_{ik}, \vec{b})$. Introducing $\Delta\Lb Y,\vec{r}_{ik},\vec{b}\Rb = 1 - \mathscr{N}\Lb Y,\vec{r}_{ik},\vec{b}\Rb$ we rewrite \eq{ME7} as follows:
\bea \label{EQXS00}
   &&\frac{\partial}{\partial Y}\,\sigma_n\Lb Y,\vec{r}_{01},\vec{b}\Rb \,\,=\,\,
 \frac{\bas}{2 \,\pi}\int
\!d^2 r_2 \,K\Lb \vec{r}_{01}| \vec{r}_{12}, \vec{r}_{02}\Rb \nn \\ &&\times\,\bigg[ \,-\,\sigma_n\Lb Y,\vec{r}_{01},\vec{b}\Rb \,\,+\,\,\sigma_n\Lb Y,\vec{r}_{12},\vec{b}\Rb\,\Delta\Lb Y,\vec{r}_{02},\vec{b}\Rb \,\,+\,\,\Delta\Lb Y,\vec{r}_{12},\vec{b}\Rb\,\sigma_n\Lb Y,\vec{r}_{02},\vec{b}\Rb\nn\\&&+\,\sum_{k=1}^{n-1}  \sigma_{n - k}\Lb Y,\vec{r}_{02},\vec{b}\Rb\,\sigma_{k}\Lb Y,\vec{r}_{12},\vec{b}\Rb\,\bigg]
\eea
The homotopy approach for finding $\Delta\Lb Y,\vec{r}_{ik},\vec{b}\Rb$ has been developed in our previous paper (see Ref.\cite{CGLM}). We wish only to point out, that  for the equation of $\sigma_n$, which we consider in this paper,    we   find $\sigma_{sd}$ with rapidity gaps  larger than $z_0$ (see Ref.\cite{CGLM}). We fix $z_0 =2$ since we can consider the Pomeron exchange only at $Y \geq 2$.

To find solutions of \eq{EQXS00} we use the simplified BFKL kernel suggested in     Ref.\cite{LETU}. This kernel describes the high energy asymptotic solution of the nonlinear BK equation and leads to the geometric scaling behaviour: 
    \bea \label{SIMKER}
\chi\Lb \gamma\Rb\,\,=\,\, \left\{\begin{array}{l}\,\,\,\displaystyle{\frac{1} {1\,-\,\gamma}}\,\,\,\,\,\,\,\,\,\,\mbox{for}\,\,z \,\geq\,\xi_0^A,\,\,\,\,\,\,\mbox{summing} \Lb z\Rb^n;\\ \\
\,\,\,\displaystyle{\frac{1}{\gamma}}\,\,\,\,\,~~~~~~\mbox{for}\,\,\,z \,<\,\xi_0^A,\,\,\,\,\,\mbox{summing}
\Lb -\xi\Rb^n.\\  \end{array}
\right.
\eea
 where $ \xi $ has been introduced in \eq{EIGENF}.
 
 Since this kernel sums log contributions it corresponds to leading twist   term of the full BFKL kernel. It has a very simple form in the coordinate representation \cite{LETU}:
 
\beq \label{K2}
\frac{\bas}{2\,\pi}\int d^2 r'\,\displaystyle{ K\Lb \vec{r}, \vec{r}'\Rb} \,\rightarrow
\,\frac{\bas}{2}\!\! \!\!\intl^{r^2}_{Q^{-2}_s(Y,b)}\!\!\!\! \frac{ d r'^2}{r'^2}\,\,+\,\,
\frac{\bas}{2}\!\!\!\! \intl^{r^2}_{Q^{-2}_s(Y, b)}\!\!\!\! \frac{ d |\vec{r} - \vec{r}'|}{|\vec{r}  - \vec{r}'|^2}\,\,
 =\,\,\frac{\bas}{2}\, \intl^{\xi}_{-\xi_s}\!\! d \xi_{02} \,\,+\,\,\frac{\bas}{2}\, \intl^{\xi}_{-\xi_s} \!\! d \xi_{12}\equiv  \intl^{\xi}_{-\xi_s}\!\! K_{LT}\Lb \xi,\xi'\Rb d \xi'\eeq
 where $\xi_{ik} \,=\,\ln \Lb r^2_{ik}\,Q_s^2(Y=Y_A,b)\Rb$
 and $\xi_s\,=\,\bas\,\kappa\,(Y-Y_A)$. 
 Note, that  the  logarithms
   originate from the decay of a large size dipole into one small
 size dipole  and one large size dipole\cite{LETU}.  However, the size of the
 small dipole is still larger than $1/Q^2_s  = \frac{b^4}{R^2}\,e^{-\xi_s} $.

The equations $\sigma_n$ of \eq{EQXS00} for the leading twist BFKL kernel \eq{K2} take the form:
\bea \label{EQXS01}
   &&\frac{\partial}{\partial Y}\,\sigma_n\Lb Y,\vec{r}_{01},\vec{b}\Rb \,\,=\,\,-\,2\,\bas\,\ln\Lb r_{01}\,Q_s(Y,\vec{b})\Rb\,\sigma_n\Lb Y,\vec{r}_{01},\vec{b}\Rb \nn\\
&&+\,\bas\,\sigma_n\Lb Y,\vec{r}_{01},\vec{b}\Rb\,\intl_{\ln\Lb Q_{s}^{-2}(Y,\vec{b})\Rb}^{\ln\Lb r_{01}^2\Rb}\!\!\!\!\!\!\!\!\!d\ln\Lb r_{02}^2\Rb\,\Delta\Lb Y,\vec{r}_{02},\vec{b}\Rb \,\,+\,\,\bas\,\Delta\Lb Y,\vec{r}_{01},\vec{b}\Rb\,\intl_{\ln\Lb Q_{s}^{-2}(Y,\vec{b})\Rb}^{\ln\Lb r_{01}^2\Rb}\!\!\!\!\!\!\!\!\!d\ln\Lb r_{02}^2\Rb\sigma_n\Lb Y,\vec{r}_{02},\vec{b}\Rb \nn\\&&+\,\bas\,\sum_{k=1}^{n-1}  \,\intl_{\ln\Lb Q_{s}^{-2}(Y,\vec{b})\Rb}^{\ln\Lb r_{01}^2\Rb}\!\!\!\!\!\!\!\!\!d\ln\Lb r_{02}^2\Rb\sigma_{n - k}\Lb Y,\vec{r}_{02},\vec{b}\Rb\,\sigma_{k}\Lb Y,\vec{r}_{01},\vec{b}\Rb \nn\\
\eea
Introducing $z \,=\, \ln\Lb  r^2_{01}\,Q^2_s\Lb Y,\vec{b}\Rb\Rb$ ($z'  \,=\,\ln\Lb  r^2_{02}\,Q^2_s\Lb Y,\vec{b}\Rb\Rb$ and $\dY\,=\,\bas\Lb Y\,-\,Y_A\Rb$ we rewrite \eq{EQXS01} as follows:
\bea \label{EQXS02}
   &&\frac{\partial}{\partial \dY}\,\sigma_n\Lb \dY,z\Rb\,\,+\,\,\kappa\,\frac{\partial}{\partial z}\,\sigma_n\Lb \dY,z\Rb\,\,=\,\,-\,z\,\sigma_n\Lb \dY,z\Rb\nn\\&&+\,\,\sigma_n\Lb \dY,z\Rb\,\intl_{\xi_0^A}^{z}\,dz'\,\Delta\Lb \dY,z'\Rb \,\,+\,\,\Delta\Lb \dY,z\Rb\,\intl_{\xi_0^A}^{z}dz'\,\sigma_n\Lb \dY,z'\Rb\,\,\nn\\&+&\,\,\sum_{k=1}^{n-1}  \,\intl_{\xi_0^A}^{z}dz' \,\sigma_{n - k}\Lb \dY,z'\Rb\,\sigma_{k}\Lb \dY,z\Rb
\eea
In the following subsections we develop the homotopy approach \cite{HE1,HE2,CLMNEW} for solving \eq{EQXS02} for small $n$.
	
\begin{boldmath}
\subsection{ $\sigma_1\Lb Y, r,b\Rb$}
\subsubsection{ Region I}
\end{boldmath}
For $z>\xi_0^A$ and $\xi<\xi_0^A$ we are in region I (see \fig{sat}). For this region we anticipate the geometric scaling (GS behaviour of $\sigma_n$ and \eq{EQXS02} for $n=1$ takes the form
\beq \label{EQXS03}
   \kappa\,\frac{d\sigma_1\Lb z\Rb}{dz} \,\,=\,\,-\,z\,\sigma_1\Lb z\Rb\,\,+\,\,\sigma_1\Lb z\Rb\,\intl_{\xi_0^A}^z d z' \,\Delta\Lb z'\Rb \,\,+\,\,\Delta\Lb z\Rb\,\intl^z _{\xi_0^A}d z' \,\sigma_{n}\Lb z'\Rb
\eeq
Splitting the integrals into two parts we have:
\bea \label{EQXS04}
   \intl_{\xi_0^A}^z dz' \,\Delta(z')\,\,=\,\,\underbrace{\intl_{\xi_0^A}^{\infty} dz' \,\Delta(z')}_{=\,z_\Delta}\,\,-\,\,\underbrace{\intl_{z}^{\infty} dz' \,\Delta(z')}_{=\,\tilde\Sigma_\Delta(z)};\,\,\,\,\,\,\,\,\,\,\intl_{\xi_0^A}^z dz' \,\sigma_{1}(z')\,\,=\,\,\underbrace{\intl_{\xi_0^A}^{\infty} dz' \,\sigma_1(z')}_{=\,\sigma_{0,1}}\,\,-\,\,\underbrace{\intl_{z}^{\infty} dz' \,\sigma_1(z')}_{=\,\tilde\Sigma_1(z)}
\eea
Using \eq{EQXS04}  we can rewrite  \eq{EQXS03} as
\bea \label{EQXS05}
   \kappa\,\frac{d\sigma_1\Lb z\Rb}{dz} \,\,&=&\,\,-\,\underbrace{\Lb z\,-\,z_\Delta\,+\,\tilde\Sigma_\Delta(z)\Rb}_{=\,T\Lb z \Rb}\,\sigma_1\Lb z\Rb \,\,+\,\,\sigma_{0,1}\,\Delta\Lb z\Rb\,\,-\,\,\Delta\Lb z\Rb\,\tilde\Sigma_1\Lb z\Rb
\eea
In the homotopy approach we rewrite 
 \eq{EQXS05} in the form of \eq{HOM1}
with the following choice of $\mathcal{L}[ \sigma_1]$ and $\mathcal{N}_{\mathcal{L}}[ \sigma_1]$:
  \begin{subequations} 
\bea
\mathcal{L}[ \sigma_1]\,\, &=&\,\, 
\kappa\,\frac{d\sigma_1\Lb z\Rb}{dz} \,\,+\,\, T(z)\,\sigma_1\Lb z\Rb \,\,-\,\,\sigma_{0,1}\,\Delta\Lb z\Rb\label{EQXS11a}\\
\mathcal{N}_{\mathcal{L}}[ \sigma_1]\,\,&= &\,\,\Delta\Lb z\Rb\,\tilde\Sigma_1\Lb z\Rb \label{EQXS11b}\eea
\end{subequations}

Following the procedure that we have discussed in the introduction (see \eq{HOM2} and \eq{HOM3})
with $\mathcal{L}[ \sigma^{(0)}_1]= 0$. \eq{HOM3}  gives  the solution to the non-linear equation at $ p$=1.  The hope is that several  terms in series of \eq{HOM3} will give a good  approximation to the solution of the non-linear  equation. This method has been applied to the solution of the BK and KL equation in Ref.\cite{CLMNEW, CGLM, CGL} with encouraging results.

First consider $p=0$.  Equation $\mathcal{L}[\sigma^{(0)}_1] = 0$ has the form
\beq \label{EQXS1P01} 
\kappa\,\frac{d\sigma_1^{(0,I)}\Lb z\Rb}{dz} \,\,=\,\, -\,T(z)\,\sigma_1^{(0,I)}\Lb z\Rb \,\,+\,\,\sigma_{0,1}\,\Delta\Lb z\Rb
\eeq
Solution of \eq{EQXS1P01} consist of a homogeneous plus a particular solution, viz.: $\sigma_1^{(0,I)}(z)\,=\,\phi_1\Lb z\Rb\,+\,\Delta_{\sigma_1}\Lb z\Rb$. The solution of the homogenouos equation is given by

\beq    \label{EQXS1P02}    
  \kappa\,\frac{d\phi_1\Lb z\Rb}{dz} \,\,=\,\, -\,T(z)\,\phi_1\Lb z\Rb;~~\phi_1\Lb z\Rb \,\,=\,\, C^{(1)}_\phi\,\exp\Lb - \frac{1}{\kappa}\intl^z_{\xi_0^A} d z' \,T\Lb z'\Rb \Rb \eeq  
and the particular solution of the non-homogenouos equation reads:

\bea    \label{EQXS1P03}   
\Delta_{\sigma_1}\Lb z\Rb\,\,&=&\,\,\frac{\sigma_{0,1}}{\kappa} \exp\Lb - \frac{1}{\kappa}\intl^z_{\xi_0^A} d z' \,T\Lb z'\Rb \Rb\,\intl^z_{\xi_0^A} d z' \,\Delta(z')\,\exp\Lb\frac{1}{\kappa}\intl^{z'}_{\xi_0^A} d z'' \, T\Lb z''\Rb\Rb
\eea 
Finally, we write the full solution for the zeroth iteration as
\beq  \label{EQXS1P03x}
\sigma^{(0,I)}_{1}\Lb z\Rb\,\,=\,\,\phi_1\Lb z\Rb\,\,+\,\,\Delta_{\sigma_1}\Lb z\Rb\,\,=\,\,\underbrace{\exp\Lb - \frac{1}{\kappa}\intl^z_{\xi_0^A} d z' \,T\Lb z'\Rb \Rb}_{\tilde {\sigma}^{(0)}_1}\,\Bigg\{ \underbrace{\frac{\sigma_{0,1}}{\kappa}\intl^z_{\xi_0^A} d z'\,\Delta\Lb z'\Rb\,\exp\Lb \frac{1}{\kappa}\intl^{z'}_{\xi_0^A} d z'' \,T\Lb z''\Rb \Rb}_{\tilde{\sigma}^{(0,I)}_1}\,\,+\,\, C^{(1)}_\phi\Bigg\}
\eeq
From the initial condition at $z=\xi_0^A$ (see \eq{MVFXS}) we find: $C_{\phi}^{(1)} \, =\,\h e^{ \xi_0^A}\exp\Lb - \h   e^{ \xi_0^A}\Rb$. We draw your attention to asymptotic behaviour of the solution of \eq{EQXS1P02}: $ \phi_1 \,\xrightarrow{z\gg 1}\,C^{(1)}_\phi\,\exp\Lb - \frac{ \Lb z \,-\, z_\Delta\Rb^2}{2 \,\kappa } \,+\,  \frac{ \Lb \xi^A_0 \,-\, z_\Delta\Rb^2}{2 \,\kappa } \Rb$. This asymptotic behaviour coincides with the solution with the general BFKL kernel \cite{LETU}.

For the next homotopy iteration we need to account for the linear terms in $p$ in \eq{HOM2}:
\beq  \label{EQXS1P04}   
\mathcal{L}[ \sigma_1^{(0)}\,+\,p\,\sigma_1^{(1)}] \,\,+\,\,p\,\mathcal{N}_{\mathcal{L}}[ \sigma_1^{(0)}] \,\,=\,\,0
\eeq
 The equation takes the form:
\beq \label{EQRINPR}
   \kappa\,\frac{d\sigma_1^{(1,I)}\Lb z\Rb}{dz} \,\,=\,\,-\,T\Lb z \Rb\,\sigma_1^{(1,I)}\Lb z\Rb \,\,-\,\,\Delta\Lb z\Rb\,\tilde\Sigma_1^{(0)}\Lb z\Rb
\eeq
where $\tilde\Sigma_1^{(0)}\Lb z\Rb\,=\,\intl_{z}^{\infty} dz' \sigma_1^{(0)}(z')$. The solution of \eq{EQRINPR} takes the form
\beq \label{EQRINPR1}
\sigma_{1}^{(1,I)}\Lb z\Rb\,\,=\,\,-\frac{1}{\kappa} \exp\Lb - \frac{1}{\kappa}\intl^z_{\xi_0^A} d z' \,T\Lb z'\Rb \Rb\,\intl^z_{\xi_0^A} d z' \Delta\Lb z'\Rb\tilde\Sigma_1^{(0)}\Lb z'\Rb\exp\Lb\frac{1}{\kappa}\intl^{z'}_{\xi_0^A} d z'' \, T\Lb z''\Rb\Rb
\eeq
In general, the equation for the $p$-iteration ($p\geq 1$) in Region I takes the form
\beq \label{EQRINPR2}
   \kappa\,\frac{d\sigma_1^{(p,I)}\Lb z\Rb}{dz} \,\,=\,\,-\,T\Lb z \Rb\,\sigma_1^{(p,I)}\Lb z\Rb \,\,-\,\,\Delta\Lb z\Rb\,\tilde\Sigma_1^{(p-1)}\Lb z\Rb
\eeq
with solution 
\beq \label{EQRINPR3}
\sigma_{1}^{(p,I)}\Lb z\Rb\,\,=\,\,-\,\frac{1}{\kappa} \exp\Lb - \frac{1}{\kappa}\intl^z_{\xi_0^A} d z' \,T\Lb z'\Rb \Rb\,\intl^z_{\xi_0^A} d z' \,\Delta\Lb z'\Rb\,\tilde\Sigma_1^{(p-1)}\Lb z'\Rb\,\exp\Lb\frac{1}{\kappa}\intl^{z'}_{\xi_0^A} d z'' \, T\Lb z''\Rb\Rb.
\eeq
~
%%%%%%%%%%%%%%%%%%%%%%%%%%%%%%%%%%%%%%%%%%%%%%%%%%%%%%%%%%

\subsubsection{ Region II }

%%%%%%%%%%%%%%%%%%%%%%%%%%%%%%%%%%%%%%%%%%%%%%%%%%%%%%%%%%
 In the region II for  $\xi>\xi_0^A$ (see \fig{sat}) there is no GS behaviour of $\sigma_n$ and we need to find a function that depends both on $\dY$ and $\xi$. In this region, \eq{EQXS02} for $n=1$ takes the form
\bea \label{EQXS03II}
   &&\dfrac{\pp \sigma_{1}\Lb \dY, z\Rb}{\pp \dY}\,\,+\,\,\kappa \,\dfrac{\pp \sigma_{1}\Lb \dY, z\Rb}{\pp z} \,\,= \nn\\ &&-\,z\,\sigma_1\Lb \dY,z\Rb\,\,+\,\,\sigma_1\Lb \dY,z\Rb\,\intl_{\xi_0^A}^{z}dz'\,\Delta\Lb \dY,z'\Rb \,\,+\,\,\Delta\Lb \dY,z\Rb\,\intl_{\xi_0^A}^{z}dz'\,\sigma_1\Lb \dY,z'\Rb
\eea
Splitting  the integrals in the following form:
\begin{subequations} 
\bea \label{EQXS04II}
   \intl_{\xi_0^A}^z dz' \,\Delta\Lb\dY, z'\Rb\,\,&=&\,\,\underbrace{\intl_{\xi_0^A}^{\infty} dz' \,\Delta\Lb \dY,z'\Rb}_{=\,z_\Delta\Lb \dY\Rb}\,\,-\,\,\underbrace{\intl_{z}^{\infty} dz' \,\Delta\Lb \dY,z'\Rb}_{=\,\tilde\Sigma_\Delta\Lb \dY,z\Rb};\\ \intl_{\xi_0^A}^z dz' \,\sigma_{1}\Lb \dY,z'\Rb\,\,&=&\,\,\underbrace{\intl_{\xi_0}^\infty dz'\,\sigma_1\Lb z'\Rb}_{=\,\sigma_{0,1}}\,\,+\,\,\underbrace{\intl_{\xi_0^A}^{\infty} dz' \,\Lb\sigma_1\Lb\dY,z'\Rb\,\,-\,\,\sigma_1\Lb z'\Rb\Rb}_{=\,\delta\Sigma_{1}\Lb\dY\Rb}\,\,-\,\,\underbrace{\intl_{z}^{\infty} dz' \sigma_1\Lb\dY,z'\Rb}_{=\,\tilde\Sigma_1\Lb\dY,z\Rb}
\eea
\end{subequations} 
we can  rewrite \eq{EQXS03II} as
\bea \label{EQXS05II}
   \dfrac{\pp \sigma_{1}\Lb \dY, z\Rb}{\pp \dY}\,\,+\,\,\kappa \,\dfrac{\pp \sigma_{1}\Lb \dY, z\Rb}{\pp z} \,\,&=&\,\,-\,\underbrace{\Lb z\,-\,z_\Delta\Lb \dY\Rb\,+\,\tilde\Sigma_\Delta\Lb \dY,z\Rb \Rb}_{=\,T\Lb \dY,z \Rb}\,\sigma_1\Lb \dY,z\Rb\,\,+\,\,\sigma_{0,1}\,\Delta\Lb \dY,z\Rb\nn\\  &-&\,\,\Delta\Lb \dY,z\Rb\,\delta\Sigma_1\Lb \dY\Rb\,\,+\,\,\Delta\Lb \dY,z\Rb\,\tilde\Sigma_1\Lb \dY,z\Rb
\eea
Developing the homotopy approach in this region in the similar way as in the region I, we choose
  \begin{subequations} 
\bea
\mathcal{L}[ \sigma_1]\,\, &=&\,\, 
\dfrac{\pp \sigma_{1}\Lb \dY, z\Rb}{\pp \dY}\,\,+\,\,\kappa \,\dfrac{\pp \sigma_{1}\Lb \dY, z\Rb}{\pp z} \,\,+\,\, T\Lb \dY,z\Rb \,\sigma_1\Lb \dY,z\Rb \,\,-\,\,\sigma_{0,1}\,\Delta\Lb \dY,z\Rb\label{EQXS11aII}\\
\mathcal{N}_{\mathcal{L}}[ \sigma_1]\,\,&= &\,\,\Delta\Lb \dY,z\Rb\,\delta\Sigma_1\Lb \dY\Rb\,\,-\,\,\Delta\Lb \dY,z\Rb\,\tilde\Sigma_1\Lb \dY,z\Rb \label{EQXS11bII}\eea
\end{subequations}
For  the zeroth iteration (i.e. for the solution of the equation  $\mathcal{L} [ \sigma^{(0)}_1]= 0$)  we search a solution in  the form:  $\sigma_1^{\Lb 0,II\Rb}(\dY,z)=\phi_1\Lb \dY,z\Rb+\Delta_{\sigma_1}\Lb \dY,z\Rb$.  The homogeneous equation for $\phi_1\Lb \dY,z\Rb$
looks as follows:
  \beq \label{SOLXS11} 
 \dfrac{\pp \phi_{1}\Lb \dY, z\Rb}{\pp \dY}\,\,+\,\,\kappa \,\dfrac{\pp \phi_{1}\Lb \dY, z\Rb}{\pp z}\,\,=\,\,- \,T\Lb\dY,z\Rb\,\phi_{1}\Lb \dY, z\Rb\eeq
with  the general solution:

   \beq \label{SOLXS12} 
\phi_1\Lb \dY, z\Rb\,\,=\,\,\Phi_1\Lb - \,\kappa\,\dY +z\Rb\,\underbrace{\exp\Lb-\,\intl_0^{\dY}d\dY' \,T\Lb\dY', - \,\kappa\,\Lb \dY\,-\,\dY'\Rb \,+\,z\Rb\Rb}_{=\,\sigma^{0 '}_1\Lb\dY,z\Rb} \eeq
with arbitrary  function $\Phi_1$. For the particular solution, searching in the form: $ \Delta_{\sigma_1}\Lb \dY,z\Rb=\sigma^{0 '}_1\Lb \dY, z\Rb\,\tilde{\sigma}^{0'}_1\Lb \dY, z\Rb$ we have the following equation for $\tilde\sigma^{0'}_1\Lb \dY, z\Rb$:
\beq
  \label{EQXS1P04} 
\dfrac{\pp \tilde{\sigma}^{0'}_{1}\Lb \dY, z\Rb}{\pp \dY}\,\,+\,\,\kappa \,\dfrac{\pp \tilde{\sigma}^{0'}_{1}\Lb \dY, z\Rb}{\pp z}\,=\,\,\underbrace{\frac{\sigma_{0,1}}{\sigma^{0'}_1\Lb \dY, z\Rb} \Delta\Lb \dY, z\Rb}_{=\,R_1(\dY, z)}\eeq
Using the $\omega$ representation for $\tilde{\sigma}^{0'}_1$:
\beq \label{OMREP0}
\tilde{\sigma}^{0'}_1\Lb\dY,  z\Rb\,\,=\,\,\intl^{\epsilon + i \infty}_{\epsilon - i \infty} \frac{ d \omega}{2\,\pi\,i} \,e^{ \omega\,\dY}\, \hat{\sigma}^{0'}_1\Lb \omega, z\Rb \eeq
we obtain:
\beq
  \label{EQXS1P05} 
\omega\, \hat{\sigma}^{0'}_{1}\Lb \omega, z\Rb\,\,+\,\,\kappa \,\dfrac{\pp \hat{\sigma}^{0'}_{1}\Lb \omega, z\Rb}{\pp z}\,=\, \underbrace{\frac{\sigma_{0,1}}{\sigma^{0'}_1\Lb \omega, z\Rb}\, \Delta\Lb \omega, z\Rb}_{=\,R_1(\omega, z)}
\eeq
As function of $\dY$ \eq{EQXS1P05} gives
\beq
  \label{EQXS1P06} 
  \tilde{\sigma}^{0'}_{1}\Lb \dY, z\Rb\,\,=\,\,\frac{1}{\kappa}\,\intl^z_{\xi_0^A}d z' \,R_1\Lb \dY - \frac{z - z'}{\kappa},z'\Rb\,\,+\,\,C_{\sigma_1}
  \eeq  
Therefore, the general solution to \eq{EQXS1P02} takes the form:
\beq \label{SOLXS121}   
    \sigma^{(0,II)}_{1}\Lb \dY, z\Rb \,\,=\,\,   \Phi_1\Lb - \,\kappa\,\dY \,+\,z\Rb\,\sigma^{0 '}_1\Lb \dY, z\Rb\,\,+\,\, \sigma^{0 '}_1\Lb \dY, z\Rb\,\tilde{\sigma}^{0'}_1\Lb \dY, z\Rb\eeq

   The solution of \eq{SOLXS121} has to satisfy the initial condition of \eq{MVFXS} at $\dY=0$ and boundary conditions that has the following form:
   
    \beq \label{SOLBC} 
    \sigma^{(0,I)}_{1}\Lb z=\kappa\,\dY+\xi_0^A\Rb   \,\,\,=\,\,\,   \sigma^{(0,II)}_{1}\Lb \dY,z=\kappa\,\dY+\xi^A_0\Rb  
    \eeq
The initial conditions can determine the function $\Phi_1\Lb - \kappa\,\dY \,+\,z\Rb =\Phi_1\Lb\xi\Rb$, viz.:
      \beq \label{SOLIC}  
    \sigma^{0 '}_1\Lb \dY=0, z=\xi\Rb\,   \Phi_1\Lb\xi \Rb\,\,=\,\,\h e^{\xi} \exp\Lb- \h e^{\xi}\Rb \,\,-\,\, \sigma^{0 '}_1\Lb \dY=0, z=\xi\Rb\,\tilde{\sigma}^{0'}_1\Lb \dY=0, z=\xi\Rb      \eeq
The boundary condition at $\xi = \xi^A_0$ allows us to determine $C_{\sigma_1}$. However, before doing this,  we need to consider the integration over $z'$ in \eq{EQXS1P06}. First the argument $\dY - \frac{z - z'}{\kappa} $ has to be larger than 0. Indeed,  only for such z' we are in the region II. For $\dY - \frac{z - z'}{\kappa}  < 0 $    we are in the region I for $\Delta$ and   $\Delta$ does not depend on $\dY$. Therefore, we can replace \eq{EQXS1P06} by the following expression:   
      \beq
  \label{EQXS1P07} 
  \tilde{\sigma}^{0'}_{1}\Lb \dY, z\Rb\,\,=\,\,\frac{1}{\kappa}\,\intl^z_{ \kappa\,\dY + \xi_0^A}\!\!\!\!\!\!\!d z' \,\frac{\sigma_{0,1}}{\sigma^{0'}_1\Lb \dY, z\Rb} \Delta\Lb \dY - \frac{z - z'}{\kappa},z'\Rb\,\,+\,\,\underbrace{\frac{1}{\kappa}\,\intl^{ \kappa\,\dY + \xi_0^A}_{\xi_0^A}\!\!\!\!\!\!\!d z'\,\frac{\sigma_{0,1}}{\tilde\sigma^{(0)}_1\Lb z'\Rb}\, \Delta \Lb z'\Rb}_{=\,\tilde\sigma_1^{(0,I)}(\kappa\,\dY\,+\,\xi_0^A)}
\,\,+\,\,C_{\sigma_1}
  \eeq  
Indeed, from \eq{SOLBC} 
      we have:
      \bea \label{SOLBC1} 
   && \sigma^{(0,I)}_{1}\Lb z=\kappa\,\dY+\xi_0^A\Rb   \,\,=\,\, \tilde{\sigma}^{(0)}_{1}\Lb \kappa \,\dY + \xi^A_0\Rb\Lb \tilde{\sigma}^{(0,I)}_{1}\Lb \kappa \,\dY + \xi_0^A\Rb\,\,+\,\,C_{\phi}^{(1)}\Rb\,=  
   \, \nn\\
   &&  \sigma^{(0,II)}_{1}\Lb \dY,z=\kappa\,\dY+\xi_0^A\Rb  \,\,=\,\,\sigma^{0'}_{1}\Lb \dY,\kappa\, \dY + \xi_0^A\Rb\,  \Big\{\Phi_1\Lb \xi_0^A\Rb\,\,+\,\,\tilde{\sigma}^{(0,I)}_{1}\Lb \kappa \,\dY + \xi^A\Rb\,\,+\,\,C_{\sigma_1}\Big\}    \eea           
Constant $C_{\sigma_1}$ can be determined from \eq{SOLBC1} with solution: $ C_{\sigma_1}\,=\,C_{\phi}^{(1)}\,-\,\Phi_1\Lb \xi^A_0\Rb $.   \\
The equation for the first iteration in the homotopy approach takes the form:
\bea \label{EQRINPR1II}
&&\dfrac{\pp \sigma^{(1,II)}_{1}\Lb \dY, z\Rb}{\pp \dY}\,\,+\,\,\kappa \,\dfrac{\pp \sigma^{(1,II)}_{1}\Lb \dY, z\Rb}{\pp z}\,\,=\nn\\ &&- \,T\Lb\dY,z\Rb\sigma^{(1,II)}_{1}\Lb \dY,z\Rb\,\,-\,\,\Delta\Lb \dY,z\Rb\,\Lb\delta\Sigma_1^{(0)}\Lb \dY\Rb\,\,-\,\,\tilde\Sigma_1^{(0)}\Lb \dY,z\Rb\Rb\eea
Searching for the solution in the form: $\sigma^{(1,II)}_{1}\Lb Y,r_{01},b\Rb\,\,=\,\,\sigma^{0'}_{1}\Lb \dY,z\Rb\,\sigma^{1'}_{1}\Lb \dY,z\Rb$ we obtain
\bea \label{EQRINPR2II}
&&\dfrac{\pp \sigma^{1'}_{1}\Lb \dY, z\Rb}{\pp \dY}\,\,+\,\,\kappa \,\dfrac{\pp \sigma^{1'}_{1}\Lb \dY, z\Rb}{\pp z}\,\,=\,\,\underbrace{-\,\frac{\Delta\Lb \dY,z\Rb\,\Lb\delta\Sigma_1^{(0)}\Lb \dY\Rb\,\,-\,\,\tilde\Sigma_1^{(0)}\Lb \dY,z\Rb\Rb}{\sigma^{0'}_{1}\Lb \dY,z\Rb}}_{=\,R_1^{\Lb 1\Rb}(\dY,z)}
\eea
Using the $\omega$ representation for $\sigma^{1'}_1$:
\beq \label{EQRINPR3II}
\sigma^{1'}_1\Lb\dY,  z\Rb\,\,=\,\,\intl^{\epsilon + i \infty}_{\epsilon - i \infty} \frac{ d \omega}{2\,\pi\,i} \,e^{ \omega\,\dY} \,\tilde{\sigma}^{1'}_1\Lb \omega, z\Rb \eeq
we obtain:
\beq
  \label{EQRINPR4II} 
\omega\, \tilde{\sigma}^{1'}_{1}\Lb \omega, z\Rb\,\,+\,\,\kappa \,\dfrac{\pp \tilde{\sigma}^{1'}_{1}\Lb \omega, z\Rb}{\pp z}\,=\, R_1^{\Lb 1\Rb}(\omega, z)
\eeq
with solution
\beq \label{EQRINPR5II}
\sigma^{1'}_{1}\Lb \dY, z\Rb\,\,=\,\,\frac{1}{\kappa}\,\intl^z_0d z' \,R_1^{\Lb 1\Rb}\Lb \dY - \frac{z - z'}{\kappa},z'\Rb\,\,\,+\,\,C^{(1)}_{\sigma_1}
\eeq

In general, for the $p$-iteration ($p\geq 1$) in Region II we have
\bea \label{EQRINPRII}
&&\dfrac{\pp \sigma^{(p,II)}_{1}\Lb \dY, z\Rb}{\pp \dY}\,\,+\,\,\kappa \,\dfrac{\pp \sigma^{(p,II)}_{1}\Lb \dY, z\Rb}{\pp z}\,\,=\nn\\ &&- \,T\Lb\dY,z\Rb\sigma^{(p,II)}_{1}\Lb \dY,z\Rb\,\,-\,\,\Delta\Lb \dY,z\Rb\,\Lb\delta\Sigma_1^{(p-1)}\Lb \dY\Rb\,\,-\,\,\tilde\Sigma_1^{(p-1)}\Lb \dY,z\Rb\Rb
\eea
Searching in the form: $\sigma^{(p,II)}_{1}\Lb\dY,z\Rb\,\,=\,\,\sigma^{0'}_{1}\Lb\dY,z\Rb\,\sigma^{p'}_{1}\Lb\dY,z\Rb$ we obtain
\beq
\sigma^{p'}_{1}\Lb\dY,z\Rb\,\,=\,\,\frac{1}{\kappa}\intl^{z}_0 d z' \,R_1^{(p)}(\dY,z')\,\,+\,\,C_{\sigma_1}^{(p)}
\eeq
where $C_{\sigma_1}^{(p)}$ is found by requiring that $\sigma_1^{(p,II)}\Lb\dY=0,z=\xi\Rb\,=\,0$. 

 \begin{figure}
 	\begin{center}
 	\leavevmode
	\begin{tabular}{c c c }
 		\includegraphics[width=6cm]{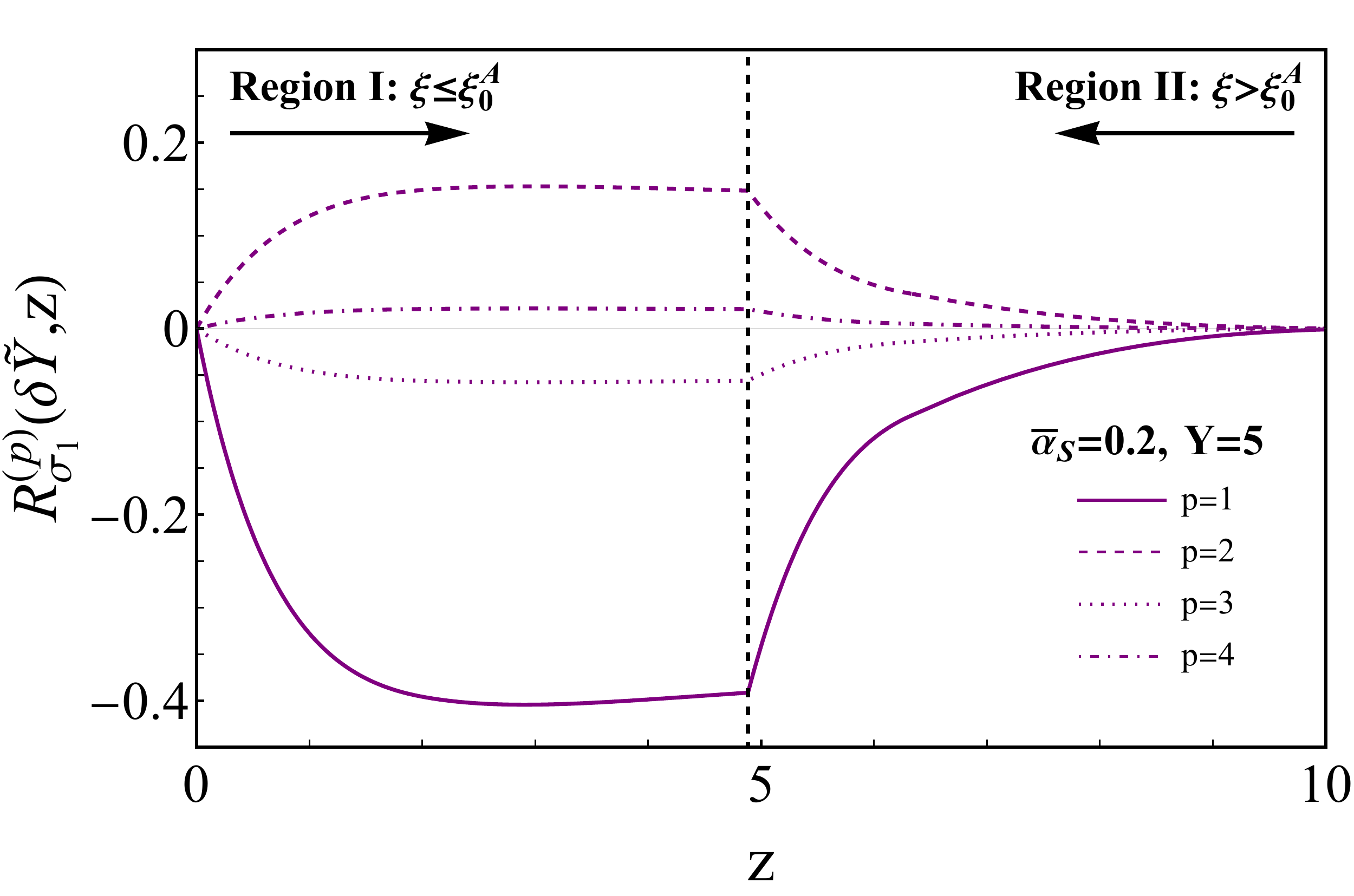}&\includegraphics[width=6cm]{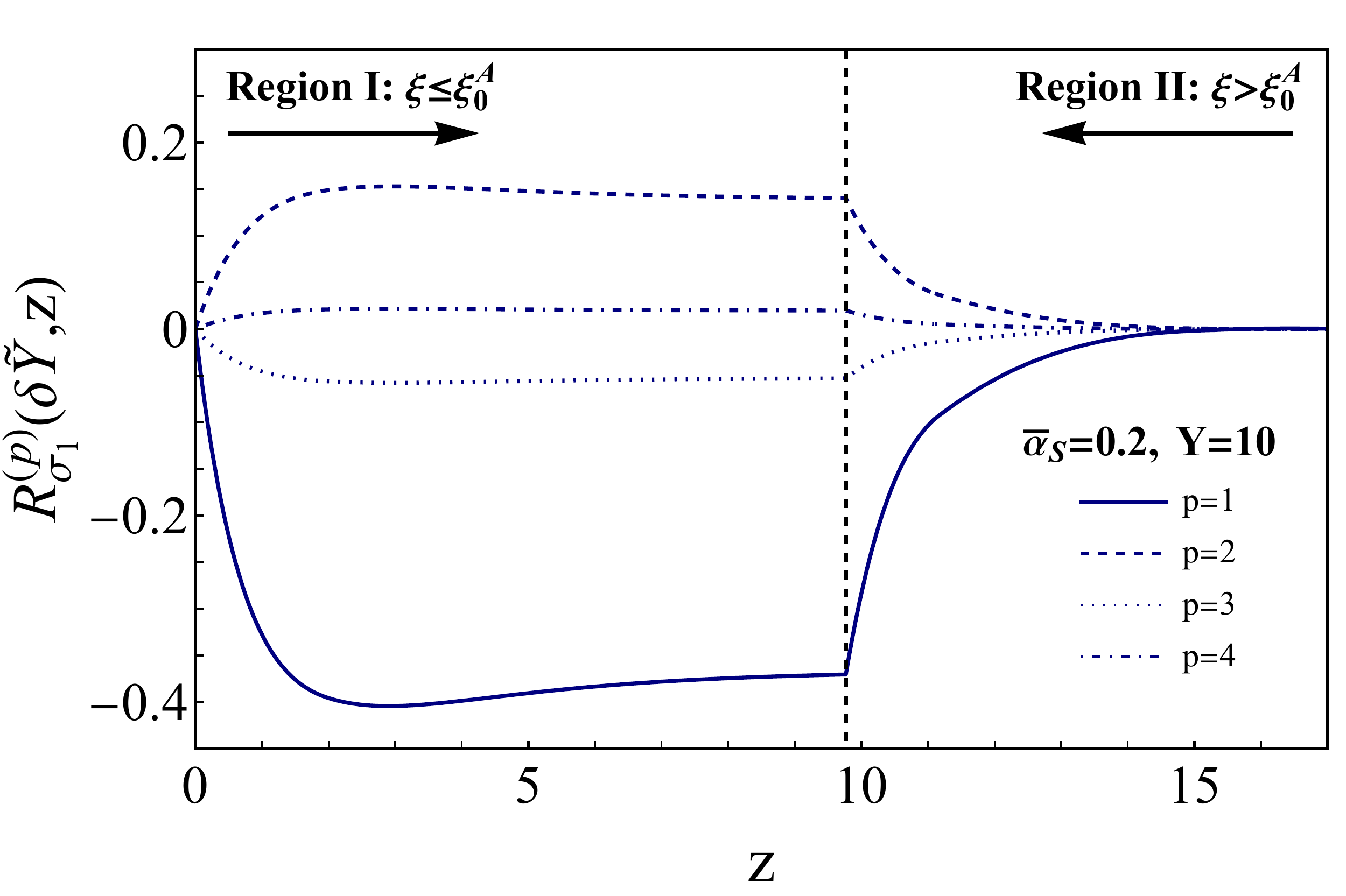}&\includegraphics[width=6.cm]{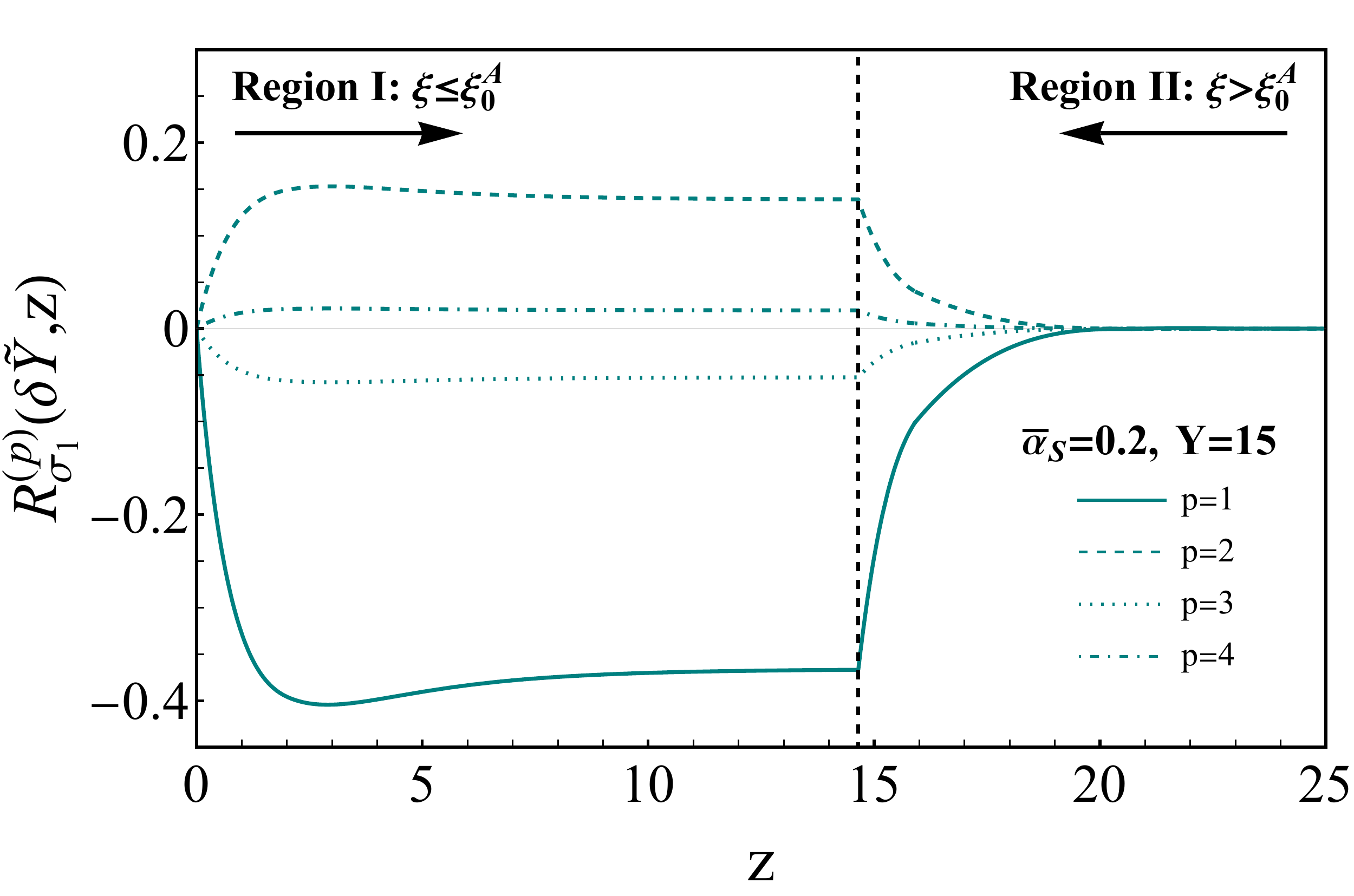} \\
		\fig{fin}-a & \fig{fin}-b&\fig{fin}-c\\
		\end{tabular}	\end{center}
	\caption{ The ratio of $R_{\sigma_1}^{(p)}\Lb \dY, z\Rb$ versus $z$ at different values of $\dY$. $N_0 =0.22$, $\bar{\gamma}=0.63$, $\xi_0^A=0$, $z_0=2$, $z_\Delta=0.87$.
}
\label{fin}
\end{figure}

In \fig{fin} we plot  $\displaystyle{R_{\sigma_1}^{(p)}\Lb \dY, z\Rb\,=\,\frac{\sigma_1^{(p,I)}\Lb \dY, z\Rb}{\sigma_1^{(0,I)}\Lb \dY, z\Rb}\,\Theta\Lb\xi_0^A\,-\,\xi\Rb\,+\,\frac{\sigma_1^{(p,II)}\Lb \dY, z\Rb}{\sigma_1^{(0,II)}\Lb \dY, z\Rb}\,\Theta\Lb\xi\,-\,\xi_0^A\Rb}$ 
up to fourth iteration. One can see that the fourth iteration gives the accuracy less than 2\%.
Therefore, our choice of $\mathcal{L}$ and $ \mathcal{N}_{\mathcal{L}}$ looks rather good making homotopy approach a regular procedure of getting the solution.

~
~

\subsubsection{ Corrections for the general BFKL kernel }

\begin{figure}[ht]
   \centering
  \leavevmode
      \includegraphics[width=10cm]{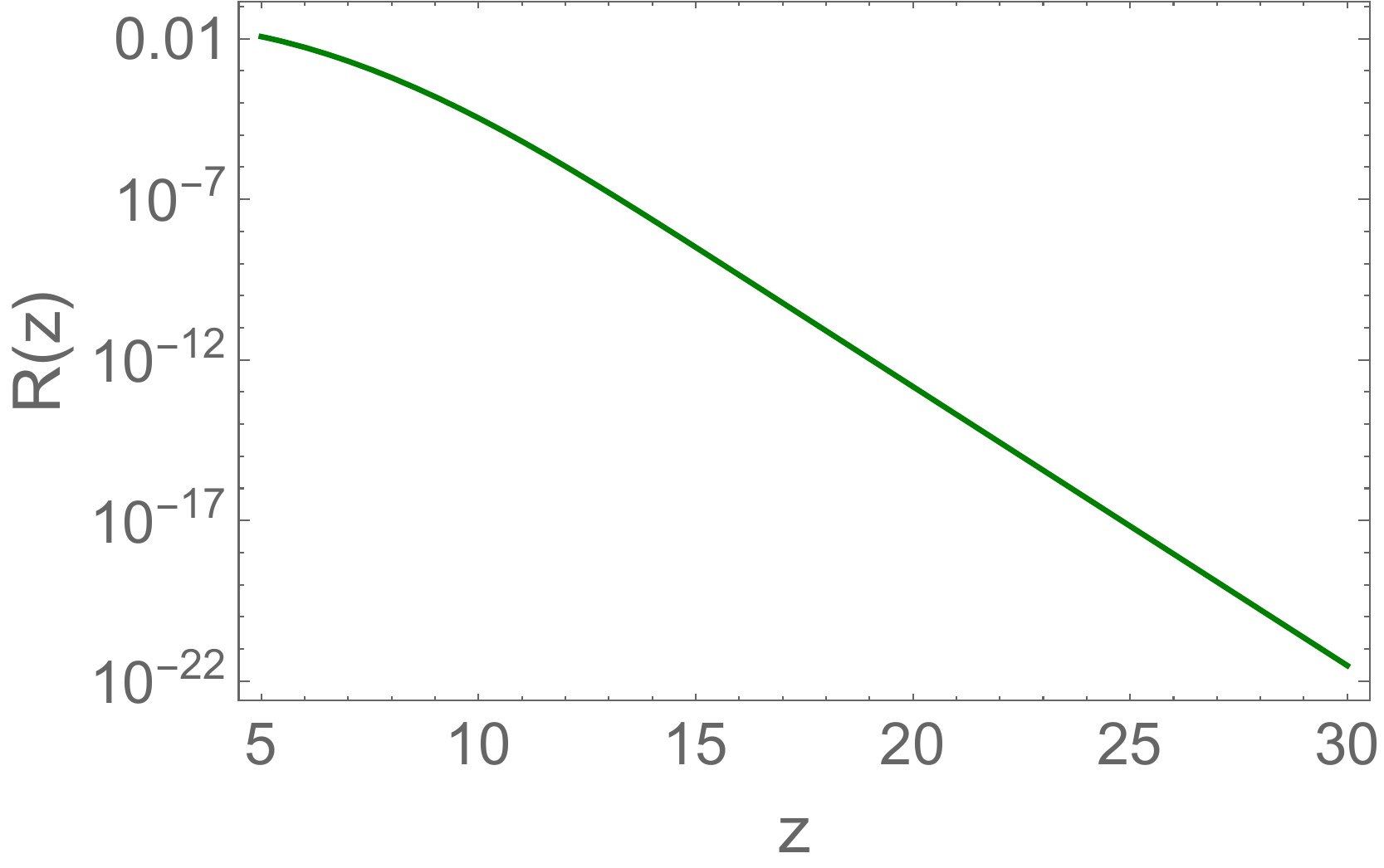}  
      \caption{ The ratio $R$ of \eq{CGK2}  versus $z$ in the region I of \fig{sat}.}
\label{r}
   \end{figure}

%%%%%%%%%%%%%%%%%%%%%%%%%%%%%%%%%%%%%%%%%%%%%%%%%%%%%%%%%%
For the general BFKL kernel we need to add  corrections for the linear equation in the first iteration. They lead  to the following contribution to the linear part of the equation for $\sigma_1$:
\beq \label{CGK1}
\Delta I\,\,=\,\,\int  d^2 r'\{ K\Lb \vec{r}, \vec{r}'\Rb - K_{LT}\Lb \vec{r},\vec{r}'\Rb\} \,\sigma^{(0)}_1\Lb \dY, \vec{r}; \vec{b}\Rb
\eeq
In \fig{r} we plot the ratio of \eq{CGK1} to the contribution in the leading twist kernel approach, viz.:

\beq \label{CGK2}
R \Lb \dY, \vec{r}, \vec{b}\Rb\,\,=\,\,\frac{\Delta I\Lb \dY, \vec{r}, \vec{b}\Rb}{\int  d^2 r'  K\Lb \vec{r}, \vec{r}'\Rb  \sigma^{(0)}_1\Lb \dY, \vec{r}', \vec{b}\Rb}
\eeq

      %%%%%%%%%%%%%%%%%%%%%%%%%%%%%%%%%%%%%%%%%
%%%%%%%%%%%%%%%%%%%%%%%%%%%%%%%%%%%%%%%%%%
One can see that  these corrections are small, especially at large values of $z$. It does not look surprisingly since the solution $\propto \exp\Lb -\,\frac{z^2}{2\,\kappa}\Rb$ stems from the general BFKL kernel.

%%%%%%%%%%%%%%%%%%%%%%%%%%%%%%%%%%%%%%%%%%%%%%%%%%%%%%%%%%
\begin{boldmath}
\subsection{  $\sigma_2\Lb Y, r,b\Rb$}
\subsubsection{Region I}
\end{boldmath}
%%%%%%%%%%%%%%%%%%%%%%%%%%%%%%%%%%%%%%%%%%%%%%%%%%%%%%%%%%   

From \eq{EQXS02} we obtain the following equation for $\sigma_2$:
\beq\label{XS20}
   \kappa\,\frac{d\sigma_2\Lb z\Rb}{dz} \,\,=\,\,-\,z\,\sigma_2\Lb z\Rb\,\,+\,\,\sigma_2\Lb z\Rb\,\intl_{\xi_0^A}^{z}dz'\,\Delta\Lb z'\Rb \,\,+\,\,\Delta\Lb z\Rb\,\intl_{\xi_0^A}^{z}dz'\,\sigma_2\Lb z'\Rb\,\,+\,\,\intl_{\xi_0^A}^{z}dz'\,\sigma_1\Lb z'\Rb\,\sigma_1\Lb z\Rb
\eeq
As discussed in the previous section, $\sigma_1\Lb Y_A, r_{12},b\Rb\,\sigma_{1}\Lb Y_A, r_{02},b\Rb \sim \sigma_2\Lb Y_A\,r_{01}, b \Rb $. Bearing this in mind we suggest to add the contribution of this term to the definition of the operator $\mathcal{L}$:
\begin{subequations} 
\bea
\mathcal{L}[ \sigma_2]\,\, &=&\,\, 
\kappa\,\frac{d\sigma_2\Lb z\Rb}{dz} \,\,+\,\, T(z)\,\sigma_2\Lb z\Rb \,\,-\,\,U_2\Lb z\Rb\label{XS202} \\
\mathcal{N}_{\mathcal{L}}[ \sigma_2]\,\,&= &\,\,\Delta\Lb z\Rb\,\tilde\Sigma_2\Lb z\Rb \label{XS203}
\eea
\end{subequations} 
where we have defined $U_2(z)\,=\,\sigma_{0,2}\,\Delta\Lb z\Rb\,+\,\intl_{\xi_0^A}^z dz'\,\sigma_1\Lb z'\Rb\,\sigma_1\Lb z\Rb$ and $\tilde\Sigma_2(z)\,=\,\intl_{z}^\infty dz'\,\sigma_2\Lb z'\Rb$. Solution to $\mathcal{L}[\sigma_2]=0$ has the form: $\sigma_2^{(0,I)}(z)=\phi_2\Lb z\Rb+\Delta_{\sigma_2}\Lb z\Rb$, where 

\beq\label{XS204}
\kappa\,\frac{d\phi_2\Lb z\Rb}{dz} \,\,=\,\, -\,T(z)\,\phi_2\Lb z\Rb;~~\phi_2\Lb z\Rb \,\,=\,\, C^{(2)}_\phi\,\exp\Lb - \frac{1}{\kappa}\intl^z_{\xi_0^A} d z' \,T\Lb z'\Rb \Rb
\eeq  
and the particular solution is given by:

\bea\label{XS205}
\Delta_{\sigma_2}\Lb z\Rb\,\,&=&\,\,\frac{1}{\kappa} \exp\Lb - \frac{1}{\kappa}\intl^z_{\xi_0^A} d z' \,T\Lb z'\Rb \Rb\,\intl^z_{\xi_0^A} d z' \,U_2\Lb z'\Rb\,\exp\Lb\frac{1}{\kappa}\intl^{z'}_{\xi_0^A} d z'' \, T\Lb z''\Rb\Rb
\eea 
Finally, 
\beq  \label{XS206}
\sigma^{(0,I)}_{2}\Lb z\Rb\,\,=\,\,\phi_2\Lb z\Rb\,\,+\,\,\Delta_{\sigma_2}\Lb z\Rb\,\,=\,\,\underbrace{\exp\Lb - \frac{1}{\kappa}\intl^z_{\xi_0^A} d z' \,T\Lb z'\Rb \Rb}_{\tilde {\sigma}^{(0)}_2}\,\Bigg\{ \underbrace{\frac{1}{\kappa}\intl^z_{\xi_0^A} d z'\,U_2\Lb z'\Rb\,\exp\Lb \frac{1}{\kappa}\intl^{z'}_{\xi_0^A} d z'' \,T\Lb z''\Rb \Rb}_{\tilde{\sigma}^{(0,I)}_2}\,\,+\,\, C^{(2)}_\phi\Bigg\}
\eeq

Using the initial condition of \eq{MVFXS} at $\xi = \xi_0^A$ we find: $C_\phi^{(2)}=\frac{1}{8} \,e^{ 2 \,\xi_0^A} \,\exp\Lb - \h e^{  \xi_0^A} \Rb$.

Performing the following iterations in a similar way as done for $\sigma_1$ we obtain ($p\geq 1$):
\beq \label{EQRINPR3}
\sigma_{2}^{(p,I)}\Lb z\Rb\,\,=\,\,-\,\frac{1}{\kappa} \exp\Lb - \frac{1}{\kappa}\intl^z_{\xi_0^A} d z' \,T\Lb z'\Rb \Rb\,\intl^z_{\xi_0^A} d z' \,U_2\Lb z'\Rb\,\tilde\Sigma_2^{(p-1)}\Lb z'\Rb\,\exp\Lb\frac{1}{\kappa}\intl^{z'}_{\xi_0^A} d z'' \, T\Lb z''\Rb\Rb.
\eeq
~
 \begin{figure}
 	\begin{center}
 	\leavevmode
	\begin{tabular}{c c c }
 		\includegraphics[width=6cm]{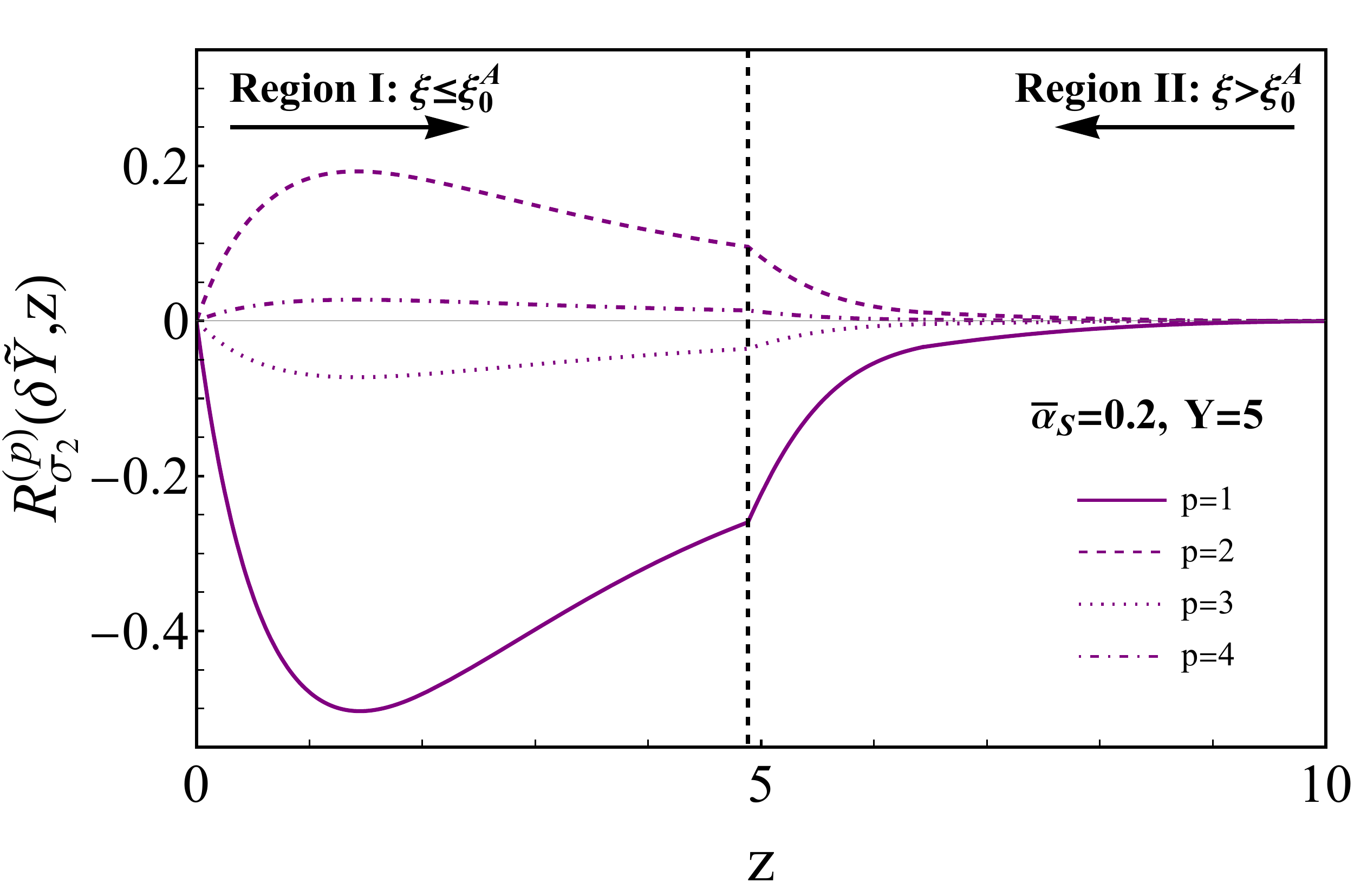}&\includegraphics[width=6cm]{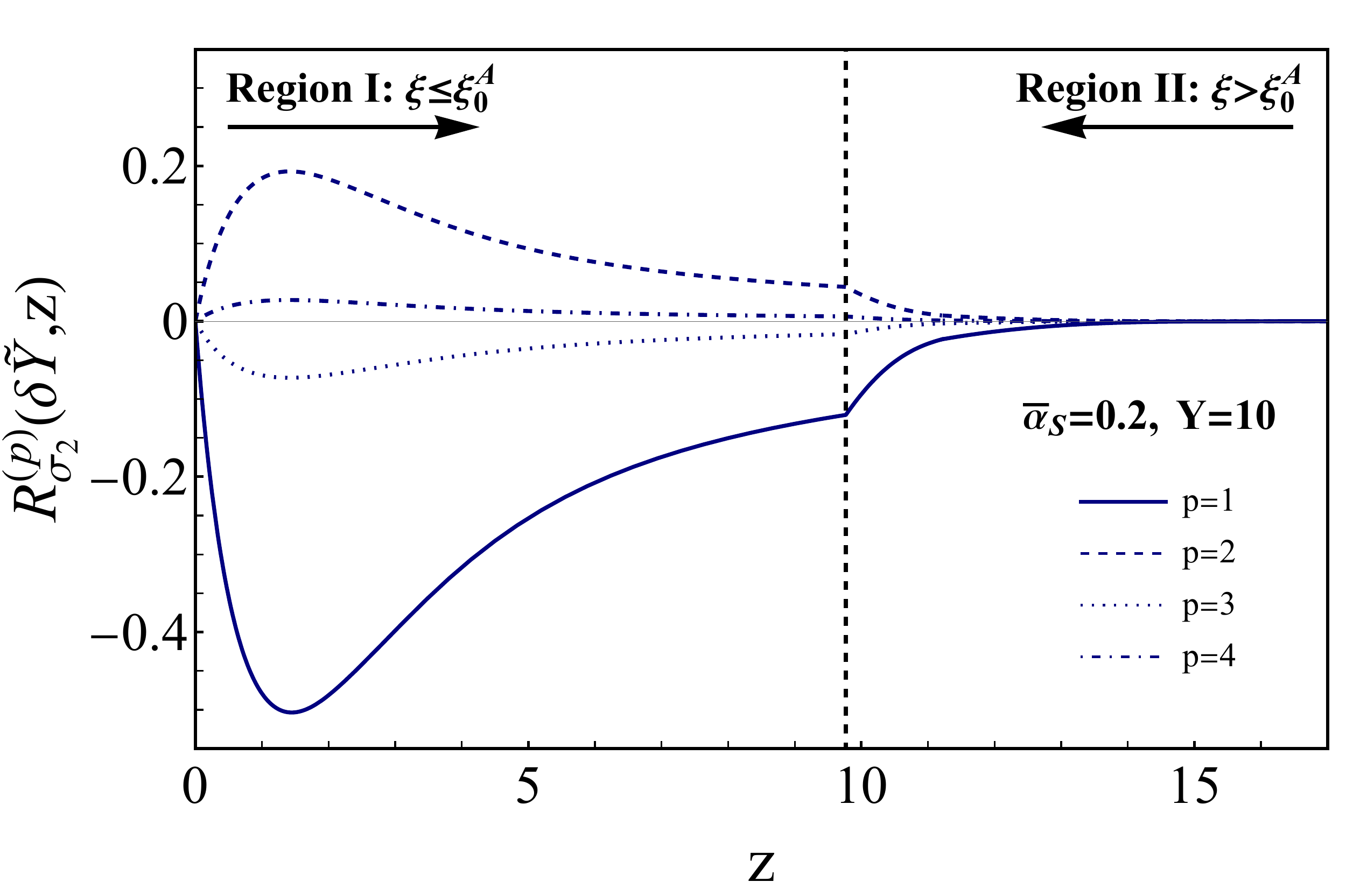}&\includegraphics[width=6.cm]{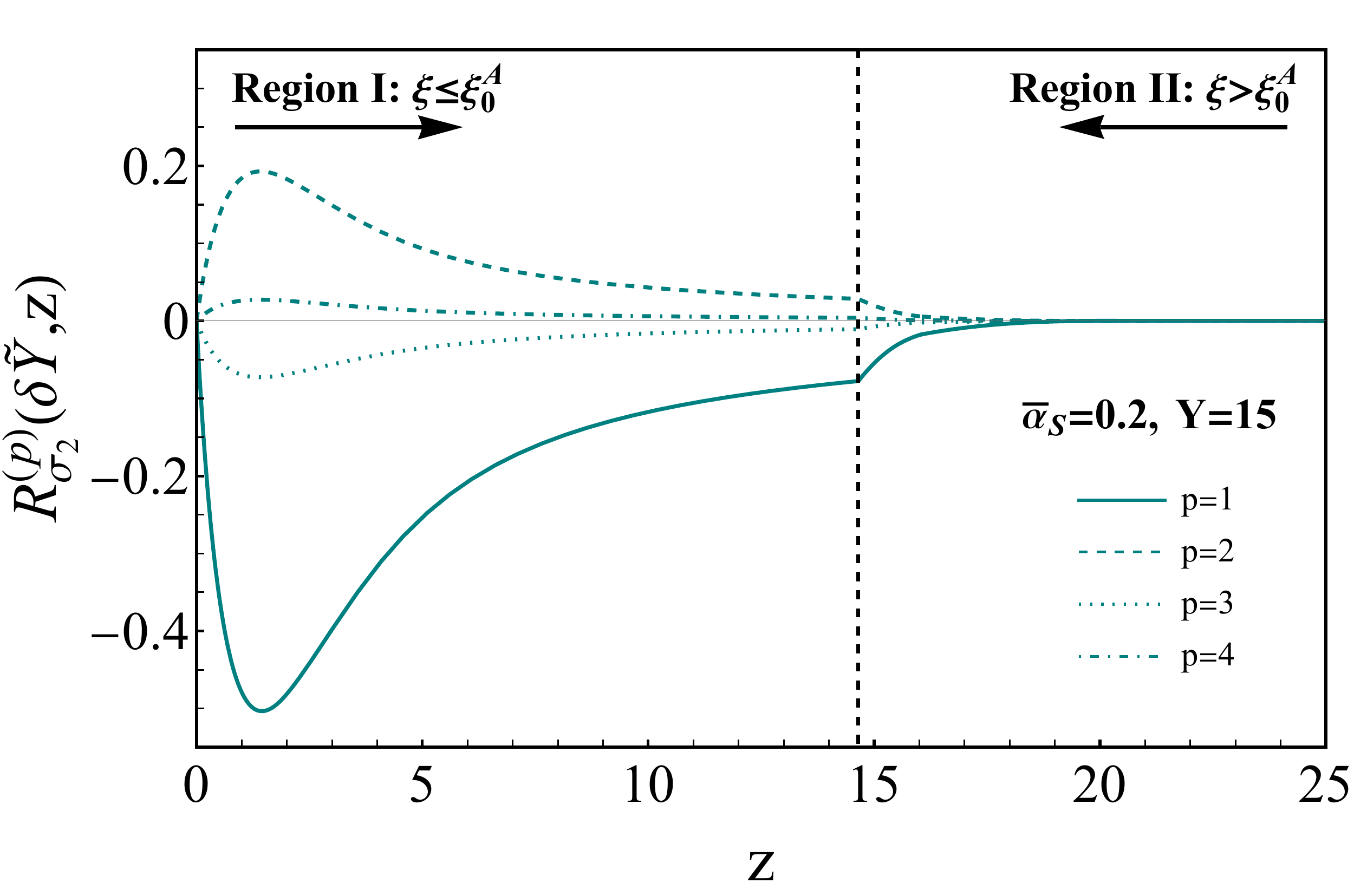} \\
		\fig{fin1}-a & \fig{fin1}-b&\fig{fin1}-c\\
		\end{tabular}	\end{center}
	\caption{ The ratio of $R_{\sigma_2}^{(p)}\Lb \dY, z\Rb$ versus $z$ at different values of $\dY$. $N_0 =0.22$, $\bar{\gamma}=0.63$, $\xi_0^A=0$, $z_0=2$, $z_\Delta=0.87$.
}
\label{fin1}
\end{figure}

%%%%%%%%%%%%%%%%%%%%%%%%%%%%%%%%%%%%%%%%%%%
\subsubsection{ Region II }

%%%%%%%%%%%%%%%%%%%%%%%%%%%%%%%%%%%%%%%%%%%%%%%%%%%%%%%%%%
In region II \eq{XS202} takes the form:

\beq \label{EQXS24} 
\dfrac{\pp \sigma^{(0,II)}_{2}\Lb\dY,  z\Rb}{\pp \dY}\,\,+\,\,\kappa\,\dfrac{\pp \sigma^{(0,II)}_{2}\Lb \dY,  z\Rb}{\pp z}\,\,=\,\,- \,T\Lb\dY,z\Rb\,\sigma^{(0,II)}_{2}\Lb \dY,z\Rb\,\,+\,\,U_2\Lb \dY,z\Rb\eeq
The solution to homogeneous part of this equation as the same form as  \eq{SOLXS12}:

   \beq \label{EQXS25} 
\Phi_2\Lb - \,\kappa\,\dY \,+\,z\Rb\,\sigma^{0 '}_2\Lb \dY, z\Rb\,\,=\,\,\Phi_2\Lb -\, \kappa\,\dY \,+\,z\Rb\,\underbrace{\exp\Lb-\,\intl_0^{\dY}d\dY' \,T\Lb\dY', - \,\kappa\,\Lb \dY\,-\,\dY'\Rb \,+\,z\Rb\Rb}_{=\,\sigma^{0 '}_2\Lb\dY,z\Rb} \eeq
with arbitrary  function $\Phi_2 $.

The general solution to \eq{EQXS24} has the form: 
 $ \sigma^{0 '}_2\Lb \dY, z\Rb\,\tilde{\sigma}^{0'}_2\Lb \dY, z\Rb$ with the following equation for $\tilde{\sigma}^{0'}_1\Lb \dY, z\Rb$:
\beq
  \label{EQXS26} 
\dfrac{\pp \tilde{\sigma}^{0'}_{2}\Lb \dY, z\Rb}{\pp \dY}\,\,+\,\,\kappa \,\dfrac{\pp \tilde{\sigma}^{0'}_{2}\Lb \dY, z\Rb}{\pp z}\,\,=\,\,\underbrace{\frac{U_2\Lb \dY, z\Rb}{\sigma^{0'}_2\Lb \dY, z\Rb}}_{R_2\Lb \dY, z\Rb} \eeq
Solution to \eq{EQXS26} can be obtained following the steps of \eq{OMREP0} and \eq{EQXS1P05}. It has the form:

\beq
  \label{EQXS27} 
  \tilde{\sigma}^{0'}_{2}\Lb \dY, z\Rb\,\,=\,\,\frac{1}{\kappa}\,\intl^z_{\xi_0^A}d z' \,R_2\Lb \dY - \frac{z - z'}{\kappa}, z'\Rb\,\,\,+\,\,C_{\sigma_2},
  \eeq  
which leads to the general solution to \eq{EQXS24}
   \beq \label{EQXS28}   
    \sigma^{(0,II)}_{2}\Lb \dY, z\Rb \,\,=\,\,   \Phi_2\Lb - \,\kappa\,\dY \,+\,z\Rb\,\sigma^{0 '}_2\Lb \dY, z\Rb\,\,+\,\, \sigma^{0 '}_2\Lb \dY, z\Rb\,\tilde{\sigma}^{0'}_2\Lb \dY, z\Rb\eeq    
   \eq{EQXS28}  has to satisfy the initial condition of \eq{MVFXS} at $\dY=0$ and boundary conditions that has the following form:
   
    \beq \label{SOLBC2} 
    \sigma^{(0,I)}_{2}\Lb z=\kappa\,\dY+\xi^A_0\Rb   \,\,=\,\,   \sigma^{(0,II)}_{2}\Lb \dY,z=\kappa\,\dY+\xi^A_0\Rb  
    \eeq  
    
The initial conditions determine the function $\Phi_2\Lb - \kappa\,\dY \,+\,z\Rb =\Phi_2\Lb\xi\Rb$,viz.:
      \beq \label{SOLIC2}  
    \sigma^{0 '}_2\Lb \dY=0, z=\xi\Rb\,   \Phi_2\Lb\xi \Rb\,\,=\,\,\frac{1}{8} e^{2\,\xi} \exp\Lb- \h e^{\xi}\Rb \,\,-\,\, \sigma^{0 '}_2\Lb \dY=0, z=\xi\Rb\,\tilde{\sigma}^{0'}_2\Lb \dY=0, z=\xi\Rb      \eeq
We need to use \eq{EQXS1P07}, which looks as follows for $\sigma_2$:
   \beq \label{EQXS29}    \tilde{\sigma}^{0'}_{2}\Lb \dY, z\Rb\,\,=\,\,\frac{1}{\kappa}\,\intl^z_{ \kappa\,\dY + \xi_0^A}\!\!\!\!\!\!\!d z' \,R_2\Lb \dY - \frac{z - z'}{\kappa},z'\Rb\,\,\,+\,\,\,\frac{1}{\kappa}\,\intl^{ \kappa\,\dY + \xi_0^A}_{\xi_0^A}\!\!\!\!\!\!\!d z'\,\frac{U_2\Lb z'\Rb}{\tilde\sigma^{(0)}_2\Lb  z'\Rb}\,\,+\,\,C_{\sigma_2}
  \eeq  
From \eq{EQXS29} our
  solution to \eq{SOLBC2}  is $ C_{\sigma_2}\,=\,C_{\phi}^{(2)}\,-\,\Phi_2\Lb \xi_0^A\Rb $. 

For iterations $p\geq1$ one finds a similar expression to that obtained for $\sigma_1^{(p,II)}$. Searching in the form $\sigma^{(p,II)}_{2}\Lb\dY,z\Rb\,\,=\,\,\sigma^{0'}_{2}\Lb\dY,z\Rb\,\sigma^{p'}_{2}\Lb\dY,z\Rb$ we obtain
\beq
\sigma^{p'}_{2}\Lb\dY,z\Rb\,\,=\,\,\frac{1}{\kappa}\intl^{z}_0 d z' \,R_2^{(p)}(\dY,z')\,\,+\,\,C_{\sigma_2}^{(p)}
\eeq
where $R_2^{(p)}(\dY,z)\,=\,-\frac{\mathcal{N}_{\mathcal{L}}[\sigma_2\Lb\dY,z\Rb]}{\sigma_2^{0'}\Lb\dY,z\Rb}$ and $C_{\sigma_2}^{(p)}$ is found by requiring that $\sigma_2^{(p,II)}\Lb\dY=0,z=\xi\Rb\,=\,0$. 

The numerical calculations of $\displaystyle{R_{\sigma_2}^{(p)}\Lb \dY, z\Rb\,=\,\frac{\sigma_2^{(p,I)}\Lb \dY, z\Rb}{\sigma_2^{(0,I)}\Lb \dY, z\Rb}\,\Theta\Lb\xi_0^A\,-\,\xi\Rb\,+\,\frac{\sigma_2^{(p,II)}\Lb \dY, z\Rb}{\sigma_2^{(0,II)}\Lb \dY, z\Rb}\,\Theta\Lb\xi\,-\,\xi_0^A\Rb}$  are shown in \fig{fin1}. One can see that they show the same pattern as the calculation for $\sigma_1$ (see \fig{fin}): The first iteration is about of 50\% of the first at low values of z but decreases more fast that the corrections of $\sigma_1$. Also we see the fourth iteration gives the accuracy of less than 3\%.

\begin{boldmath}
\subsection{ $\sigma_3\Lb Y, r,b\Rb$  and $\sigma_n \Lb Y, r,b\Rb $} 
\subsubsection{ Region I }
\end{boldmath}
%%%%%%%%%%%%%%%%%%%%%%%%%%%%%%%%%%%%%%%%%%%%%%%%%%%%     
 \eq{ME7} for $\sigma_3(Y, r_{01}; b)$ has the form:

\bea \label{XS30}
\frac{\partial \sigma_3 (Y,r_{01},b)}{\partial Y} \,\,&=&\,\,
\frac{\bas}{2 \pi} \!\!\!\!\! \int\limits_
{\begin{subarray}{l}
r_{02} \,\gg\,1/Q_s(Y)\\
r_{12}\,\gg\,1/Q_s(Y)
\end{subarray}}
\!\!\!\!\!\!\!\!\!\!\!\!\!d^2 r_2 \,K\Lb r_{01}| r_{12}, r_{02}\Rb \Bigg\{ \sigma_3\Lb Y,r_{12},b\Rb \,+\,\sigma_3\Lb Y,r_{02},b\Rb\,-\,\sigma_3\Lb Y,r_{01},b\Rb\,\,\nn\\
&+&\,\,\sigma_3\Lb Y,r_{12},b\Rb\,\sigma_{sd} \Lb Y,r_{02},b\Rb\,
\,+\,\sigma_3\Lb Y,r_{02},b\Rb\,\sigma_{sd}\Lb Y,r_{12},b\Rb\nn\\
&+& \sigma_2\Lb Y,r_{12},b\Rb\,\sigma_{1}\Lb Y,r_{02},b\Rb 
\,\,+\,\,\, \sigma_2\Lb Y,r_{02},b\Rb\,\sigma_{1}\Lb Y,r_{12},b\Rb \nn\\
&-&\,2\,\sigma_3\Lb Y,r_{12},b\Rb\,N \Lb Y,r_{02},b\Rb\,-\,2\,
\sigma_3\Lb Y,r_{02},b\Rb\,N\Lb Y,r_{12},b\Rb\Bigg\}
\eea
Our choice of $\mathcal{L}(\sigma_3)$ follows the same reasoning as in the previous sections, but with the inclusion of an additional term $\propto 2\,\sigma_2(Y,r;b)\,\sigma_1(Y,r;b)$. The equation in Region I takes the form:
\beq \label{XS31}
\kappa\,\dfrac{\pp \sigma^{(0,I)}_{3}\Lb z\Rb}{\pp z}\,\,=\,\,- \,T\Lb z\Rb\,\sigma^{(0,I)}_{3}\Lb z\Rb\,\,+\,\,\underbrace{\sigma_{0,3} \,\Delta\Lb  z\Rb \,\,+\,\,\sigma_1\Lb z\Rb\,\intl^z_{\xi_0^A} d z'\,\sigma_2\Lb z'\Rb \,+\, \sigma_2\Lb z\Rb\,\intl^z_{\xi_0^A} d z'\,\sigma_1\Lb z'\Rb}_{U_3\Lb z\Rb}
\eeq
with solution
\beq \label{XS32} \sigma^{(0,I)}_{3}\Lb z\Rb\,\,=\,\,\phi_3\Lb z\Rb\,\,+\,\,\Delta_{\sigma_3}\Lb z\Rb\,\,=\,\,\underbrace{\exp\Lb - \frac{1}{\kappa}\intl^z_{\xi_0^A} d z' \,T\Lb z'\Rb \Rb}_{\tilde {\sigma}^{(0)}_3}\,\Bigg\{ \underbrace{\frac{1 }{\kappa}\intl^z_{\xi_0^A} d z'\,U_3\Lb z'\Rb\,\exp\Lb \frac{1}{\kappa}\intl^{z'}_{\xi_0^A} d z'' \,T\Lb z''\Rb \Rb}_{\tilde{\sigma}^{(0,I)}_3}\,\,+\,\, C^{(3)}_\phi\Bigg\}\eeq
Using the initial condition of \eq{MVFXS} at $\xi = \xi_0^A$ we find:
$C^{(3)}_\phi\,\,=\,\,\frac{1}{48} \,e^{ 3\, \xi_0^A} \exp\Lb - \h \,e^{ \xi_0^A} \Rb$. 

\eq{XS32} can be easily generalized 
 for $\sigma_n^{(0,I)}(Y,r;b)$  and it takes the form: 
\beq \label{XSN2}
\kappa\,\dfrac{\pp \sigma^{(0,I)}_{n}\Lb z\Rb}{\pp z}\,\,=\,\,- \,T\Lb z\Rb\,\sigma^{(0,I)}_{n}\Lb z\Rb\,\,+\,\,\underbrace{\sigma_{0,n} \,\Delta\Lb  z\Rb \,\,+\,\,\sum_{k=1}^{n-1}  \,\intl_{\xi_0^A}^zdz'\,\sigma_{n - k}\Lb z'\Rb\,\sigma_{k}\Lb z\Rb }_{U_n\Lb z\Rb}\eeq
with the following solution 
\beq  \label{XSN3}\sigma^{(0,I)}_{n}\Lb z\Rb\,\,=\,\,\phi_n\Lb z\Rb\,\,+\,\,\Delta_{\sigma_n}\Lb z\Rb\,\,=\,\,\underbrace{\exp\Lb - \frac{1}{\kappa}\intl^z_{\xi_0^A} d z' \,T\Lb z'\Rb \Rb}_{\tilde {\sigma}^{(0)}_n}\,\Bigg\{ \underbrace{\frac{1 }{\kappa}\intl^z_{\xi_0^A} d z'\,U_n\Lb z'\Rb\,\exp\Lb \frac{1}{\kappa}\intl^{z'}_{\xi_0^A} d z'' \,T\Lb z''\Rb \Rb}_{\tilde{\sigma}^{(0,I)}_n}\,\,+\,\, C^{(n)}_\phi\Bigg\}\eeq
where the constant  $C^{(n)}_\phi$ is fixed from the initial condition of \eq{MVFXS}: 
\beq \label{CN1}
C^{(n)}_\phi\,\,=\,\,\frac{\Lb  \h e^{\xi^A_0}\Rb^n}{n!} e^{ - \h e^{\xi_0^A}}
\eeq
For $p\geq 1$ it takes the form 
\beq  \label{XSN4}\sigma^{(p,I)}_{n}\Lb z\Rb\,\,=\,\,\exp\Lb - \frac{1}{\kappa}\intl^z_{\xi_0^A} d z' \,T\Lb z'\Rb \Rb\,\Bigg\{ \frac{1 }{\kappa}\intl^z_{\xi_0^A} d z'\,U_n\Lb z'\Rb\,\tilde\Sigma_n^{(p-1)}\Lb z'\Rb\,\exp\Lb \frac{1}{\kappa}\intl^{z'}_{\xi_0^A} d z'' \,T\Lb z''\Rb \Rb\,\,+\,\, C^{(n)}_\phi\Bigg\}\eeq
 One can see that \eq{XSN3} leads to the cross section that decreases as $e^{-\frac{\left(z-z_{\Delta }^I\right){}^2}{2 \,\kappa }}$. We need to find the solution to $\sigma_n$ at large $n$ to see how the summation over $n$ will result in the inelastic cross section of about a constant.  We will do this in section IV. 
 
 \begin{figure}
 	\begin{center}
 	\leavevmode
	\begin{tabular}{c c c }
 		\includegraphics[width=6cm]{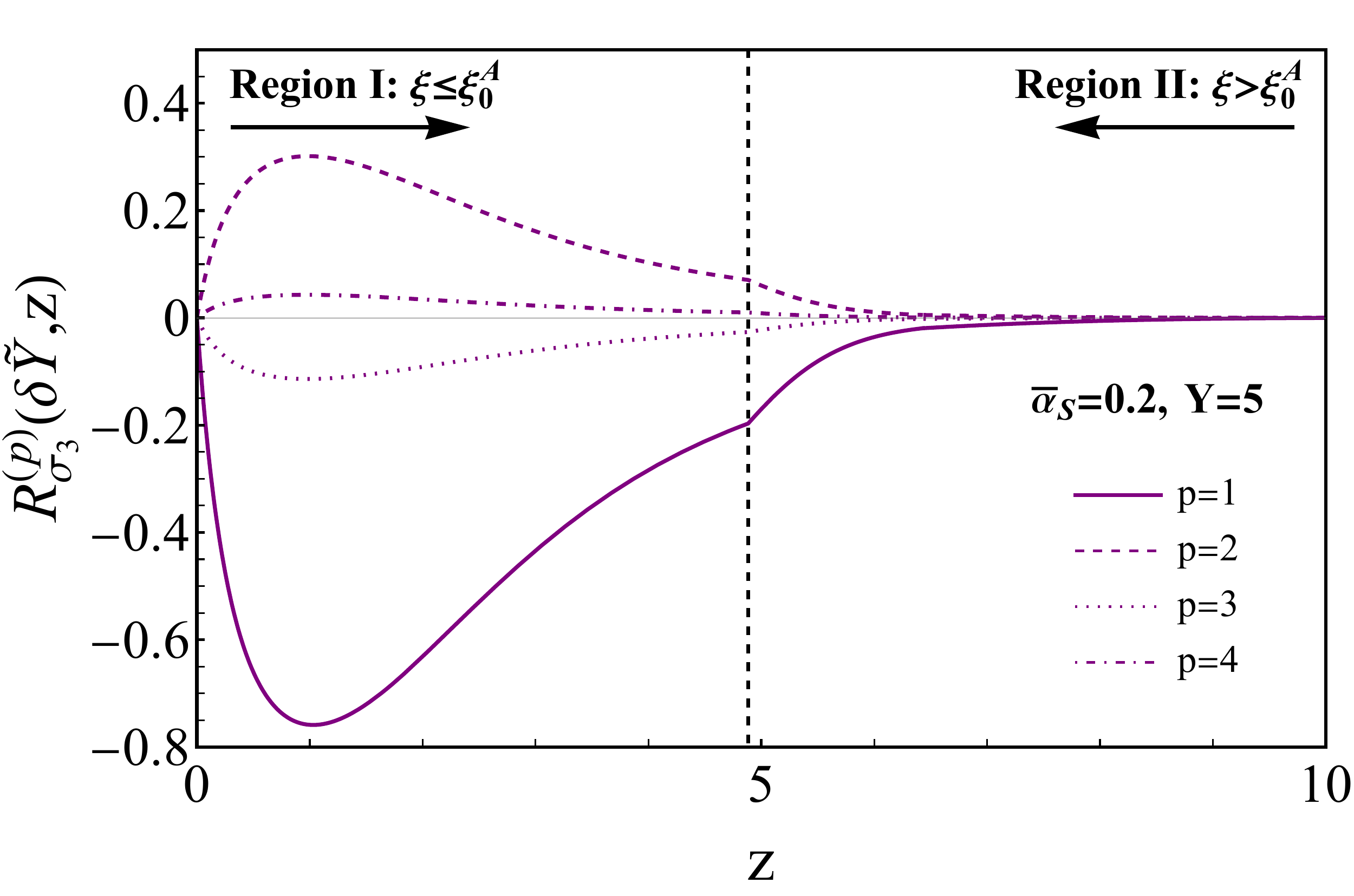}&\includegraphics[width=6cm]{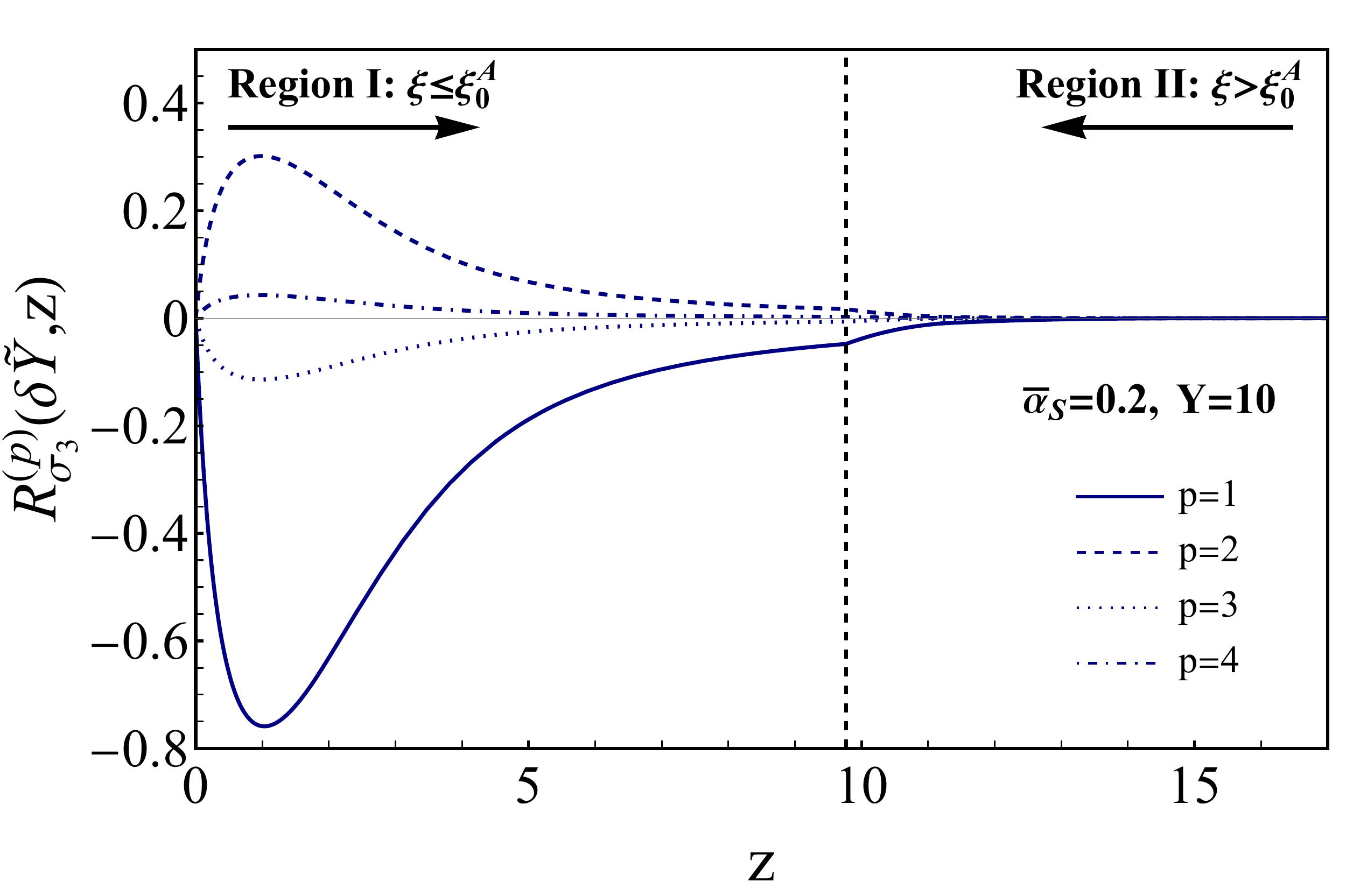}&\includegraphics[width=6.cm]{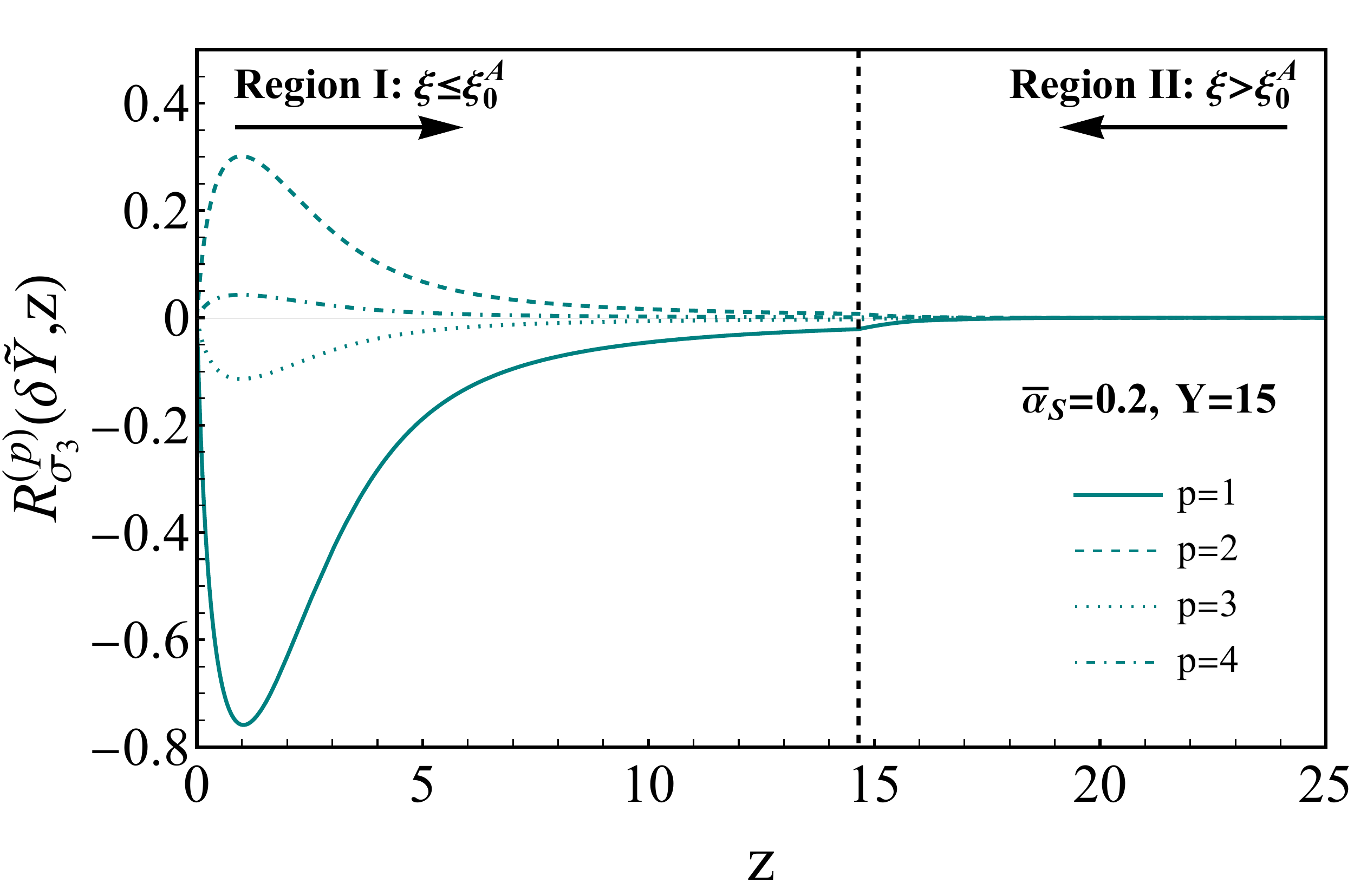} \\
		\fig{fin2}-a & \fig{fin2}-b&\fig{fin1}-c\\
		\end{tabular}	\end{center}
	\caption{ The ratio of $R_{\sigma_3}^{(p)}\Lb \dY, z\Rb$ versus $z$ at different values of $\dY$. $N_0 =0.22$, $\bar{\gamma}=0.63$, $\xi_0^A=0$, $z_0=2$, $z_\Delta=0.87$.
}
\label{fin2}
\end{figure}
 %%%%%%%%%%%%%%%%%%%%%%%%%%%%%%%%%%%%%%%%%%%%%%%%%%%%%%%%%% 
~

%%%%%%%%%%%%%%%%%%%%%%%%%%%%%%%%%%%%%%%%%%%%%%%%%%%%%%%%%%

\subsubsection{ Region II }

In region II \eq{XS30} takes the form:

\bea\label{XS33}
&&\dfrac{\pp \sigma^{(0,II)}_{3}\Lb\dY,  z\Rb}{\pp \dY}\,\,+\,\,\kappa\,\dfrac{\pp \sigma^{(0,II)}_{3}\Lb \dY,  z\Rb}{\pp z}\,\,=\,\,- \,T\Lb\dY,z\Rb\,\sigma^{(0,II)}_{3}\Lb z\Rb\nn\\ 
&&+\,\,\underbrace{\sigma_{0,3} \,\Delta\Lb  \dY,z\Rb \,\,+\,\,\sigma_1\Lb \dY,z\Rb\,\intl^z_{\xi_0^A} d z'\,\sigma_2\Lb\dY,z'\Rb \,\,+\,\, \sigma_2\Lb\dY,z\Rb\,\intl^z_{\xi_0^A} d z'\,\sigma_1\Lb\dY,z'\Rb}_{U_3\Lb \dY,z\Rb}\eea
with the general solution
   \beq  \label{XS34}    \sigma^{(0,II)}_{3}\Lb \dY, z\Rb \,\,=\,\,   \Phi_3\Lb - \,\kappa\,\dY \,+\,z\Rb\,\sigma^{0 '}_3\Lb \dY, z\Rb\,\,+\,\, \sigma^{0 '}_3\Lb \dY, z\Rb\,\tilde{\sigma}^{0'}_3\Lb \dY, z\Rb\eeq   
where 
\beq \label{XX35}  \tilde{\sigma}^{0'}_{3}\Lb \dY, z\Rb\,\,=\,\,\frac{1}{\kappa}\,\intl^z_{\xi_0^A}d z' R_3\Lb \dY - \frac{z - z'}{\kappa}, z'\Rb\,\,+\,\,C_{\sigma_3},
  \eeq  
We need to use the same procedure as was described for $\sigma_2^{(0,II)}\Lb\dY,z\Rb$ to find all unknown functions.
For the $p\geq1$ iteration we search in the form $\sigma^{(p,II)}_{3}\Lb\dY,z\Rb\,\,=\,\,\sigma^{0'}_{3}\Lb\dY,z\Rb\,\sigma^{p'}_{3}\Lb\dY,z\Rb$ with solution
\beq\label{XX36} 
\sigma^{p'}_{3}\Lb\dY,z\Rb\,\,=\,\,\frac{1}{\kappa}\intl^{z}_0 d z' \,R_3^{(p)}(\dY,z')\,\,+\,\,C_{\sigma_3}^{(p)}.
\eeq
The numerical calculations of $\displaystyle{R_{\sigma_3}^{(p)}\Lb \dY, z\Rb\,=\,\frac{\sigma_3^{(p,I)}\Lb \dY, z\Rb}{\sigma_3^{(0,I)}\Lb \dY, z\Rb}\,\Theta\Lb\xi_0^A\,-\,\xi\Rb\,+\,\frac{\sigma_3^{(p,II)}\Lb \dY, z\Rb}{\sigma_3^{(0,II)}\Lb \dY, z\Rb}\,\Theta\Lb\xi\,-\,\xi_0^A\Rb}$ are shown in \fig{fin2}. One can see that they show the same pattern as the calculation for $\sigma_1$ (see \fig{fin}) and for $\sigma_2$: The second iteration is about of 30\% of the first but the fourth iteration gives the accuracy of less than 3\%.    
   
The common feature of $\sigma_1, \sigma_2 $ and $\sigma_3$ calculations (see \fig{fin}-\fig{fin2}) is that corrections at large $z$ are very small.
The equation can be generalized to arbitrary $n$ as follows:
\bea \label{XSN4}
\dfrac{\pp \sigma^{(0,II)}_{n}\Lb\dY,  z\Rb}{\pp \dY}\,\,+\,\,\kappa\,\dfrac{\pp \sigma^{(0,II)}_{n}\Lb \dY,  z\Rb}{\pp z}\,\,&=&\,\,- \,T\Lb\dY,z\Rb\,\sigma^{(0,II)}_{n}\Lb \dY,z\Rb\nn\\ &+&\,\,\underbrace{\sigma_{0,n}\,\Delta\Lb  \dY,z\Rb \,\,+\,\,\sum_{k=1}^{n-1}  \,\intl_{\xi_0^A}^zdz'\,\sigma_{n - k}\Lb \dY,z'\Rb\,\sigma_{k}\Lb \dY,z\Rb }_{U_n\Lb \dY,z\Rb}
\eea
with the solution 
   \beq   \label{XSN5}    \sigma^{(0,II)}_{n}\Lb \dY, z\Rb \,\,=\,\,   \Phi_n\Lb - \kappa\dY \,+\,z\Rb\,\sigma^{0 '}_n\Lb \dY, z\Rb\,\,+\,\, \sigma^{0 '}_n\Lb \dY, z\Rb\,\tilde{\sigma}^{0'}_n\Lb \dY, z\Rb\eeq   
where 
\beq\label{XSN6}
  \tilde{\sigma}^{0'}_{n}\Lb \dY, z\Rb\,\,=\,\,\frac{1}{\kappa}\,\intl^z_{\xi_0^A}d z' \,R_n\Lb \dY - \frac{z - z'}{\kappa}, z'\Rb\,\,+\,\,C_{\sigma_n},
  \eeq  
with $R_n\,=\,U_n/\sigma_n^{0'}$ and the functions $C_{\sigma_n}$ and $\Phi_n$ are determined to satisfy the initial condition given by \eq{MVFXS}.
For the $p\geq1$ iteration the solution takes the form
 $\sigma^{(p,II)}_{n}\Lb\dY,z\Rb\,\,=\,\,\sigma^{0'}_{n}\Lb\dY,z\Rb\,\sigma^{p'}_{n}\Lb\dY,z\Rb$ where
\beq\label{XSN7} 
\sigma^{p'}_{n}\Lb\dY,z\Rb\,\,=\,\,\frac{1}{\kappa}\intl^{z}_0 d z' \,R_n^{(p)}(\dY,z')\,\,+\,\,C_{\sigma_n}^{(p)}.
\eeq
where $R_n^{(p)}(\dY,z)\,=\,-\frac{\mathcal{N}_{\mathcal{L}}[\sigma_n\Lb\dY,z\Rb]}{\sigma_n^{0'}\Lb\dY,z\Rb}$ and $C_{\sigma_n}^{(p)}$ is found by requiring that $\sigma_n^{(p,II)}\Lb\dY=0,z=\xi\Rb\,=\,0$. 
 
  \begin{figure}
 	\begin{center}
 	\leavevmode
	\begin{tabular}{c c c }
 		\includegraphics[width=6cm]{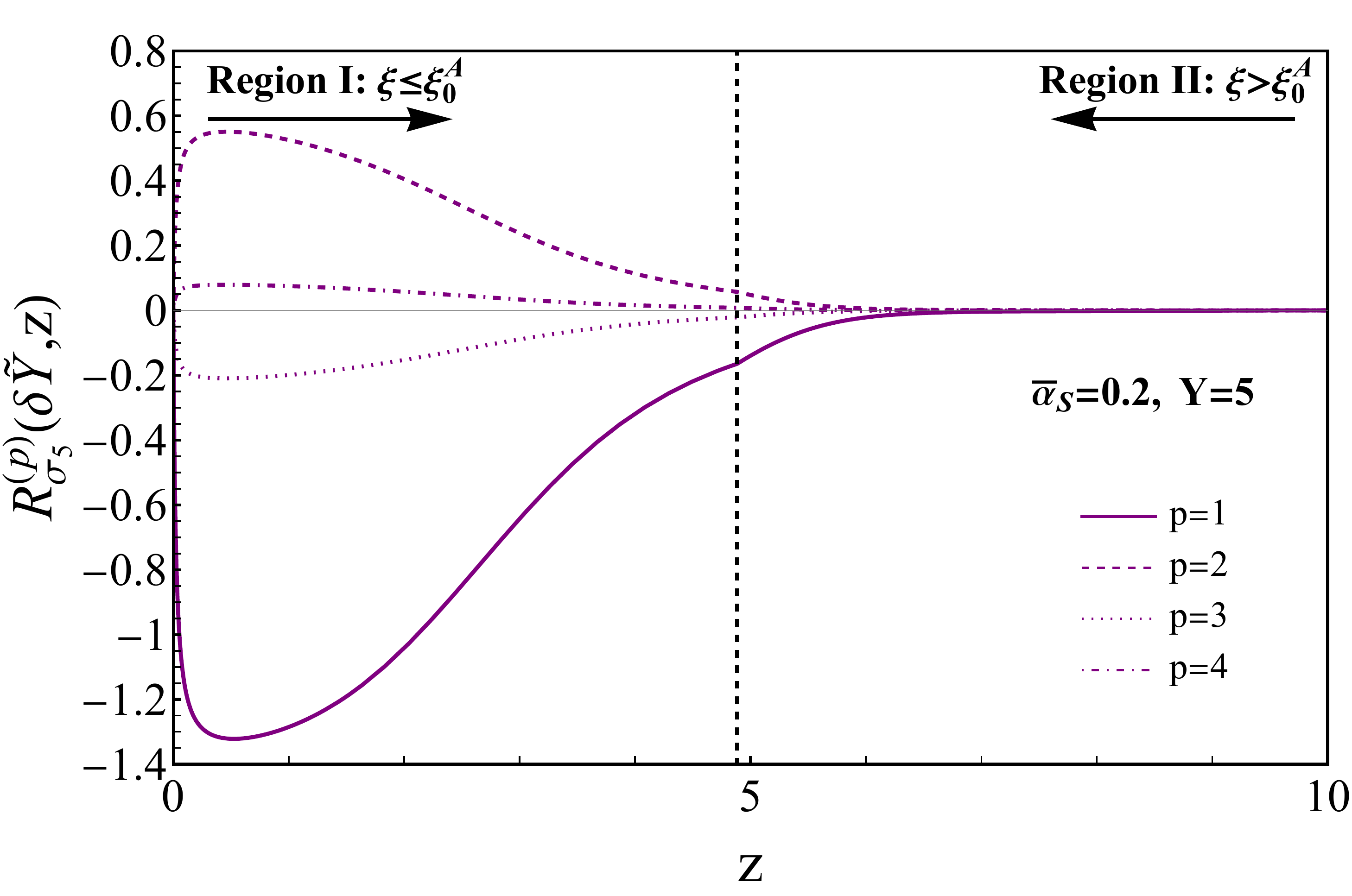}&\includegraphics[width=6cm]{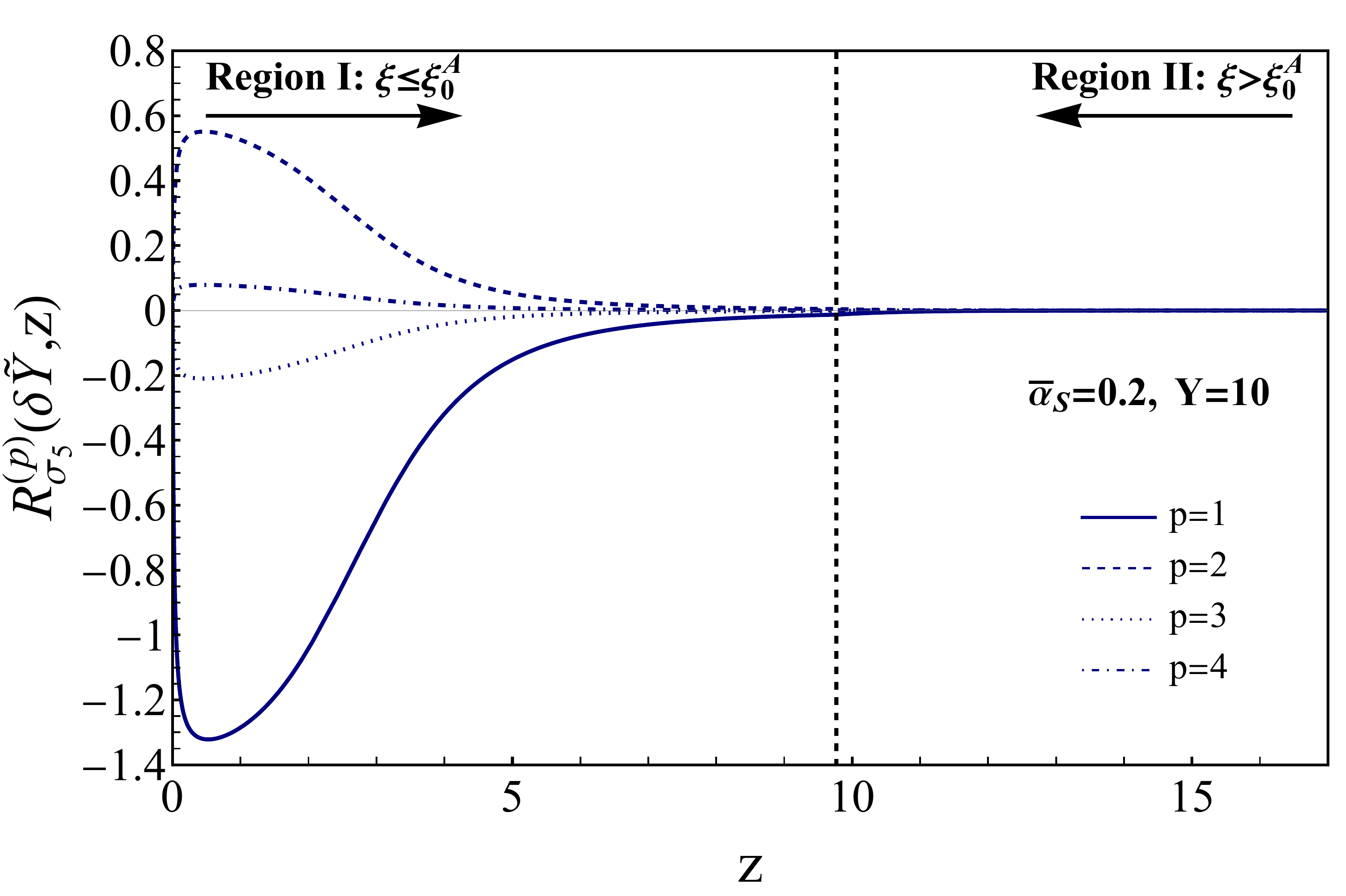}&\includegraphics[width=6.cm]{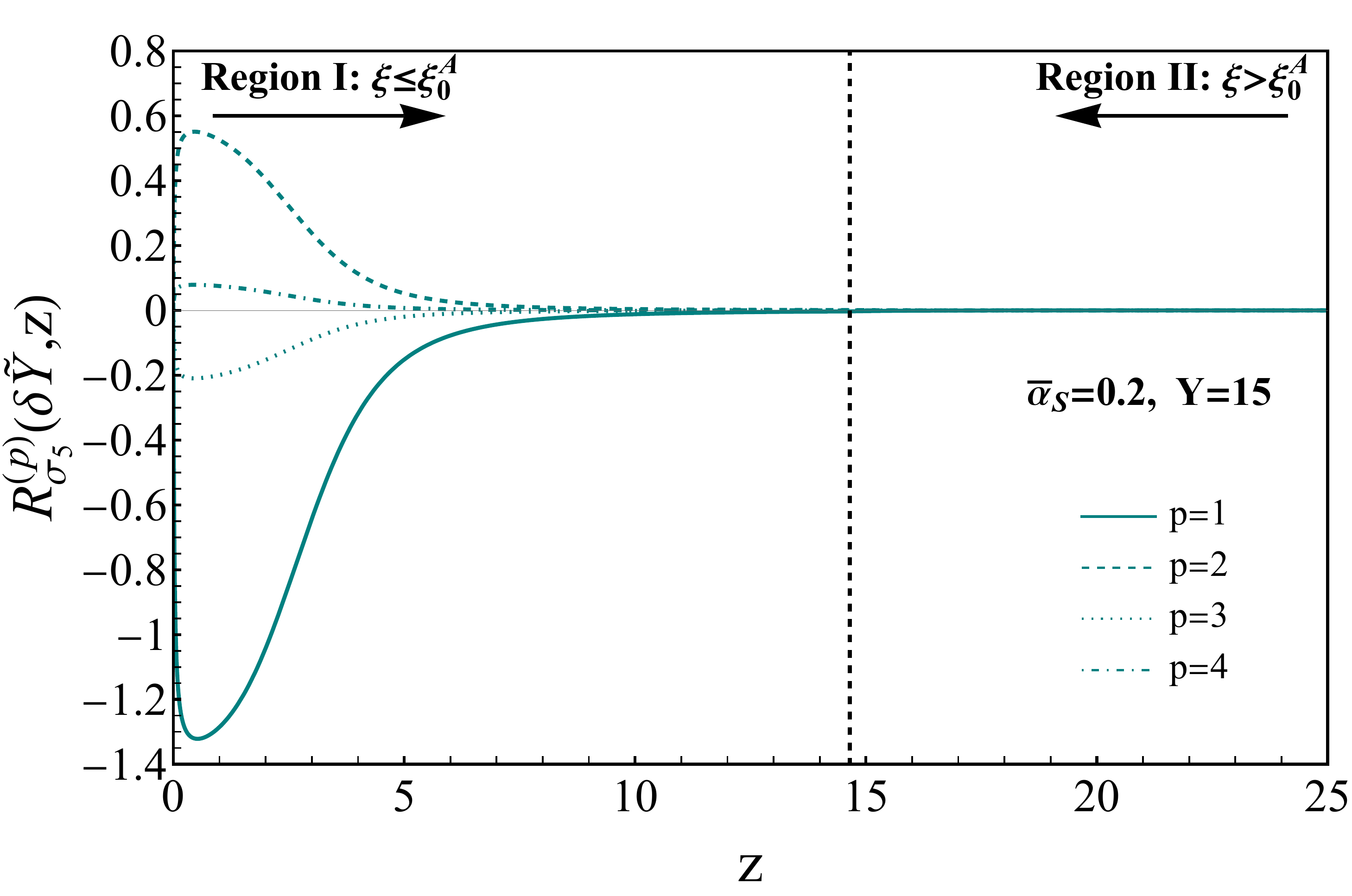} \\
		\fig{fin2}-a & \fig{fin2}-b&\fig{fin1}-c\\
		\end{tabular}	\end{center}
	\caption{ The ratio of $R_{\sigma_5}^{(p)}\Lb \dY, z\Rb$ versus $z$ at different values of $\dY$. $N_0 =0.22$, $\bar{\gamma}=0.63$, $\xi_0^A=0$, $z_0=2$, $z_\Delta=0.87$.
}
\label{fin3}
\end{figure}
 %%%%%%%%%%%%%%%%%%%%%%%%%%%%%%%%%%%%%%%%%%%%%%%%%%%%%%%%%% 
 
As an application for the results of general $n$, we plot the quantity (see \fig{fin3})
\beq\label{XSN8} 
R_{\sigma_5}^{(p)}\Lb \dY, z\Rb\,=\,\frac{\sigma_5^{(p,I)}\Lb \dY, z\Rb}{\sigma_5^{(0,I)}\Lb \dY, z\Rb}\,\Theta\Lb\xi_0^A\,-\,\xi\Rb\,+\,\frac{\sigma_5^{(p,II)}\Lb \dY, z\Rb}{\sigma_5^{(0,II)}\Lb \dY, z\Rb}\,\Theta\Lb\xi\,-\,\xi_0^A\Rb
\eeq
One can see that the first iteration turns out to be rather large in the small-$z$ region which corresponds to region I in \fig{sat}. Note, that the behaviour in the region II of the same figure gives a small contributions. Despite the fact that after an appropriate number of iterations the corrections  become small, the approach may break down for a sufficiently large $n$. For this reason, we use this method only for small values of $n$. In the next section we solve the equation in the large-$n$ limit, and study the possible matching with this solution. We will show that actually at small values of $z$ the solution at large $n$ is prevailed, and therefore,  the large corrections in this region for $\sigma_5$, for example, does not look important.

~

 ~

%%%%%%%%%%%%%%%%%%%%%%%%%%%%%%%%%%%%%%%%%%%%%%%%%%%%%%%%%%
  \begin{boldmath}

\section{ Multiplicity distribution at large $n$}

  %%%%%%%%%%%%%%%%%%%%%%%%%%%%%%%%%%%%%%%%%%%%%%%%%%%%%%%%%%

\subsection{ Solutions to the equations.}
\end{boldmath}
%%%%%%%%%%%%%%%%%%%%%%%%%%%%%%%%%%%%%%%%%%%%%%%%%%%%%%%%%%
In this section we solve the equation for  $\sigma_n$ at large $n$:  $n\,\gtrsim\,N\Lb z\Rb$, where  the typical multiplicity $N\Lb z \Rb $ we have to find from the solution. The equation takes the form:

\beq \label{MDLN}
\kappa \frac{ d \,\sigma_{n}\Lb z \Rb}{d\,z}\,\,=\,\,-\,z\,\sigma_{n} + \Delta\Lb z \Rb \Sigma_{n}\Lb z\Rb + \intl^z_0 d\,  z'\,\Delta\Lb z'\Rb \sigma_{n}\Lb z\Rb\,\,+\,\,\sum_{k=1}^{n-1}\Sigma_{n-k}\Lb z \Rb \sigma_{k}\Lb z \Rb
\eeq
where $\Sigma_{n}\Lb z\Rb = \intl^z _{0}d z' \sigma_{n}\Lb z'\Rb$.

 We suggest that the solution has the following form:
\beq \label{MDLN1}
\Sigma_n\Lb z\Rb\, = \,\phi\Lb z\Rb \exp\Lb - n \,\Phi\Lb z \Rb\Rb
\eeq
Functions $\Phi\Lb z \Rb$ and $\phi\Lb z \Rb$ we will find from \eq{MDLN}. First, we note that
\beq \label{MDLN2}
\sum_{k=1}^{n-1}\Sigma_{n-k}\Lb z \Rb \sigma_{k}\Lb z \Rb= \h \frac{d}{d z} \sum_{k=1}^{n-1}\Sigma_{n-k}\Lb z \Rb \Sigma_{k}\Lb z \Rb =  \frac{n-1}{2} \frac{d}{d z} \phi^2\Lb z\Rb\exp\Lb - n \Phi\Lb z\Rb\Rb
\eeq
\eq{MDLN} can be rewritten as
\beq \label{MDLN3}
\kappa \frac{d^2 \Sigma_n}{d z^2}\,\,=\,\,- \frac{d}{d\,z} \Lb z \Sigma_n\Lb z\Rb\Rb + \Sigma_n\Lb z\Rb\,\,+\,\,\frac{d}{d\,z} \Lb \Sigma_n\Lb z\Rb\Sigma_\Delta\Lb z \Rb\Rb\,\, +\,\,\frac{n-1}{2} \frac{d}{d z} \Lb\phi\Lb z\Rb\,\Sigma_n\Lb z\Rb\Rb
\eeq
with $\Sigma_\Delta\Lb z \Rb = \intl^z d z' \Delta\Lb z'\Rb$. Introducing $\mathcal{S}_n\Lb z\Rb =\intl^z d z' \,\Sigma_n\Lb z'\Rb$ we reduces \eq{MDLN3} to the following equation:
\beq \label{MDLN4}
\kappa \frac{d^2 \Sigma_n}{d z^2}\,\,=\,\,- \frac{d}{d\,z} \Lb z \Sigma_n\Lb z\Rb\Rb + \frac{d}{d z} \mathcal{S}_n\Lb z\Rb\,\,+\,\,\frac{d}{d\,z} \Lb  \Sigma_n\Lb z\Rb\Sigma_\Delta\Lb z \Rb\Rb\,\,+\,\, \frac{n-1}{2} \frac{d}{d z} \Lb\phi\Lb z\Rb\,\Sigma_n\Lb z\Rb\Rb
\eeq
Assuming that all functions decreases at large $z$ we have:
\bea \label{MDLN5}
\kappa \frac{d \Sigma_n}{d z}\,\,&=&\,\,-   z \Sigma_n\Lb z\Rb + \mathcal{S}_n\Lb z\Rb\,\,+\,\, \Sigma_n\Lb z\Rb \Sigma_\Delta\Lb z \Rb\,\,+\,\, \frac{n-1}{2}  \phi\Lb z\Rb\,\Sigma_n\Lb z\Rb\nn\\
&=& -  \Lb z \,-\,z_\Delta\Rb \Sigma_n\Lb z\Rb + \mathcal{S}_n\Lb z\Rb\,\,-\,\, \Sigma_n\Lb z\Rb \widetilde{\Sigma}_\Delta\Lb z \Rb\,\,+\,\, \frac{n-1}{2}  \phi\Lb z\Rb\,\Sigma_n\Lb z\Rb\eea
with $\widetilde{\Sigma}_\Delta\Lb z \Rb\,\,=\,\,\intl^\infty_z d z' \Delta\Lb z'\Rb$.

Using that $\frac{d \Sigma_n}{d z} \,=\,\Lb \phi'\Lb z\Rb\,-\,n\,\phi\Lb z\Rb \Phi'\Lb z\Rb\Rb \exp\Lb - n \,\Phi\Lb z \Rb\Rb$ and equating terms that are proportional to $n$ we obtain:
\beq    \label{MDLN6}   
   \Phi'\Lb z\Rb\,\,=\,\,- \frac{1}{2\,\kappa}\phi\Lb z\Rb;~~~~~\Phi\Lb z \Rb = - \frac{1}{2\,\kappa}\intl^z d z' \phi\Lb z'\Rb + C
   \eeq
   
   From this equation $\mathcal{S}_n\Lb z\Rb\,\,=\,\frac{2\,\kappa}{n}\,\exp\Lb -  n\,\Phi\Lb z\Rb\Rb$ and \eq{MDLN5} takes the form:
  \beq    \label{MDLN7}  
   \kappa\,\phi'\Lb z \Rb\,=\,- ( z - z_\Delta) \phi\Lb z \Rb \,-\,\widetilde{\Sigma}_\Delta\Lb z \Rb\phi\Lb z \Rb\,+\,\frac{2\,\kappa}{n } - \h \phi^2\Lb z \Rb
   \eeq
   
   Neglecting the term $\frac{2\,\kappa}{n }$ at large $n$    we solve the homogenous equation:
     \beq    \label{MDLN8}  
   \kappa\,\phi'\Lb z \Rb\,=\,- \underbrace{\Lb z - z_\Delta \,+\,\widetilde{\Sigma}_\Delta\Lb z \Rb\Rb}_{T\Lb z \Rb}\phi\Lb z \Rb\, -\, \h \phi^2\Lb z \Rb
   \eeq   
   Introducing $\phi\Lb z \Rb = 2 \,\kappa \frac{U'\Lb z \Rb}{U\Lb z \Rb}$ we obtain for function $U$ the following equation:
     \beq    \label{MDLN9}    
  \kappa\, U''\Lb z \Rb \,=\,- T\Lb z \Rb  U'\Lb z \Rb;~U'\Lb z \Rb = C^{(1)}_\phi\exp\Lb - \frac{1}{\kappa} \intl^z\!\!  d z' T\Lb z'\Rb \Rb; ~U\Lb z \Rb =C^{(1)}_\phi \intl^z\!\! d z' \exp\Lb - \frac{1}{\kappa} \intl^{z'} \!\! d z'' T\Lb z''\Rb \Rb\,+\,C^{(2)}_\phi;  \eeq  
  Therefore, for the solution of the homogenouos equation $\phi_h\Lb z \Rb$ we have:
    \bea    \label{MDLN10}   
  \phi\Lb z \Rb\,&=&\,2\,\kappa \frac{\exp\Lb - \frac{1}{\kappa} \intl^z d z' \,T\Lb z'\Rb \Rb}{ 
    \intl^z \!d z' \exp\Lb - \frac{1}{\kappa} \intl^{z'} d z''\, T\Lb z''\Rb \Rb\,+\,C^{(2)}_\phi/C^{(1)}_\phi}\\
    &\xrightarrow{z \,\gg\,1}& \,\,2 \,\kappa 
   \frac{ e^{ - \frac{(z - z_\Delta)^2}{2\,\kappa}}}{ \intl^z_{z_\Delta}d z'\, e^{ - \frac{(z' - z_\Delta)^2}{2\,\kappa}} \,+\,C^{(2)}_\phi/C^{(1)}_\phi} \,=\,2 \,\kappa \frac{ e^{ - \frac{(z - z_\Delta)^2}{2\,\kappa}}}{\sqrt{\frac{\pi \kappa }{2}} \, \text{erf}\left(\frac{z-z_\Delta}{\sqrt{2 \kappa }}\right) \,+\,C^{(2)}_\phi/C^{(1)}_\phi}  \,\,\xrightarrow{z \,\gg\,1}  C_\phi e^{ - \frac{(z - z_\Delta)^2}{2\,\kappa}} \equiv \phi_{asym} \Lb z\Rb\nn\eea    
  
 In \eq{MDLN10} the arbitrary constants $C_\phi,C^{(1)}_\phi$ and  $C^{(2)}_\phi$
 have to be chosen from the matching of our solution with the initial conditions of \eq{MVFXS}.

  Plugging      \eq{MDLN6} and \eq{MDLN10} into \eq{MDLN1} we obtain  $\Sigma_n$, which is equal to
  \bea \label{MDLN14}  
  \Sigma_n\Lb z\Rb\, &=& \,\phi\Lb z\Rb \exp\Lb - n \,\Phi\Lb z \Rb\Rb\,\,
  \xrightarrow  {z\,\gg\,1} \,\, \,C_\phi \, e^{ - \frac{(z - z_\Delta)^2}{2\,\kappa}} \exp \Bigg( -n\,\frac{C_\phi}{2\,\kappa}\intl^\infty_z d\,z'e^{ - \frac{(z' - z_\Delta)^2}{2\,\kappa}} \Bigg)\\
  &=& \,\,C_\phi\, \, e^{ - \frac{(z - z_\Delta)^2}{2\,\kappa}}     \exp\Bigg(- n\, \,\frac{C_\phi}{2} \sqrt{\frac{\pi}{2\,\kappa}} \text{erfc}\left(\frac{z- z_\Delta}{\sqrt{2 \kappa }}\right) \Bigg)\,\,  \xrightarrow  {z\,\gg\,1} \,\, \, \, C_\phi\, \, e^{ - \frac{(z - z_\Delta)^2}{2\,\kappa}} \,\exp\Bigg(-n\, \, C_\phi \frac{1}{2\,z}\, e^{ - \,\frac{(z - z_\Delta)^2}{2\,\kappa}} \Bigg)\nn
   \eea       
   
   From this equation we can calculate the value of $\sigma_n$ at large $n$, which is equal to:
    \beq \label{MDLN15}  
          \sigma_n\Lb z \Rb\,=\,-\phi\Lb z\Rb\,n\,\Phi'\Lb z\Rb \exp\Bigg( - n\,\Phi\Lb z \Rb\Bigg) \,\,\xrightarrow  {z\,\gg\,1} \,\, \phi_{asym}\Bigg(- n\,\Phi'_{asym}\Lb z\Rb\Bigg) \,\exp\Bigg\{ -\,n\,\Phi_{asym}\Lb z\Rb\Bigg\}
 \eeq         
  From \eq{MDLN14}  one can see that   $\Phi_{asym}\Lb z\Rb\,=\, \frac{ \phi_{asym}}{2\,z}$
  and, therefore, at large $z$ the multiplicity distribution follows the KNO\cite{KNO,KNO1,KNO2} scalling  behaviour:
      \beq \label{MDLN16}    
  \sigma_n\Lb z \Rb\,\,=\,\,\Bigg( \phi'\Lb z \Rb - n \phi\Lb z \Rb \Phi'\Lb z\Rb\Bigg)\exp\Lb - n \Phi\Lb z\Rb\Rb  \xrightarrow{n \geq N(z)}
   \,\frac{2}{ \kappa}\frac{z^2}{N\Lb z\Rb} \Psi\Lb\xi\,=\,\frac{n}{N\Lb z\Rb}\Rb   ;~~~~~~\Psi\Lb \xi\Rb= \xi\,e^{- \xi};
  \eeq
  
  with 
    \beq \label{MDLN170}    
      N\Lb z\Rb \,=\, 1/ \Phi_{asym}\Lb z\Rb =  \frac{2\,z}{C_\phi}\,e^{ \frac{\Lb z-z_\Delta\Rb^2}{2\,\kappa}}
    \eeq

      ~
      
      ~  
                %%%%%%%%%%%%%%%%%%%%%%%%%%%%%%%%%%%%%%%%%%%%%%%%%%%%
\subsection{Entropy of produced gluons }
%%%%%%%%%%%%%%%%%%%%%%%%%%%%%%%%%%%%%%%%%%%%%%%%%%%%%

Since $\sigma_{in} =\sum^\infty_{n=1} \sigma_n$ at large $z$ tends to 1 (see \eq{ME72} and the next subsection), the probability to have $n$-cut Pomerons in the final sate from \eq{MDLN16} is equal to 
   \beq\label{MDLN17}
{\cal P}_n^{\mbox{\tiny AGK}}\Lb z\Rb\,\,\equiv\,\,\frac{\sigma_n^{\mbox{\tiny AGK}}\Lb z\Rb}{\sigma_{in}^{\mbox{\tiny AGK}}\Lb z\Rb}\,\,=\,\,\frac{2}{ \kappa}\,\frac{z^2}{N\Lb z\Rb} \,\Psi\Lb \frac{n}{N\Lb z\Rb}\Rb 
\eeq 
The entropy content of multiplicity distributions is defined by 
   \beq\label{MDLN171}
S_E\Lb z\Rb \,\,= \,\,-\,\sum_n \mathcal{P}_n^{\mbox{\tiny AGK}}\Lb z\Rb\,\ln\Lb\mathcal{P}_n^{\mbox{\tiny AGK}}\Lb z\Rb\Rb 
\eeq 
which we interpret as the entropy for the produced gluons. Using \eq{MDLN17} in \eq{MDLN171} we find that the von Neumann entropy at large $z$ is equal to
   \bea \label{MDLN18}
   S_E\Lb z\Rb \,\,&= &\,\, \underbrace{\sum_{n=n_0}^\infty  \frac{2}{\kappa}\,\frac{z^2}{N\Lb z\Rb}\,\Psi  \Lb \frac{n}{N\Lb z\Rb}\Rb}_{1}\,\ln\Lb \frac{\kappa}{2}\,\frac{N\Lb z\Rb}{z^2}\Rb \,\,-\,\,\underbrace{\sum_{n=n_0}^\infty \frac{2}{\kappa}\,\frac{z^2}{N\Lb z\Rb}\,\Psi  \Lb \frac{n}{N\Lb z\Rb}\Rb\,\ln\Lb \Psi\Lb \frac{n}{N\Lb z\Rb}\Rb  \Rb}_{\rm Const}  \nn\\
   &=&\,\,\ln\Lb  N\Lb z\Rb\Rb\,\,-\,\,\ln\Lb\frac{z^2}{2\,\kappa}\Rb\,\,-\,\,2\,\ln 2
    \,\,+\,\, \frac{2}{\kappa}\,\frac{n_0}{N\Lb z\Rb} \,\underbrace{2 \,\exp\Lb -\,\frac{n_0}{N\Lb z\Rb}\Rb}_{1 \,\,see \,\,\eq{MDLN25}}\nn\\ &=&\,\,\frac{(z - z_\Delta)^2}{2 \,\kappa} \,\,-\,\,\ln z \,\, -\,\, \ln \Lb \frac{C_\phi}{\kappa}\Rb\,\,+\,\,\frac{2}{\kappa}\,\ln 2 
    \,\,\xrightarrow{z\,\gg\,1} \,\, \frac{z^2}{2 \,\kappa} 
 \eea
 In this equation we introduce $n_0 \sim N\Lb z\Rb$, which  we will explain below (see \eq{MDLN25}). Choosing the value of $n_0$ from condition that $\sum_{n=n_0} \sigma_n =Const$ we get the second term in \eq{MDLN18} is equal to $\ln 2$.
 Therefore, in QCD  $S_E\Lb z\Rb \,\,=\,\, \ln N\Lb z\Rb = \frac{z^2}{2 \,\kappa}   $ in accordance with the result of Ref.\cite{KHLE,LEMULT,GOLEMULT}. 
It should be noted that we assume that 
   \eq{MDLN17}   described all $n$ having $\sum_{n=1}^{\infty} {\cal P}_n^{\mbox{\tiny AGK}}\Lb z\Rb =1$.  As we will discuss below in section IV-E  the main contribution to the total inelastic cross section come from $n \geq N(z)$. Based on this fact we can safely consider  \eq{MDLN17} for calculation of $S_E$.
   
   ~   
      %%%%%%%%%%%%%%%%%%%%%%%%%%%%%%%%%%%%%%%%%%%%%%%%%%%%
\subsection{Inelastic cross section }
%%%%%%%%%%%%%%%%%%%%%%%%%%%%%%%%%%%%%%%%%%%%%%%%%%%%%
Inelastic cross section can be found as
\beq \label{INXS}
\sigma_{in}\Lb z\Rb= \sum^\infty_{n=1} \sigma_n\Lb z \Rb\,=\,  \frac{d}{d z}  \sum^\infty_{n=1} \Sigma_n\Lb z\Rb 
\eeq
Using \eq{MDLN14} we obtain
\beq \label{INXS0}
\Sigma_{in}\Lb z\Rb\,=\, \sum^\infty_{n=1}\Sigma_n\Lb z \Rb = 2\,z\,\sum^\infty_{n=1} \frac{\phi_{asym}\Lb z \Rb}{2\,z} \exp\Lb - n\,\frac{\phi_{asym}\Lb z \Rb}{2\,z}\Rb\,=\,\frac{\phi_{asym}\Lb z \Rb}{\exp\Lb \frac{\phi_{asym}\Lb z \Rb}{2\,z}\Rb\,-\,1}\,\xrightarrow{z \gg 1}\,\,2\,z - \h \phi_{asym}\Lb z \Rb
\eeq
Finally from \eq{INXS0} $ \sigma_{in} \Lb z \Rb \,\,=\,\,\Sigma_{in}'\Lb z\Rb = 2$ at large $z$. 
It is instructive to note that this  cannot be  obtained  using \eq{MDLN16} at large $n$. We need to use the  first equation (see \eq{MDLN16}) which is correct at any values of $n$  for our solution and expand $\Phi\Lb z \Rb $ at large $z$ within accuracy $1/z^3$.

The limit $\sigma_{in} = 2$ at large $z$  contradicts the unitarity constraints that lead to $\sigma_{in} \Lb z \Rb\,\to\,1$ at large $z$. The unitarity limit  
 can be found from
from \eq{ME72}.
Indeed, the inelastic cross section:  $\sigma_{in} = \sum_{n=1}^\infty \sigma_n$, satisfies the BK equation as has been demonstrated in \eq{ME72}.  For  leading twist BFKL  kernel   
it  takes the form:
\beq \label{INXS1}
\kappa \frac{ d \,\sigma_{in}\Lb z \Rb}{d\,z}\,\,=\,\,-\,z\,\sigma_{in} + \Delta\Lb z \Rb \Sigma_{in}\Lb z\Rb + \intl^z_0 d\,  z'\,\Delta\Lb z'\Rb \sigma_{in}\Lb z\Rb\,\,+\,\,\Sigma_{in}\Lb z \Rb \sigma_{in}\Lb z \Rb
\eeq
where $\Sigma_{in}\Lb z\Rb = \intl^z _{0}d z' \sigma_{in}\Lb z'\Rb$.

Introducing $\sigma_{in}\Lb z\Rb= 1 - \Delta_\sigma\Lb z \Rb$ and $ \Sigma_{in}\Lb z \Rb = z- \Delta_\Sigma\Lb z \Rb$ with $ \Delta_\Sigma\Lb z \Rb \,\,=\,\,\intl^z_{0} d z'  \Delta_\sigma\Lb z '\Rb$  we reduce \eq{INXS1} to the following form:
\beq \label{INXS2}
\kappa \,\frac{ d\,\Delta_\sigma\Lb z \Rb}{d\,z}\,\,=\,\,\Delta_\Sigma\Lb z \Rb\Lb1\,-\,\Delta_\sigma\Lb z \Rb\Rb   \,\,-\,\, z\,\Delta\Lb z\Rb   \,\,+\,\,\Delta\Lb z \Rb\,\Delta_\Sigma\Lb z \Rb \,\,-\,\,\intl^z_0 d z' \,\Delta\Lb z'\Rb \,\Lb 1 \,-\, \Delta_\sigma\Lb z\Rb \Rb\eeq

Solution of this equation is $\Delta_\sigma \Lb z \Rb = \Delta\Lb z\Rb$. Indeed, for this solution \eq{INXS2} reduces to
\beq \label{INXS3}
\kappa \,\frac{ d\,\Delta_\sigma\Lb z \Rb}{d\,z}\,\,=\,\,- z\,\Delta_\sigma\Lb z \Rb\,\,+\,\, \Delta_\sigma\Lb z \Rb\,\Delta_\Sigma\Lb z \Rb
\eeq
which is the BK equation for our kernel.

Finally

\beq \label{INXS4}
\sigma_{in}\Lb z \Rb\,\,\xrightarrow{z\gg 1}\,\, 1\,\,\,-\,\,\,\Delta\Lb z\Rb
\eeq

 Comparing this expression with \eq{MDLN1} we can see that solution of \eq{MDLN1} deciphers  what means 1 in \eq{INXS4} from the point of production processes.
 
 ~

~

      %%%%%%%%%%%%%%%%%%%%%%%%%%%%%%%%%%%%%%%%%%%%%%%%%%%%
\subsection{Corrections 1/n }
%%%%%%%%%%%%%%%%%%%%%%%%%%%%%%%%%%%%%%%%%%%%%%%%%%%%%
Plugging solutions of \eq{MDLN14} and \eq{MDLN15} into \eq{MDLN4} we can see that our assumptions that $\mathcal{S}\Lb z\Rb$ decreases at large $z$ is not correct:  $\mathcal{S}\Lb z\Rb\, \xrightarrow{z\,\gg\,1} \,\frac{2\,\kappa}{n}$. Hence \eq{MDLN5} takes the form:

\beq \label{MDLN19}
\kappa \frac{d \Sigma_n\Lb z\Rb}{d z}\,\,=\,\, - \Lb z \,-\,z_\Delta\Rb \Sigma_n\Lb z\Rb + \Lb \mathcal{S}_n\Lb z\Rb\,\,-\,\,\frac{2\,\kappa}{n}\Rb-\,\, \Sigma_n\Lb z\Rb \widetilde{\Sigma}_\Delta\Lb z \Rb\,\,+\,\, \frac{n-1}{2}  \phi\Lb z\Rb\,\Sigma_n\Lb z\Rb\eeq
   For solving \eq{MDLN19} we are searching the solution in the form: $\phi\Lb z \Rb = \phi_h\Lb z\Rb \,+\,\Delta_\phi\Lb z\Rb$, and neglecting $ \Delta^2_\phi\Lb z\Rb $ contribution. The equation for $ \Delta_\phi\Lb z\Rb$ takes the form:
   \beq    \label{MDLN20}   
    \kappa\,\Delta'_\phi\Lb z \Rb\,=\,- \Lb T\Lb z\Rb +\phi_h\Lb z\Rb\Rb \Delta_\phi\Lb z\Rb\,  + \,\frac{2\,\kappa}{n}\Lb \exp\Lb  n\,\Phi\Lb z \Rb\Rb\,-\,1\Rb
    \eeq 
        For large $z$ we can rewrite this equation as
\beq    \label{MDLN21}   
    \kappa\,\Delta'_\phi\Lb z \Rb\,=\,- \Lb T\Lb z\Rb +\phi_h\Lb z\Rb\Rb \Delta_\phi\Lb z\Rb\,  + \,\intl^\infty_z d z' \phi_h\Lb z'\Rb
    \eeq

        Hence, the particular solution of the non-homogenouos equation is equal to:
          \bea    \label{MDLN22}    
\Delta_\phi\Lb z \Rb\,\,&=&\,\,\frac{1}{\kappa} \exp\Lb - \frac{1}{\kappa}\intl^z_{z_\Delta} d z' \Lb T\Lb z'\Rb +\phi_h\Lb z'\Rb\Rb \Rb\intl^z_{z_\Delta} d z' \Bigg(\intl^\infty_{z'} d \bar{z}\, \phi_h\Lb \bar{z}\Rb\Bigg) \exp\Lb\frac{1}{\kappa}\intl^{z'}_{z_\Delta} d z'' \Lb T\Lb z''\Rb +\phi_h\Lb z''\Rb\Rb \Rb\nn\\
&\xrightarrow  {z\,\gg\,1} & C_\phi\,\ln \Lb z\Rb  \exp\Lb - \frac{\Lb z - z_\Delta\Rb^2}{2\, \kappa}\Rb
\eea 
      Finally, the solution to \eq{MDLN19}  is
      
      \beq \label{MDLN23}
      \phi_{n.h.}\Lb z \Rb  \,=\,\phi_h \Lb z\Rb +\Delta_\phi\Lb z \Rb\,\,\xrightarrow  {z\,\gg\,1}  \,\,  \phi_{asym} \Lb z\Rb\Lb 1 + \ln z\Rb
      \eeq
      
      Plugging \eq{MDLN12} in to \eq{MDLN6} we see that
       \bea \label{MDLN24}
       \Phi\Lb z \Rb \,&=&\,\frac{1}{2\,\kappa}\intl^\infty_{z} dz'\,\phi_{n.h.} \Lb z'\Rb\nn\\
       &\xrightarrow  {z\,\gg\,1} & \,\,   \frac{C_\phi}{2\,\kappa}\intl^\infty_z dz'e^{ - \frac{(z' - z_\Delta)^2}{2\,\kappa}} \Lb 1 + \ln\Lb z'\Rb\Rb\,=\,\Phi_{asym}\Lb z \Rb \Lb 1 + \ln z\Rb     \eea
Therefore these corrections does not change our solution adding only weak $\ln z $ dependence of $N\Lb z \Rb$.
 ~
   
   ~

      %%%%%%%%%%%%%%%%%%%%%%%%%%%%%%%%%%%%%%%%%%%%%%%%%%%%
\subsection{Discussion of the solution }
%%%%%%%%%%%%%%%%%%%%%%%%%%%%%%%%%%%%%%%%%%%%%%%%%%%%% 

  Concluding this section we infer that each $\sigma_n $ of our solution decreases as $\exp\Lb - z^2/(2\,\kappa)\Rb$ as it follows from the solutions of section III. We consider this property as an attractive feature of the solution. On the other hand 
  the inelastic cross section is equal to 2 at large $z$ instead of 1 as  expected from the unitarity constraints.
  
   This shortcoming is closely  related to that we do not know from which value of $n$ we can trust our solution.  Coming back to \eq{INXS0}, let us introduce  $\sigma^{n_0}_{in}=
   \sum^\infty_{n=n_0} \sigma_n$. For $\sigma^{n_0}_{in}$ \eq{INXS0} takes the form:
    \beq \label{MDLN25}     
  \sigma^{n_0}_{in}\Lb z \Rb\,\,  \xrightarrow{z\gg 1}\,\,2\,\, \exp\Lb-\, \frac{n_0}{N\Lb z \Rb}\Rb
  \eeq     
    From this equation it follows that we can trust our solution for $n_0 \sim N\Lb z \Rb$. Indeed, for such $n_0$  sum of all $\sigma_n$ with $n\,>\,n_0$ will give the cross section smaller than 1.
    
    Hence, we conclude that we need  to solve a difficult problem of matching our solution with the solution for $n \,\leq\,N\Lb z \Rb$, which we have not found. To illustrate the difficulties  we could face in finding such a matching, we note that our solution at large $n$ leads to $\sigma_n$ at $z \to 0$ which are constants with quite different $n$ dependence that the initial conditions of  \eq{MVFXS}.  Indeed, $n$-dependence of our solution at z=0 is $\sigma_n \propto n \exp\Lb - Const\,n\Rb$ instead of  $\propto\,\,\frac{ Const^n}{n!}$.

\begin{figure}[ht]
   \centering
  \leavevmode
      \includegraphics[width=10cm]{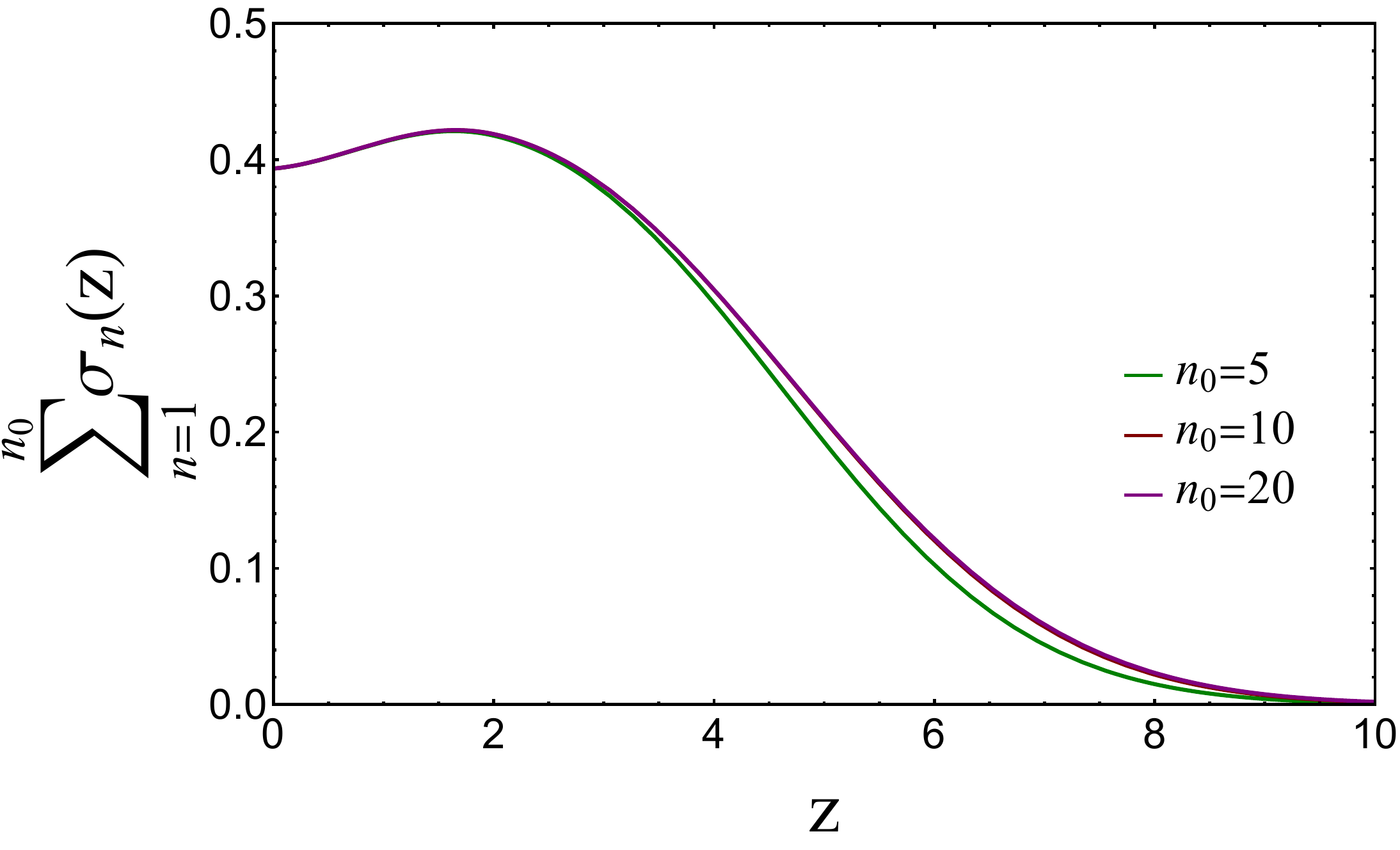}  
      \caption{Plot of $\sum_{n=1}^{n_0}\sigma_n\Lb z\Rb$ for different values of $n_0$.}
      \label{Sigmainn0vsZ}
   \end{figure}

The question arises whether $\sum_{n=1}^{n_0}\sigma_n$ could give the substantial part of the inelastic cross section. As has been discussed in section III $\sigma_n$ at small $n$ in the homotopy approach takes the following form for large $z$.

  \beq \label{MDLN26}  
  \sigma_n\Lb z\Rb \,\,=\,\,\sigma_n^{\Lb 0\Rb}\Lb z,\eq{XSN3}\Rb\Lb1\,\,+\,\,\sum_{p=1}^{p_{\mathrm{max}}}\,\frac{\sigma_n^{\Lb p\Rb}\Lb z,\eq{XSN4}\Rb}{\sigma_n^{\Lb 0\Rb}\Lb z,\eq{XSN3}\Rb}\Rb\,\,\xrightarrow  {z\,\gg\,1}  \,\,\sigma_n^{\Lb 0\Rb}\Lb z,\eq{XSN3}\Rb\eeq
where $p_{\mathrm{max}}$ is the maximum number of iterations taken in the homotopy approach. Using this equation we compute $\sum_{n=1}^{n_0}\sigma_n\Lb z\Rb$ in order to see if it is much less than $1$. As can be seen in \fig{Sigmainn0vsZ}, it  approaches  $0$ for sufficiently large $z$.
   Therefore, it is  expected that  sum over $n \leq n_0$ leads to the  negligible  contribution to the inelastic cross section.  
Hence  in spite of the fact  that we do not know solutions for $n < N\Lb z \Rb$ we obtain correct estimates for the entropy of the produced gluons since this result is based on  the expectation   that the main contributions come from $n \sim N\Lb z\Rb$.

\begin{figure}[ht]
   \centering
  \leavevmode
      \includegraphics[width=10cm]{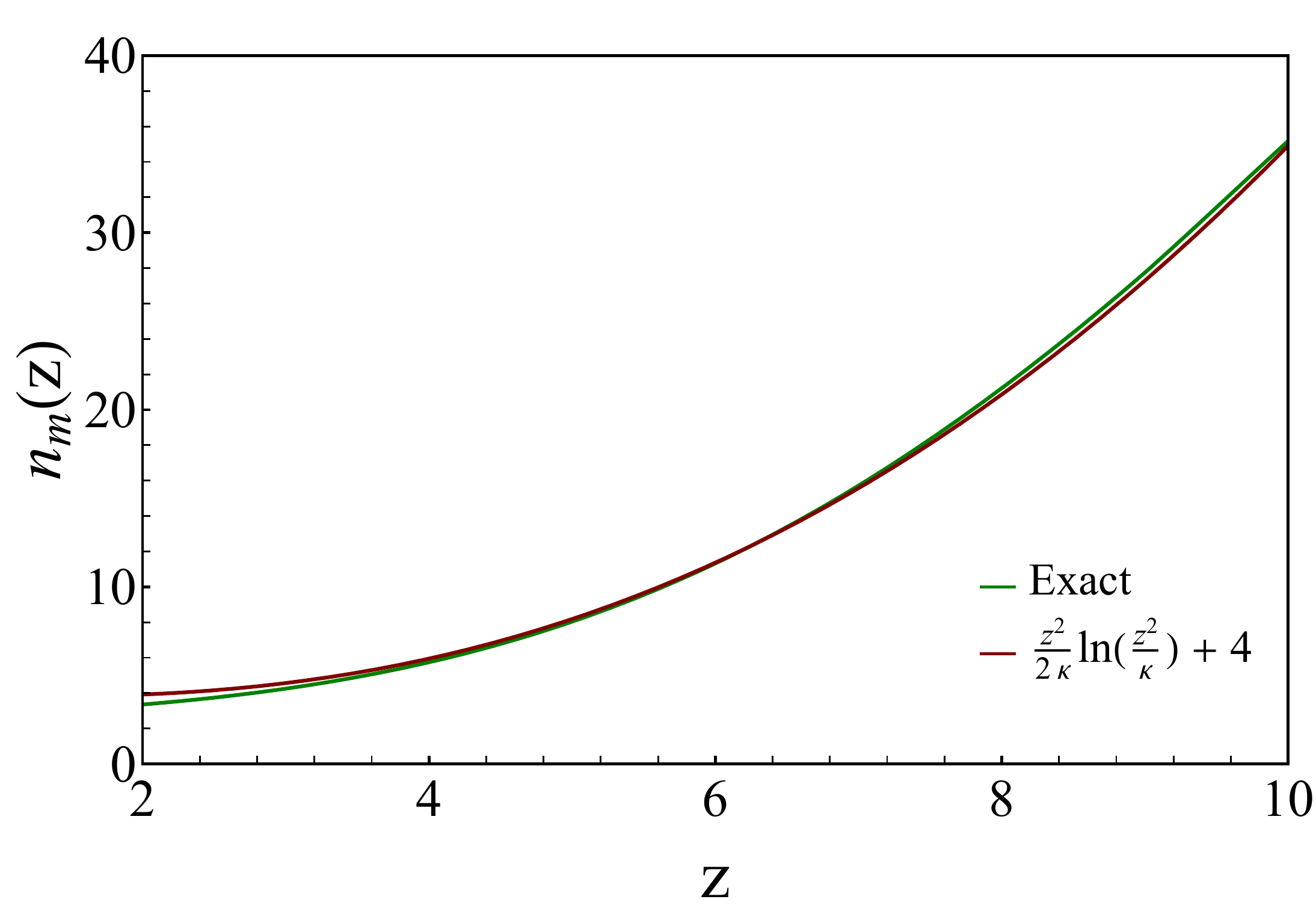}  
      \caption{Solution of \eq{MDLN28} for $\xi^A_0 = 0$ and $z_\Delta=0.87$ (blue line), together with the approximating function (purple line).}
      \label{match}
   \end{figure}

Coming back to the problem of matching of our solution at large $n$ with the solution of \eq{MDLN15} we could estimate the value of $n=n_m$ at which we expect the matching occurs. This value is the solution to the following equation:
 \beq \label{MDLN28}
 \sigma_{\mbox{small $n$}}\,\,= \,\,\sigma_{n_m}^{\Lb 0\Rb}\Lb z,\eq{XSN3}\Rb\Lb1\,\,+\,\,\sum_{p=1}^{p_{\mathrm{max}}}\,\frac{\sigma_{n_m}^{\Lb p\Rb}\Lb z,\eq{XSN4}\Rb}{\sigma_{n_m}^{\Lb 0\Rb}\Lb z,\eq{XSN3}\Rb}\Rb\,\,=\,\,\sigma_{\mbox{large $n$}} \,=\,\frac{2}{ \kappa}\frac{z^2}{N\Lb z\Rb} \Psi\Lb\frac{n_m}{N\Lb z\Rb}\Rb  
 \eeq
\fig{match} shows the value of $n_m(z)$, which is the solution to \eq{MDLN28} for  $\xi^A_0 = 0$ and $z_\Delta=0.87$ (Exact in \fig{match}). This solution is compared in this figure with the approximate analytic solution: $\frac{z^2}{2 \,\kappa}\ln\Lb \frac{z^2}{\kappa}\Rb + 4$.
 
 %%%%%%%%%%%%%%%%%%%%%%%%%%%%%%%%%%%%%%%%%
%%%%%%%%%%%%%%%%%%%%%%%%%%%%%%%%%%%%%%%%%%

Therefore, we see that we can hope for finding a matching procedure at rather large $n$ but much smaller than $N(z)$.

\begin{figure}[ht]
   \centering
  \leavevmode
      \includegraphics[width=10cm]{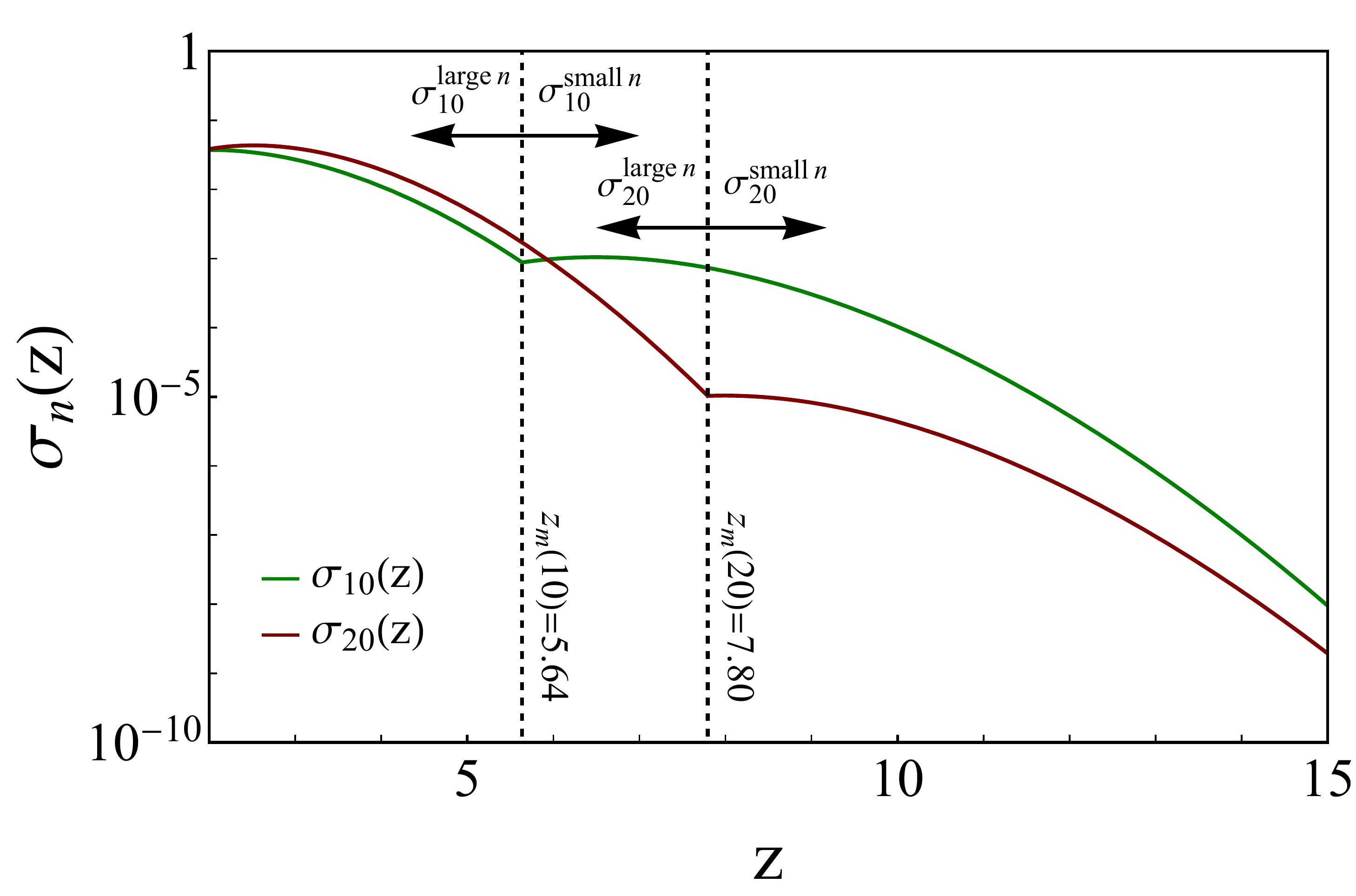}  
      \caption{The cross sections $\sigma_{10}, \sigma_{20}$ showing the transition at $n=n_m$ (see \eq{MDLN29}).}
      \label{fig12}
   \end{figure}

Considering 
\beq\label{MDLN29}
\sigma_{n}(z)\,\,=\,\,\sigma_{\mbox{large  $n$}}(z)\,\Theta(z_m \,-\, z)\,\,+\,\,\sigma_{\mbox{small $n$}}(z)\,\Theta(z \,-\, z_m)
\eeq
in \fig{fig12} we plot $\sigma_{10}$, $\sigma_{20}$ to show explicitly the transition. One can see that we have not solved the problem of matching the solutions at large and small $n$. Note, that $z_m\Lb n\Rb$ in \eq{MDLN29} is the inverse function of $n =  n_m\Lb z\Rb$.

~

      %%%%%%%%%%%%%%%%%%%%%%%%%%%%%%%%%%%%%%%%%%%%%%%%%%%%
\section{Conclusions }
%%%%%%%%%%%%%%%%%%%%%%%%%%%%%%%%%%%%%%%%%%%%%%%%%%%%%

In this paper we discuss the multiplicity distribution in the deep inelastic processes in the framework of high energy QCD. We obtained three results. 
First, we give a new derivation of the equations for the cross sections of productions of $n$-cut BFKL Pomerons in the final states ($\sigma_n$, see \eq{ME7} and \fig{eveq}). This derivation is carried out  in the framework of the dipole approach to high energy QCD  and it  is not based on the Abramovsky, Gribov and Kancheli cutting rules but reproduces them.

 Second, we developed the homotopy approach for finding the solutions to these equations. 
 It consists of the analytic solution for the first iteration and the convergent  procedure of calculating the next iterations using computing.  We demonstrated that we need three iterations to obtain the solution within 0.2\% accuracy. Using this approach we can calculate several $\sigma^{\mbox{\tiny AGK}}_n$ with small $n = 1,2, \dots$. Having these cross sections we can calculate the moments of our multiplicity distributions \cite{MUMULT,MGF} and obtain the multiplicity distribution  $P_n\Lb z\Rb$ since
 \beq \label{C1}
P_n\Lb z\Rb\,\,=\,\,\oint\frac{d \lambda}{2\,\pi\,i} \frac{e^{ f\Lb \lambda\Rb}}{\lambda^{n + 1}}
\eeq 
 where the contour of  integration  is the circle around the point  $\lambda =0$  and  $f\Lb \lambda\Rb$ is the cumulant generating function, which  is defined as
\beq \label{CGF}
f\Lb \lambda\Rb\,\,=\,\,\sum^\infty_{n=1} \frac{C_n }{n!} \Lb \lambda - 1\Rb^n
\eeq
where $C_n$ are cumulants \cite{STAR}. The first cumulant $C_1 =\bar{n}$ is the average number of produced gluons, the second is the variance and the third is the third central moment of our distributions (see  Ref.\cite{MGF,STAR} for more information).  All cumulants can be calculated in the BFKL Pomeron approach, using that $\sigma^{\mbox{\tiny AGK}}_k\Lb z \Rb $ \footnote{Recall that $\sigma_k$ is given at fixed $b$ which enters the definition of $z$.} gives the $k$-th moment of the multiplicity distribution ($ \sigma^{\mbox{\tiny AGK}}_k\Lb z \Rb 
= < n^k>n_{\pom}^k$\footnote{$n_{\pom}$ is the multiplicity of the produced gluons in the BFKL Pomeron.})\cite{AGK,MUMULT}.
The distribution with the first two cumulants are given in Ref.\cite{MUMULT}.  Hence, using the procedure   suggested in this paper,  we can describe the experimental data using several $\sigma^{\mbox{\tiny AGK}}_n$.  This approach will be the subject of our future publications.

 Third, we found an analytical solution for  $\sigma_n$  at large  $n\,\gtrsim\,N\Lb z\Rb =    
     \frac{2\,z}{C_\phi}\,e^{ \frac{\Lb z-z_\Delta\Rb^2}{2\,\kappa}}$ (see  \eq{MDLN17})  . This solution has two remarkable features: (i) it decreases as $\exp\Lb -\frac{\Lb  z - z_\Delta\Rb^2}{2\,\kappa}\Rb$ at large $z$; and (ii) it  gives $\sigma_{in} \to Const$ at large $z$. In spite of  this,  $Const$ turns out to be larger than 1 that  comes from the $s$ channel unitarity constraints, we argue that we can trust this solution for $n \geq N\Lb z \Rb$.The first property we observed in section III from the homotopy approach to finding the solution to the main equations.

  The problem that has not been solved in the paper is the matching of  our solution with the initial condition of \eq{MVFXS}. $ \sigma_n$ in  our solution tends to the value
   at $z \to 0$  which has a different $n$ dependence than the initial conditions. 
   In section IV-E we argued that the matching occurs at large multiplicities  proportional to $z^2/(2\,\kappa)$, which however are much smaller than $N(z)$. Therefore, we need to find a solution for small $n$, which has been discussed in section III, at rather large $n$.  We expect that this  will be a difficult problem.  
  
Finally, our  solution leads to the value of the entropy of the produced gluons $S_E = \ln \Lb N\Lb z\Rb\Rb  $ which confirms the result of Ref.\cite{KHLE,LEMULT,GOLEMULT}.

   We hope that this paper will contribute to the  gradual but inevitable understanding the main features of the Pomeron calculus approach to high energy QCD.

~

~
 
{\bf  Acknowledgements}
 
   We thank our colleagues at Tel Aviv University and UTFSM for encouraging discussions. This research was
supported by Fondecyt (Chile) grants No. 1231829 and 1231062. J. G. acknowledges support of the fellowship 
``Beca de T\'ermino de Tesis de Doctorado $N^o$ 079/2025'' of Direcci\'on de Postgrado (DP), UTFSM, and expresses his gratitude to the Institute of Physics of PUCV (IFIS).


\begin{thebibliography}{99} \frenchspacing
 
\bibitem{CLMNEW}
C.~Contreras, E.~Levin and R.~Meneses,
%``Homotopy solution to non-linear evolution for heavy nuclei,''
\href{https://journals.aps.org/prd/abstract/10.1103/PhysRevD.107.094030}{Phys. Rev. D \textbf{107} (2023) 094030},
[\href{https://arxiv.org/abs/2302.10497}{2302.10497}].

\bibitem{CGLM}
C.~Contreras, J. ~Garrido, ~E.~Levin and R.~Meneses,
%``Modified homotopy approach for diffractive production in the saturation region,"
\href{https://journals.aps.org/prd/abstract/10.1103/PhysRevD.110.054045}{Phys. Rev. D \textbf{110} (2024) 054045},
[\href{https://arxiv.org/abs/2406.11673}{2406.11673}].

%\cite{Contreras:2025zsc}
\bibitem{CGL}
C.~Contreras, J.~Garrido and E.~Levin,
%``Homotopy approach for scattering amplitude for running QCD coupling,''
\href{https://journals.aps.org/prd/abstract/10.1103/PhysRevD.111.096025}{Phys. Rev. D \textbf{111} (2025) 096025}, [\href{https://arxiv.org/abs/2503.19771}{2503.19771}].

\bibitem{KUT}
K.~Kutak,
%{\it ``Gluon saturation and entropy production in proton-proton collisions,''}
\href{https://www.sciencedirect.com/science/article/pii/S0370269311012135?via%3Dihub}{Phys. Lett. \textbf{B705} (2011) 217--221}, 
[\href{https://arxiv.org/abs/1103.3654}{1103.3654}].

\bibitem{PES}
R.~Peschanski,
%{\it ``Dynamical entropy of dense QCD states,''}
\href{https://journals.aps.org/prd/abstract/10.1103/PhysRevD.87.034042}{Phys. Rev. D \textbf{87} (2013) 034042}, 
[\href{https://arxiv.org/abs/1211.6911}{1211.6911}].

\bibitem{KOLU1}
A.~Kovner and M.~Lublinsky,
%{\it ``Entanglement entropy and entropy production in the Color Glass Condensate framework,''}
\href{https://journals.aps.org/prd/abstract/10.1103/PhysRevD.92.034016}{Phys. Rev. D \textbf{92} (2015) 034016}, 
[\href{https://arxiv.org/abs/1506.05394}{1506.05394}].

\bibitem{PESE}
R.~Peschanski and S.~Seki,
%{\it ``Entanglement Entropy of Scattering Particles,''}
\href{https://www.sciencedirect.com/science/article/pii/S0370269316301423?via%3Dihub}{Phys. Lett. \textbf{B758} (2016) 89--92}, 
[\href{https://arxiv.org/abs/1602.00720}{1602.00720}].

\bibitem{KHLE}
D.~E.~Kharzeev and E.~M.~Levin,
%{\it ``Deep inelastic scattering as a probe of entanglement,''}
\href{https://journals.aps.org/prd/abstract/10.1103/PhysRevD.95.114008}{Phys. Rev. D \textbf{95} (2017) 114008}, 
[\href{https://arxiv.org/abs/1702.03489}{1702.03489}].

\bibitem{BAKH}
O.~Baker and D.~Kharzeev,
%{\it ``Thermal radiation and entanglement in proton-proton collisions at energies available at the CERN Large Hadron Collider,''}
\href{https://journals.aps.org/prd/abstract/10.1103/PhysRevD.98.054007}{Phys. Rev. D \textbf{98} (2018) 054007}, 
[\href{https://arxiv.org/abs/1712.04558}{1712.04558}].

\bibitem{BFV}
J.~Berges, S.~Floerchinger and R.~Venugopalan,
%{\it ``Dynamics of entanglement in expanding quantum fields,''}
\href{https://link.springer.com/article/10.1007/JHEP04(2018)145}{JHEP \textbf{04} (2018) 145}, 
[\href{https://arxiv.org/abs/1712.09362}{1712.09362}].

\bibitem{HHXY}
Y.~Hagiwara, Y.~Hatta, B.~W.~Xiao and F.~Yuan,
%{\it ``Classical and quantum entropy of parton distributions,''}
\href{https://journals.aps.org/prd/abstract/10.1103/PhysRevD.97.094029}{Phys. Rev. D \textbf{97} (2018) 094029}, 
[\href{https://arxiv.org/abs/1801.00087}{1801.00087}].

\bibitem{KOV1}
N.~Armesto, F.~Dominguez, A.~Kovner, M.~Lublinsky and V.~Skokov,
%{\it ``The Color Glass Condensate density matrix: Lindblad evolution, entanglement entropy and Wigner functional,''}
\href{https://link.springer.com/article/10.1007/JHEP05(2019)025}{JHEP \textbf{05} (2019) 025}, 
[\href{https://arxiv.org/abs/1901.08080}{1901.08080}].

\bibitem{GOLE1}
E.~Gotsman and E.~Levin,
%{\it ``Thermal radiation and inclusive production in the CGC/saturation approach at high energies,''}
\href{https://link.springer.com/article/10.1140/epjc/s10052-019-6923-0}{Eur. Phys. J. C \textbf{79} (2019) 415}, 
[\href{https://arxiv.org/abs/1902.07923}{1902.07923}].

\bibitem{GOLE2}
E.~Gotsman and E.~Levin,
%{\it ``Thermal radiation and inclusive production in the Kharzeev-Levin-Nardi model for ion-ion collisions,''}
\href{https://journals.aps.org/prd/abstract/10.1103/PhysRevD.100.034013}{Phys. Rev. D \textbf{100} (2019) 034013}, 
[\href{https://arxiv.org/abs/1905.05167}{1905.05167}].

\bibitem{KOV2}
A.~Kovner, M.~Lublinsky and M.~Serino,
%{\it ``Entanglement entropy, entropy production and time evolution in high energy QCD,''}
\href{https://www.sciencedirect.com/science/article/pii/S0370269318308141}{Phys. Lett. \textbf{B792} (2019) 4--15}, 
[\href{https://arxiv.org/abs/1806.01089}{1806.01089}].

\bibitem{NEWA}
D.~Neill and W.~J.~Waalewijn,
%{\it ``Entropy of a Jet,''}
\href{https://journals.aps.org/prl/abstract/10.1103/PhysRevLett.123.142001}{Phys. Rev. Lett. \textbf{123} (2019) 142001}, 
[\href{https://arxiv.org/abs/1811.01021}{1811.01021}].

\bibitem{LIZA}
Y.~Liu and I.~Zahed,
%{\it ``Entanglement in Regge scattering using the AdS/CFT correspondence,''}
\href{https://journals.aps.org/prd/abstract/10.1103/PhysRevD.100.046005}{Phys. Rev. D \textbf{100} (2019) 046005}, 
[\href{https://arxiv.org/abs/1803.09157}{1803.09157}].

\bibitem{FPV}
X.~Feal, C.~Pajares and R.~Vazquez,
%{\it ``Thermal behavior and entanglement in Pb-Pb and p-p collisions,''}
\href{https://journals.aps.org/prc/abstract/10.1103/PhysRevC.99.015205}{Phys. Rev. C \textbf{99} (2019) 015205}, 
[\href{https://arxiv.org/abs/1805.12444}{1805.12444}].

\bibitem{TKU}
Z.~Tu, D.~E.~Kharzeev and T.~Ullrich,
%{\it ``Einstein-Podolsky-Rosen Paradox and Quantum Entanglement at Subnucleonic Scales,''}
\href{https://journals.aps.org/prl/abstract/10.1103/PhysRevLett.124.062001}{Phys. Rev. Lett. \textbf{124} (2020) 062001}, 
[\href{https://arxiv.org/abs/1904.11974}{1904.11974}].

\bibitem{KOV3}
H.~Duan, C.~Akkaya, A.~Kovner and V.~V.~Skokov,
%{\it ``Entanglement, partial set of measurements, and diagonality of the density matrix in the parton model,''}
\href{https://journals.aps.org/prd/abstract/10.1103/PhysRevD.101.036017}{Phys. Rev. \textbf{D101} (2020) 036017}, 
[\href{https://arxiv.org/abs/2001.01726}{2001.01726}].

\bibitem{KOV4}
C.~Akkaya and A.~Kovner, [\href{https://arxiv.org/abs/2007.15970}{2007.15970}].

\bibitem{DVA1}
G.~Dvali,
%``Entropy Bound and Unitarity of Scattering Amplitudes,''
\href{https://link.springer.com/article/10.1007/JHEP03(2021)126}{JHEP \textbf{03} (2021) 126}, 
[\href{https://arxiv.org/abs/2003.05546}{2003.05546}].

\bibitem{HE1}
	J.H. He, 
	%\emph{Homotopy perturbation technique},
	\href{https://www.sciencedirect.com/science/article/abs/pii/S0045782599000183}{Comput. Methods Appl. Mech. Engrg. {\bf 178} (1999) 257--262}.  
	
\bibitem{HE2}
	J.H. He, 
	%\emph{A coupling method of homotopy technique and a perturbation technique for nonlinear problems},
	 \href{https://www.sciencedirect.com/science/article/abs/pii/S0020746298000857?via%3Dihub}{Int. J. Nonlinear Mech. {\bf 35} (2000) 37--43}.
 
\bibitem{AGK}
V.~A.~Abramovsky, V.~N.~Gribov and O.~V.~Kancheli,
%``Character of Inclusive Spectra and Fluctuations Produced in Inelastic Processes by Multi - Pomeron Exchange,''
Yad. Fiz. \textbf{18} (1973) 595--616, [Sov. J. Nucl. Phys. \textbf{18} (1974) 308--317].

\bibitem{KNO} A.M. Polyakov, 
        Zh. Eksp. Teor. Fiz. {\bf 59} (1970) 542--552.
        
\bibitem{KNO1} Z. Koba, H.B. Nielsen and P. Olesen, 
	\href{https://www.sciencedirect.com/science/article/pii/0550321372905512?via%3Dihub}{Nucl. Phys. {\bf B40} (1972) 317--334}. 
	
\bibitem{KNO2} Z. Koba,
	in Proc. of the \href{https://cds.cern.ch/record/186268?ln=en}{1973 CERN School of Physics,
	p.~171, CERN Yellow Report CERN-73-12 (1973)}.

 \bibitem{KLP}
A.~Kormilitzin, E.~Levin and A.~Prygarin,
%``Multiparticle production in the mean field approximation of high density QCD,''
\href{https://www.sciencedirect.com/science/article/abs/pii/S0375947408006957?via%3Dihub}{Nucl. Phys. \textbf{A813} (2008) 1--13}, 
[\href{https://arxiv.org/abs/0807.3413}{0807.3413}].

\bibitem{KOLEB}
Yuri V. Kovchegov and Eugene Levin, \href{https://www.cambridge.org/us/universitypress/subjects/physics/particle-physics-and-nuclear-physics/quantum-chromodynamics-high-energy?format=HB}{{\it Quantum Chromodynamics at High Energies}}, Cambridge Monographs on Particle Physics, Nuclear Physics and Cosmology, Cambridge University Press, Cambridge, England, 2012.  

\bibitem{BK}
I.~Balitsky,
%``Factorization and high-energy effective action,''
\href{https://journals.aps.org/prd/abstract/10.1103/PhysRevD.60.014020}{{Phys.\ Rev.} {\bf D60} (1999) 014020}, [\href{https://arxiv.org/abs/hep-ph/9812311}{hep-ph/9812311}];\,\,
Y.~V.~Kovchegov,
%``Small x F(2) structure function of a nucleus including multiple pomeron exchanges,''
\href{https://journals.aps.org/prd/abstract/10.1103/PhysRevD.60.034008}{{Phys.\ Rev.}  {\bf D60} (1999) 034008}, [\href{https://arxiv.org/abs/hep-ph/9901281}{hep-ph/9901281}].

\bibitem{MUT}
A.~H.~Mueller and D.~N.~Triantafyllopoulos, \href{https://www.sciencedirect.com/science/article/abs/pii/S0550321302005813?via%3Dihub}{{Nucl.\ Phys.} {\bf B640} (2002) 331}, [\href{https://arxiv.org/abs/hep-ph/0205167}{hep-ph/0205167}];\,\,D.~N.~Triantafyllopoulos,
\href{https://www.sciencedirect.com/science/article/abs/pii/S0550321302010003?via%3Dihub}{{Nucl.\ Phys.}  {\bf B648} (2003) 293}, 
[\href{https://arxiv.org/abs/hep-ph/0209121}{hep-ph/0209121}].

\bibitem{MUPE}
S.~Munier and R.~B.~Peschanski,
%``Traveling wave fronts and the transition to saturation,''
\href{https://journals.aps.org/prd/abstract/10.1103/PhysRevD.69.034008}{Phys. Rev. D \textbf{69} (2004) 034008}, 
[\href{https://arxiv.org/abs/hep-ph/0310357}{hep-ph/0310357}];\,\,
%``Geometric scaling as traveling waves,''
\href{https://journals.aps.org/prl/abstract/10.1103/PhysRevLett.91.232001}{Phys. Rev. Lett. \textbf{91} (2003) 232001}, 
[\href{https://arxiv.org/abs/hep-ph/0309177}{hep-ph/0309177}].      
                
\bibitem{GLR} 
L.~V.~Gribov, E.~M.~Levin and M.~G.~Ryskin,
  %{\it ``Semihard Processes in QCD,''}
  \href{https://www.sciencedirect.com/science/article/abs/pii/0370157383900224?via%3Dihub}{Phys.\ Rept.\  {\bf 100} (1983) 1--150}.

\bibitem{BALE}
J.~Bartels and E.~Levin,
%``Solutions to the Gribov-Levin-Ryskin equation in the nonperturbative region,''
\href{https://www.sciencedirect.com/science/article/abs/pii/055032139290209T?via%3Dihub}{Nucl. Phys. \textbf{B387} (1992) 617--637}.

\bibitem{GS}
A.~M.~Stasto, K.~J.~Golec-Biernat and J.~Kwiecinski,
%``Geometric scaling for the total gamma* p cross-section in the low x region,''
\href{https://journals.aps.org/prl/abstract/10.1103/PhysRevLett.86.596}{Phys. Rev. Lett. \textbf{86} (2001) 596--599}, 
[\href{https://arxiv.org/abs/hep-ph/0007192}{hep-ph/0007192}];\,\,
L.~McLerran and M.~Praszalowicz,
%``Saturation and Scaling of Multiplicity, Mean $p_T$ and $p_T$ Distributions from 200 GeV \ensuremath{<} sqrt{s} \ensuremath{<} 7 TeV - Addendum,''
\href{https://www.actaphys.uj.edu.pl/index_n.php?I=R&V=42&N=1#99}{Acta Phys. Polon. \textbf{B42} (2011) 99--103}, 
[\href{https://arxiv.org/abs/1011.3403}{1011.3403}];\,\,
%``Geometrical Scaling and the Dependence of the Average Transverse Momentum on the Multiplicity and Energy for the ALICE Experiment,''
\href{https://www.sciencedirect.com/science/article/pii/S0370269314009277?via%3Dihub}{Phys. Lett. \textbf{B741} (2015) 246--251}, 
[\href{https://arxiv.org/abs/1407.6687}{1407.6687}].

\bibitem{LETU}
E.~Levin and K.~Tuchin,
  %``Solution to the evolution equation for high parton density QCD,''
  \href{https://www.sciencedirect.com/science/article/abs/pii/S0550321399008251?via%3Dihub}{Nucl.\ Phys.\ {\bf B573} (2000) 833--852}, 
  [\href{https://arxiv.org/abs/hep-ph/9908317}{hep-ph/9908317}];\,\,
%``New scaling at high-energy DIS,''
  \href{https://www.sciencedirect.com/science/article/abs/pii/S0375947401005905?via%3Dihub}{{\bf A691} (2001) 779--790}, 
  [\href{https://arxiv.org/abs/hep-ph/0012167}{hep-ph/0012167}];\,\, %``Nonlinear evolution and saturation for heavy nuclei in DIS,''
  \href{https://www.sciencedirect.com/science/article/abs/pii/S0375947401008806?via%3Dihub}{{\bf A693} (2001) 787--798}, 
  [\href{https://arxiv.org/abs/hep-ph/0101275}{hep-ph/0101275}].

 \bibitem{LIP}
 L.~N.~Lipatov,
% {\it ``The Bare Pomeron in Quantum Chromodynamics,''}
  Sov.\ Phys.\ JETP {\bf 63}, 904 (1986)
  [Zh.\ Eksp.\ Teor.\ Fiz.\  {\bf 90}, 1536 (1986)].
  
 \bibitem{LIPREV}
                L.~N.~Lipatov,
  %``Small x physics in perturbative QCD,''
  \href{https://www.sciencedirect.com/science/article/pii/S0370157396000452?via%3Dihub}{Phys.\ Rept.\  {\bf 286} (1997) 131--198}, [\href{https://arxiv.org/abs/hep-ph/9610276}{hep-ph/9610276}].
 
\bibitem{GOST}
K.~J.~Golec-Biernat and A.~M.~Stasto,
%``On solutions of the Balitsky-Kovchegov equation with impact parameter,''
\href{https://www.sciencedirect.com/science/article/abs/pii/S0550321303005935?via%3Dihub}{Nucl. Phys. \textbf{B668} (2003) 345--363}, 
[\href{https://arxiv.org/abs/hep-ph/0306279}{hep-ph/0306279}].

\bibitem{BEST}
J.~Berger and A.~Stasto,
%``Numerical solution of the nonlinear evolution equation at small x with impact parameter and beyond the LL approximation,''
\href{https://journals.aps.org/prd/abstract/10.1103/PhysRevD.83.034015}{Phys. Rev. D \textbf{83} (2011) 034015}, 
[\href{https://arxiv.org/abs/1010.0671}{1010.0671}].

\bibitem{BFKL}
   V.~S. Fadin, E.~A. Kuraev and L.~N. Lipatov,
%{\it ``On the pomeranchuk singularity in asymptotically free theories"},
\newblock \href{https://www.sciencedirect.com/science/article/abs/pii/0370269375905249?via%3Dihub}{Phys. Lett. {\bf B60} (1975) 50--52};\,\,
E.~A. Kuraev, L.~N. Lipatov and V.~S. Fadin,
%{\it``The Pomeranchuk Singularity in Nonabelian Gauge Theories"}
\newblock Sov. Phys. JETP {\bf 45} (1977) 199--204,
\newblock [Zh. Eksp. Teor. Fiz. 72 (1977) 377--389];\,\,
%%CITATION = SPHJA,45,199;%
%%CITATION = PHLTA,B60,50;%%
I.~I. Balitsky and L.~N. Lipatov,
%  {\it ``The Pomeranchuk Singularity in Quantum Chromodynamics,''}
\newblock Sov. J. Nucl. Phys. {\bf 28} (1978) 822--829,
\newblock [Yad. Fiz. 28 (1978) 1597--1611].       

 \bibitem{KOLE}
  Y.~V.~Kovchegov and E.~Levin,
  %``Diffractive dissociation including multiple pomeron exchanges in high parton density QCD,''
  \href{https://www.sciencedirect.com/science/article/abs/pii/S0550321300001255?via%3Dihub}{Nucl.\ Phys.\ {\bf B577} (2000) 221--239}, [\href{https://arxiv.org/abs/hep-ph/9911523}{hep-ph/9911523}].
  
  \bibitem{LEHP}
E.~Levin,
%``'Hard' pomeron approach to 'soft' processes at high-energy,''
\href{https://journals.aps.org/prd/abstract/10.1103/PhysRevD.49.4469}{Phys. Rev. D \textbf{49} (1994) 4469--4480}.

\bibitem{GBW}
 K. Golec-Biernat and M. Wuesthoff, \href{https://journals.aps.org/prd/abstract/10.1103/PhysRevD.59.014017}{Phys. Rev. D 59 (1998) 014017}, [\href{https://arxiv.org/abs/hep-ph/9807513}{hep-ph/9807513}].

  \bibitem{MV}
L. McLerran and R. Venugopalan, 
%{\it ``Computing quark and gluon distribution functions for very large nuclei"},
\href{https://journals.aps.org/prd/abstract/10.1103/PhysRevD.49.2233}{Phys. Rev. {\bf D49} (1994) 2233--2241}, [\href{https://arxiv.org/abs/hep-ph/9309289}{hep-ph/9309289}];\,\,
%{\it ``Gluon distribution functions for very large nuclei at small transverse momentum"}, 
\href{https://journals.aps.org/prd/abstract/10.1103/PhysRevD.49.3352}{{\bf D49} (1994) 3352--3355}, [\href{https://arxiv.org/abs/hep-ph/9311205}{hep-ph/9311205}];\,\,
%  {\it `Green's function in the color field of a large nucleus"}, 
\href{https://journals.aps.org/prd/abstract/10.1103/PhysRevD.50.2225}{{\bf D50} (1994) 2225--2233}, [\href{https://arxiv.org/abs/hep-ph/9402335}{hep-ph/9402335}];\,\,
% {\it ``Fock space distributions, structure functions, higher twists, and small $x$"} ,
 \href{https://journals.aps.org/prd/abstract/10.1103/PhysRevD.59.094002}{{\bf D59} (1999) 09400}, [\href{https://arxiv.org/abs/hep-ph/9809427}{hep-ph/9809427}]. 

   
\bibitem{MUDI}
  A.~H.~Mueller,
  %``Soft gluons in the infinite momentum wave function and the BFKL pomeron,''
  \href{https://www.sciencedirect.com/science/article/pii/0550321394901163?via%3Dihub}{Nucl.\ Phys. {\bf B415} (1994) 373--385};\,\,
  %``Unitarity and the BFKL pomeron,''
  \href{https://www.sciencedirect.com/science/article/pii/0550321394004803?via%3Dihub}{{\bf B437} (1995) 107--126}, [\href{https://arxiv.org/abs/hep-ph/9408245}{hep-ph/9408245}];\,\, A.~H.~Mueller and B.~Patel, 
      %{\it ``Single and double BFKL pomeron exchange and a dipole picture of high-energy hard processes",}      
      \href{https://www.sciencedirect.com/science/article/pii/0550321394902844?via%3Dihub}{Nucl. Phys. {\bf B425} (1994) 471--488}, [\href{https://arxiv.org/abs/hep-ph/9403256}{hep-ph/9403256}].
      
  \bibitem{MUSA}
A.~H.~Mueller and G.~P.~Salam,
  \href{https://www.sciencedirect.com/science/article/abs/pii/0550321396003367?via%3Dihub}{Nucl.\ Phys. {\bf B475} (1996) 293--317},  [\href{https://arxiv.org/abs/hep-ph/9605302}{hep-ph/9605302}];\,\,
  G.~P.~Salam,
 % {\it ``Studies of unitarity at small x using the dipole formulation,''}
  \href{https://www.sciencedirect.com/science/article/abs/pii/0550321395006583?via%3Dihub}{Nucl.\ Phys. {\bf B461} (1996) 512--538}, [\href{https://arxiv.org/abs/hep-ph/9509353}{hep-ph/9509353}].

\bibitem{MUQI}
A. H. Mueller and J. Qiu, 
%{\it ``  Gluon recombination and shadowing at small values of $x$",} 
\href{https://www.sciencedirect.com/science/article/pii/0550321386901641?via%3Dihub}{Nucl. Phys. {\bf B268} (1986) 427--452}.

 \bibitem{RY}
I. Gradstein and I. Ryzhik, {\it  Table of Integrals, Series, and Products},
Fifth Edition, Academic Press, London, 1994.

  \bibitem{CHMU}
Z.~Chen and A.~H.~Mueller,
%``The Dipole picture of high-energy scattering, the BFKL equation and many gluon compound states,''
\href{https://www.sciencedirect.com/science/article/pii/0550321395003502?via%3Dihub}{Nucl. Phys. \textbf{B451} (1995) 579--604}. 

\bibitem{LEPRI}
  E.~Levin and A.~Prygarin,
  %``The BFKL Pomeron Calculus in zero transverse dimension: Summation of the Pomeron loops and the generating functional for the multiparticle production processes,''
  \href{https://link.springer.com/article/10.1140/epjc/s10052-007-0458-5}{Eur.\ Phys.\ J.\ C {\bf 53} (2008) 385--399}, [\href{https://arxiv.org/abs/hep-ph/0701178}{hep-ph/0701178}].
  
  \bibitem{LEUTM}
E.~Levin,
%``Particle production in a toy model: Multiplicity distribution and entropy,''
\href{https://journals.aps.org/prd/abstract/10.1103/PhysRevD.111.016019}{Phys. Rev. D \textbf{111} (2025) 016019}, 
[\href{https://arxiv.org/abs/2412.02504}{2412.02504}] and references therein.

\bibitem{KO01}
  Y.~V.~Kovchegov,
  %``Diffractive gluon production in proton nucleus collisions and in DIS,''
  \href{https://journals.aps.org/prd/abstract/10.1103/PhysRevD.64.114016}{Phys.\ Rev.\  D {\bf 64} (2001) 114016}
  [\href{https://journals.aps.org/prd/abstract/10.1103/PhysRevD.68.039901}{Erratum-ibid.\  D {\bf 68} (2003) 039901}], 
  [\href{https://arxiv.org/abs/hep-ph/0107256}{hep-ph/0107256}].

\bibitem{KOTU01}
  Y.~V.~Kovchegov and K.~Tuchin,
  %``Inclusive gluon production in DIS at high parton density,''
  \href{https://journals.aps.org/prd/abstract/10.1103/PhysRevD.65.074026}{Phys.\ Rev.\  D {\bf 65} (2002) 074026}, 
  [\href{https://arxiv.org/abs/hep-ph/0111362}{hep-ph/0111362}].

\bibitem{JMKO}
  J.~Jalilian-Marian and Y.~V.~Kovchegov,
  %``Inclusive two-gluon and valence quark-gluon production in DIS and p A,''
  \href{https://journals.aps.org/prd/abstract/10.1103/PhysRevD.70.114017}{Phys.\ Rev.\  D {\bf 70} (2004) 114017}
  [\href{https://journals.aps.org/prd/abstract/10.1103/PhysRevD.71.079901}{Erratum-ibid.\  D {\bf 71} (2005) 079901}],
  [\href{https://arxiv.org/abs/hep-ph/0405266}{hep-ph/0405266}].

%\cite{Braun:2006wj}
\bibitem{BRAGK}
  M.~A.~Braun,
  %``On the inclusive gluon jet production from the triple pomeron vertex in
  %the perturbative QCD,''
  \href{https://link.springer.com/article/10.1140/epjc/s10052-006-0030-8}{Eur.\ Phys.\ J.\  C {\bf 48} (2006) 501--510},
  [\href{https://arxiv.org/abs/hep-ph/0603060}{hep-ph/0603060}].

\bibitem{MAAGK}
  C.~Marquet,
  %``A QCD dipole formalism for forward-gluon production,''
  \href{https://www.sciencedirect.com/science/article/abs/pii/S0550321304008946?via%3Dihub}{Nucl.\ Phys. {\bf B705} (2005) 319--338}, 
  [\href{https://arxiv.org/abs/hep-ph/0409023}{hep-ph/0409023}].

\bibitem{KOLUAGK}
  A.~Kovner and M.~Lublinsky,
  %``One gluon, two gluon: Multigluon production via high energy evolution,''
  JHEP {\bf 0611} (2006) 083, 
  [\href{https://iopscience.iop.org/article/10.1088/1126-6708/2006/11/083}{hep-ph/0609227}].

\bibitem{LEPRAGK}
E.~Levin and A.~Prygarin,
%``Inclusive gluon production in the dipole approach: AGK cutting rules,''
\href{https://journals.aps.org/prc/abstract/10.1103/PhysRevC.78.065202}{Phys. Rev. C \textbf{78} (2008) 065202},
[\href{https://arxiv.org/abs/0804.4747}{0804.4747}].

\bibitem{LEMULT}
E.~Levin,
%``Multiplicity distribution and entropy of produced gluons in deep inelastic scattering at high energies,''
\href{https://link.springer.com/article/10.1140/epjc/s10052-024-13008-w}{Eur. Phys. J. C \textbf{84} (2024) no.7, 662}, [\href{https://arxiv.org/abs/2306.12055}{2306.12055}].

\bibitem{GOLEMULT}
E.~Gotsman and E.~Levin,
%``High energy QCD: multiplicity dependence of quarkonia production,''
\href{https://link.springer.com/article/10.1140/epjc/s10052-020-08775-1}{Eur. Phys. J. C \textbf{81} (2021) no.2, 99},
[\href{https://arxiv.org/abs/2008.10911}{2008.10911}].

 \bibitem{MUMULT}
A.~H.~Mueller,
%{\it``Multiplicity Distributions in Regge Pole Dominated Inclusive Reactions,''}
\href{https://journals.aps.org/prd/abstract/10.1103/PhysRevD.4.150}{Phys. Rev. D \textbf{4} (1971) 150--155}.

\bibitem{MGF}
Beran, J. and Ghosh, S. (2011), \href{https://link.springer.com/rwe/10.1007/978-3-642-04898-2_375}{{\it Moment Generating Function}}, In: Lovric M. (eds) International Encyclopedia of Statistical Science, Springer Berlin, Heidelberg.

\bibitem{STAR}
M.~Abdallah \textit{et al.} [STAR],
%``Higher-order cumulants and correlation functions of proton multiplicity distributions in sNN=3~GeV~Au+Au collisions at the RHIC STAR experiment,''
\href{https://journals.aps.org/prc/abstract/10.1103/PhysRevC.107.024908}{Phys. Rev. C \textbf{107} (2023) 024908}, 
[\href{https://arxiv.org/abs/2209.11940}{2209.11940}].

\end{thebibliography}
\end{document}